\documentclass[fleqn,usenatbib]{mnras}

\usepackage{newtxtext,newtxmath}
\usepackage[T1]{fontenc}
\DeclareRobustCommand{\VAN}[3]{#2}
\let\VANthebibliography\thebibliography
\def\thebibliography{\DeclareRobustCommand{\VAN}[3]{##3}\VANthebibliography}

\usepackage{graphicx}	% Including figure files
\usepackage{amsmath}	% Advanced maths commands
\usepackage[colorinlistoftodos, color=green!40, prependcaption]{todonotes}
\usepackage{subcaption}
\usepackage{tabularx}
\usepackage{multirow}
\usepackage{hhline}
\usepackage{float}

\newcommand{\HL}{Hubble-Lema\^{i}tre }
\newcolumntype{P}[1]{>{\centering\arraybackslash}p{#1}}
\newcommand{\msun}{\mathrm{\, M_\odot}}

\title[Gray sirens to resolve $H_0$ tension]{Using Gray Sirens to Resolve the Hubble-Lema\^{i}tre Tension}

\author[I. Gupta]{
Ish Gupta\thanks{E-mail: ishgupta@psu.edu}
\\
Institute for Gravitation and the Cosmos, Department of Physics, Pennsylvania State University, University Park, PA 16802, USA\\
}

\date{Accepted 2023 July 10. Received 2023 May 9; in original form 2023 January 31}

\pubyear{2023}

\begin{document}
\label{firstpage}
\pagerange{\pageref{firstpage}--\pageref{lastpage}}
\maketitle

\begin{abstract}
The measurement of the Hubble-Lema\^{i}tre constant $(H_0)$ from the cosmic microwave background and the Type IA supernovae are at odds with each other. One way to resolve this tension is to use an independent way to measure $H_0$. This can be accomplished by using gravitational-wave (GW) observations. Previous works have shown that with the onset of the next generation of GW detector networks, it will be possible to constrain $H_0$ to better than 2\% precision (which is enough to resolve the tension) with binary black hole systems that are extremely well localized in the sky, also called golden dark sirens. Bright sirens like binary neutron star systems can also help resolve the tension if both the GW and the following electromagnetic counterpart are detected. In this work, we show that neutron star-black hole (NSBH) mergers can act both as golden dark sirens as well as bright sirens, thus, assigning them the term gray sirens. We assess the potential of using NSBH mergers to measure $H_0$ and find that the Voyager network might be able to resolve the tension in an observation span of 5 years. The next generation networks which include the Cosmic Explorer detectors and the Einstein Telescope will be able to measure $H_0$ to sub-percent level just by using NSBH mergers.
\end{abstract}

\begin{keywords}
cosmological parameters -- gravitational waves --  black hole - neutron star
mergers
\end{keywords}

%%%%%%%%%%%%%%%%%%%%%%%%%%%%%%%%%%%%%%%%%%%%%%%%%%

%%%%%%%%%%%%%%%%% BODY OF PAPER %%%%%%%%%%%%%%%%%%

\section{Introduction}
The expansion rate of the Universe is given by the \HL constant $(H_0)$. Thus, precise measurement of $H_0$ is of fundamental importance to cosmology. Two of the most accurate measurements of $H_0$ come from the Planck Collaboration and the SH0ES Program. The Planck Collaboration obtains constraints on $H_0$ using the cosmic microwave background measurements, reporting a value of $H_0 = 67.4\pm0.5\,\,\mathrm{km}\,\mathrm{s}^{-1}\,\mathrm{Mpc}^{-1}$ \citep{Planck:2018vyg} . The SH0ES Program utilizes the Cepheid variables and the Type-Ia supernovae to obtain bounds on $H_0$. The reported value from this approach is $H_0 = 73.30 \pm 1.04$ \citep{Riess:2021jrx}---a $5\sigma$ discrepancy in the two measurements. Several other analyses have reported measurements of $H_0$ that are $4\sigma-6\sigma$ discrepant between the early and late Universe approaches \citep{Verde:2019ivm, DiValentino:2021izs}. This discrepancy is known as the \HL tension.

Gravitational wave (GW) observations can be used to estimate the luminosity distance $(D_L)$ associated with a compact binary \citep{Schutz:1986gp}, without the need for an external distance calibrator (like Cepheid variables). Thus, GW sources are referred to as \textit{standard sirens} \citep{Holz:2005df}. With the $D_L$ measurement from GW observations, if one can also measure the redshift associated with the source, then cosmological parameters can be inferred with GW observations. Hence, GWs can be used as a tool to measure $H_0$. This measurement is completely independent of the ones performed by the Planck and the SH0ES projects and can help resolve the \HL tension.

There are several ways of obtaining the redshift associated with GW sources. For coalescing compact binaries containing one or more neutron stars (NS), the tidal disruption of the NS can lead to the generation of electromagnetic (EM) counterparts which can be detected by telescopes \citep{Li:1998bw,Metzger:2019zeh,Goodman:1986az,Eichler:1989ve}. For events where the EM counterpart is detected, the source can be localized precisely in the sky and the host galaxy can be identified. The measurement of the redshift of the host galaxy gives the redshift associated with the GW source. The sources for which the redshift can be obtained in this way are called \textit{bright sirens}. This was first realized with the detection of GW170817 \citep{LIGOScientific:2017vwq}, which was the first binary neutron star (BNS) GW event and also the first GW event for which EM counterparts were detected \citep{LIGOScientific:2017ync,LIGOScientific:2017zic,LIGOScientific:2017pwl}. Using the EM detection, the host galaxy was identified as NGC4993 and $H_0$ was measured to be $70^{+12}_{-8}\,\,\mathrm{km}\,\mathrm{s}^{-1}\,\mathrm{Mpc}^{-1}$ \citep{LIGOScientific:2017adf}. It has been shown that $\sim 50$ BNS mergers with counterparts will be enough to measure $H_0$ to better than $2\%$ precision and resolve the \HL tension \citep{Chen:2017rfc,Mortlock:2018azx}.

Another approach, first discussed in \citet{Schutz:1986gp}, allows the use of GW signals that are not followed by EM counterpart detections for the measurement of $H_0$. Without the counterpart, one can look for galaxies in the sky patch acquired from the localization of events from GW observations. Each galaxy/galaxy cluster in that sky patch will have an associated redshift and give a value of $H_0$, with the true value of $H_0$ among them. Combining $H_0$ measurements from multiple such GW detections should help isolate the true value of $H_0$ from the noise. This is often referred to as the \textit{statistical standard siren} approach, or, more commonly, as the \textit{dark siren} approach. The statistical approach is implemented using Bayesian frameworks developed in \citet{DelPozzo:2011vcw} and  \citet{Chen:2017rfc}. It was employed in \citet{LIGOScientific:2018gmd} to obtain a dark siren measurement of $H_0$ using GW170817, i.e., without using the host galaxy identification. Among the galaxies in the \texttt{GLADE} catalog \citep{Dalya:2018cnd} that lie within the localization volume, \citet{LIGOScientific:2018gmd} obtained $H_0 = 76^{+48}_{-23}\,\,\mathrm{km}\,\mathrm{s}^{-1}\,\mathrm{Mpc}^{-1}$. \citet{DES:2019ccw} applied the statistical method to GW170814 \citep{LIGOScientific:2017ycc}, marking the first $H_0$ measurement using a binary black hole (BBH) merger. Using the Dark Energy Survey data \citep{DES:2018gui} for redshift information, they found $H_0 = 75^{+40}_{-32}\,\,\mathrm{km}\,\mathrm{s}^{-1}\,\mathrm{Mpc}^{-1}$. While these measurements are worse than the one obtained by the bright siren, they are expected to improve with more BBH detections. With Gravitational-Wave Transient Catalog-2 (GWTC-2) \citep{LIGOScientific:2020ibl}, without GW170817, $H_0$ is constrained to $67.3^{+27.6}_{-17.9}\,\,\mathrm{km}\,\mathrm{s}^{-1}\,\mathrm{Mpc}^{-1}$ \citep{Finke:2021aom}. \citet{Palmese:2021mjm} combine $8$ well-localized dark siren events from GWTC-3 \citep{LIGOScientific:2021djp} to get $H_0 = 79.8^{+19.1}_{-12.8}\,\,\mathrm{km}\,\mathrm{s}^{-1}\,\mathrm{Mpc}^{-1}$, which is closer to the $H_0$ estimate from the bright siren approach with GW170817. The measurement of $H_0$ from GW170817 alone is still better than the combined measurement using GWTC-2 and well-localized GWTC-3 events, excluding GW170817. In fact, \citet{Chen:2017rfc} showed that BNS systems with detected counterparts will constrain $H_0$ better than the combined estimates from BBH systems without counterparts.

There have also been studies that have combined the $H_0$ measurements from dark and bright siren approaches. Using GWTC-3 events, $H_0$ is estimated to be $68^{+12}_{-8}\,\,\mathrm{km}\,\mathrm{s}^{-1}\,\mathrm{Mpc}^{-1}$ \citep{LIGOScientific:2021aug}. Similarly, \citet{Finke:2021aom} includes the $H_0$ measurement using GW170817 to that obtained using dark sirens from GWTC-2, to report an improved measurement of $H_0 = 72.2^{+13.9}_{-7.5}\,\,\mathrm{km}\,\mathrm{s}^{-1}\,\mathrm{Mpc}^{-1}$. Similarly, with the inclusion of GW170817, \citet{Palmese:2021mjm} obtains an improved measurement of $H_0 = 72.77^{+11.0}_{-7.55}\,\,\mathrm{km}\,\mathrm{s}^{-1}\,\mathrm{Mpc}^{-1}$.

BBH systems are ideal dark siren candidates for two reasons. Firstly, BBH systems are, in general, heavier and lead to louder, i.e., higher signal-to-noise ratio (SNR) events compared to BNS systems. Higher SNRs lead to better estimation of binary parameters, especially the luminosity distance and the arrival times of the signal in the detector. An improved measurement of arrival times leads to better localization in the sky, which, along with the improved $D_L$ measurement, results in better $H_0$ estimates. Secondly, BBH systems can be asymmetric (mass ratio greater than 2), leading to the activation of sub-dominant higher-order multipoles \citep{Roy:2019phx}. The presence of these higher-order modes causes modulations in the amplitude and the phase of the GW waveform, breaking the degeneracy between distance and inclination angle in the GW waveform when only the dominant $(2,2)$ mode is considered. Various studies have shown that their presence results in the improved estimation of binary parameters, including $D_L$ and sky localization \citep{VanDenBroeck:2006ar,Arun:2007hu,Graff:2015bba,Ajith:2009fz}. This is also validated by the fact that two of the best localized dark siren events, GW190412 \citep{LIGOScientific:2020stg} and GW190814 \citep{LIGOScientific:2020zkf}, have mass ratios $\sim 4$ and $\sim 9$, respectively. With more sensitive GW detectors in the future, the sky position for a fraction of these dark sirens can be measured well enough such that, on average, only one galaxy will lie in that sky patch, allowing one to uniquely identify the host galaxy \citep{Chen:2016tys,Nishizawa:2016ood,Borhanian:2020vyr}. In such cases, the statistical dark siren approach coincides with the bright siren method in the sense that a unique host galaxy can be identified for the purpose of redshift measurement. Such systems are called \textit{golden} dark sirens. Using this technique, \citet{Borhanian:2020vyr} claim to resolve the \HL tension with two years of observation using a planned $5$ detector network \texttt{HLVKI} at \texttt{A+} sensitivities \citep{Miller:2014kma,KAGRA:2013rdx}, which is expected to be functional in the late $2020$s.

In January 2020, the LIGO-Virgo detectors recorded the first detection of a neutron star-black hole (NSBH) merger \citep{LIGOScientific:2021qlt}, adding another class of compact binary systems to the set of detected events. Depending on the characteristics of the system, the NS in the NSBH system can get tidally disrupted by the BH before crossing the innermost stable circular orbit, leading to the generation of EM counterparts (see \citet{Kyutoku:2021icp} for a review on NSBH mergers and factors that affect tidal disruption of the NS). Thus, NSBH systems can be used as bright sirens for $H_0$ measurement. \citet{Feeney:2020kxk} have shown that for an NSBH merger rate density of $610\,\mathrm{Gpc}^{-3}\,\mathrm{yr}^{-1}$, the \texttt{HLVKI} network with \texttt{A+} sensitivities might be able to resolve the \HL tension in $5$ years. However, the merger rate density chosen by this analysis has been ruled out by the event-based local merger rate density reported in GWTC-3 \citep{LIGOScientific:2021psn}, with the new upper limit on the NSBH merger rate being $140\,\mathrm{Gpc}^{-3}\,\mathrm{yr}^{-1}$. This will affect the constraints on $H_0$ reported by the work. NSBH mergers can also act as effective dark siren candidates. In general, the mass ratio $q = m_{\mathrm{BH}}/m_{\mathrm{NS}}$ for NSBH systems is greater than 2, which leads to the activation of the higher-order modes. Further, NSBH systems are heavier than BNS systems and lead to signals with higher SNR (if present at the same $D_L$). Both these factors aid in the improved estimation of $D_L$ and sky localization with NSBH systems compared to BNS systems. The constraints on $H_0$ from NSBH events that qualify as dark sirens and the ones that qualify as bright sirens can be combined to calculate $H_0$ to better precision compared to only using the dark siren or the bright siren approach. Due to the ability to act both as dark and as bright sirens, we refer to NSBH systems as \textit{gray sirens}.

In this work, we will assess the potential of NSBH systems to resolve the \HL tension using GW observations and EM detections. In particular, we will consider NSBH events that qualify as golden dark sirens, as well as the ones that qualify as bright sirens, and present results for the individual and combined (i.e., golden gray siren) bounds on $H_0$ from the two approaches. This study uses results from our previous work on analyzing the detectability, measurement ability, and the science that can be done with next generation ground-based GW detector networks with NSBH mergers \citep{Gupta:2023evt}. We consider NSBH events that can be detected by six future GW observatories. These networks include advancements like \texttt{A+} and Voyager sensitivities \citep{LIGO:2020xsf} for the three LIGO detectors (we denote LIGO-Hanford by \texttt{H}, LIGO-Livingston by \texttt{L}, and the planned detector in Aundha, India \citep{LIGO-India} by \texttt{I}), Virgo detector \citep{VIRGO:2014yos} in Italy (denoted by \texttt{V}), KAGRA detector \citep{Somiya:2011np,KAGRA:2020agh,Aso:2013eba} in Japan (denoted by \texttt{K}), two $40$ km long Cosmic Explorer (denoted by CE or \texttt{C} and \texttt{S}) detectors \citep{Evans:2021gyd,LIGOScientific:2016wof,Reitze:2019iox} and  triangular-shaped Einstein Telescope (denoted by ET or \texttt{E}) with $10$ km arms \citep{Punturo:2010zz,Hild:2010id}. The combinations of detectors making up the six networks are listed in Table \ref{tab:net}.
\begin{table} 
  \centering
  \caption{\label{tab:net}The six next generation ground-based GW detector networks that are included in the analysis, with the abbreviation used to refer to the network. In the text, \texttt{HLVKI+} and \texttt{VK+HLIv} are also referred to as the \texttt{A+} network and the Voyager network, respectively.}
  \renewcommand{\arraystretch}{1.5} 
  {
    \begin{tabular}{ l c }
    \hhline{--}
    Network & Detectors \\
    \hhline{--}
    HLVKI\texttt{+} & LIGO (HL\texttt{+}), Virgo\texttt{+}, KAGRA\texttt{+}, LIGO-I\texttt{+}\\
    VK\texttt{+}HLIv & Virgo\texttt{+}, KAGRA\texttt{+}, LIGO (HLI-Voyager)\\
    VKI\texttt{+}C & Virgo\texttt{+}, KAGRA\texttt{+}, LIGO-I\texttt{+}, CE-North\\
    HLKI\texttt{+}E & LIGO (HL\texttt{+}), KAGRA\texttt{+}, LIGO-I\texttt{+}, ET \\
    KI\texttt{+}EC & KAGRA\texttt{+}, LIGO-I\texttt{+}, ET, CE-North\\
    ECS & ET, CE-North, CE-South\\
    \hhline{--}
    \end{tabular}
    }
\end{table}

Following our previous work \citep{Gupta:2023evt}, we construct two NSBH populations for an observation time of $10$ years. The population parameters are described in Section \ref{sec:pop}. In Section \ref{sec:Method}, we explain the use of the Fisher information matrix in calculating the measurement errors on $H_0$ using the errors in luminosity distance. In Sections \ref{sec:dark_sirens} and \ref{sec:bright_sirens}, we discuss the performance of NSBH systems as golden dark sirens and as bright sirens, respectively. Using the results in these two sections, we justify the treatment of NSBH systems as (golden) gray sirens, showing results for the same in Section \ref{sec:gray_sirens}. In Section \ref{sec:sys}, we discuss the systematic uncertainties that affect this analysis. Our conclusions are summarized in Section \ref{sec:concl}.

%%%%%%%%%%%%%%%%%%%%%%%%%%%%%%%%%%%%%%%%%%%%%%%%%%

%%%%%%%%%%%%%%%%% SECTION BREAK %%%%%%%%%%%%%%%%%%

\section{Population Characteristics} \label{sec:pop}
Due to the low number of NSBH detections, the population characteristics of NSBH systems are uncertain. While there have been attempts at deriving the mass and spin distributions for the NS and the BH using the available events \citep{Zhu:2021jbw,Biscoveanu:2022iue}, these models can change drastically with more detections in the coming years. From a theoretical standpoint, different formation channels can give distinct predictions regarding the masses and spins of the two components in NSBH systems \citep{1976ApJ...207..574S,PortegiesZwart:1999nm,Clausen:2012zu,Downing:2009ag,Santoliquido:2020bry,Rastello:2020sru,Belczynski:2016ieo}. To gauge these uncertainties, we consider two population models for NSBH systems- \textit{Pop-1} and \textit{Pop-2}. 
\begin{table*}
    \centering
    \renewcommand{\arraystretch}{1.5}
    \begin{tabular}{l| P{2cm} P{2.4cm} P{2cm} P{4cm}}
    \hhline{-----}
    \multirow{2}[2]{*}{Parameter} & \multicolumn{2}{c}{Pop-1} & \multicolumn{2}{c}{Pop-2} \\
    \hhline{~----}         & Neutron Star & Black Hole & Neutron Star & Black Hole \\
    \hhline{-----}
    Mass $m$ & [1,2.9] $\msun$ & [3,100] $\msun$ & [1.26,2.50] $\msun$ & [2.6,39.2] $\msun$ \\
    %\hhline{-----}
    Mass Model & Uniform & \texttt{POWER+PEAK} \citep{LIGOScientific:2021psn} & \multicolumn{2}{c}{Derived from the fiducial model \citep{Broekgaarden:2021iew}}\\
    %\hhline{-----}
    Spin $\chi$ & [-0.05,0.05] & [-0.75,0.75] & 0 & [0,1] \\
    %\hhline{-----}
    \multirow{2}[2]{*}{Spin Model} & \multicolumn{2}{c}{\multirow{2}[2]{*}{Aligned Uniform}} & \multirow{2}[2]{*}{Aligned} & Eqs. (2) and (3) in \\
      & & & & \citet{Chattopadhyay:2022cnp} \\
    %\hhline{-----}
    $z$ & \multicolumn{4}{c}{Uniform in six bins: [0.02,0.05], [0.05,1], [1,2], [2,4] and [4,20]}\\
    %\hhline{-----}
    $D_L$ & \multicolumn{4}{c}{$z$ converted using \texttt{ASTROPY.Planck18}}  \\
    %\hhline{-----}
    $\cos(\iota)$ & \multicolumn{4}{c}{Uniform in [-1,1]} \\
    %\hhline{-----}
    $\alpha$ & \multicolumn{4}{c}{Uniform in [0,2$\pi$]} \\
    %\hhline{-----}
    $\cos(\delta)$ & \multicolumn{4}{c}{Uniform in [-1,1]} \\
    %\hhline{-----}
    $\psi$ & \multicolumn{4}{c}{Uniform in [0,2$\pi$]} \\
    %\hhline{-----}
    $t_c$, $\phi_c$ & \multicolumn{4}{c}{0} \\
    \hhline{-----}
    \end{tabular}
    \caption{\label{tab:pop_par}The table summarizes the parameters that characterize the two population models considered in this study. The parameters are used to generate injections and carry out the analysis for the constraints that can be put on the \HL constant using GW observations of NSBH systems.}
\end{table*}
Pop-1 considers broad mass and spin distributions for NSBH systems. The BH mass follows the \texttt{POWER+PEAK} \citep{LIGOScientific:2021psn} distribution between $[3\msun,100\msun]$ and the NS mass is sampled from a uniform distribution between $[1\msun,2.9\msun]$. The spins for both the NS and the BH are assumed to be aligned with the orbital angular momentum of the binary system, i.e., $\chi_{\mathrm{1x}} = \chi_{\mathrm{1y}} = \chi_{\mathrm{2x}} = \chi_{\mathrm{2y}} = 0$. Here, the orbital angular momentum is assumed to be along the $z$ direction and $(\boldsymbol{\chi_1}, \boldsymbol{\chi_2})$ denote the dimensionless spin vectors of the BH and the NS, respectively. The NS spins are chosen from a uniform distribution between $[-0.05,0.05]$ and the BH spins are taken from a uniform distribution between $[-0.75,0.75]$. For Pop-2, the masses and spins for the NS and the BH are taken from the \textit{fiducial} model in \citet{Broekgaarden:2021iew}, which is a binary population synthesis model for NSBH systems which are formed through the isolated binary formation channel \citep[see][Table 1]{Broekgaarden:2021iew}. The NS are assumed to be non-spinning, whereas BH have aligned spins and their magnitude is calculated using Eqs. (2) and (3) in \citet{Chattopadhyay:2022cnp}. The equations only apply to systems where the NS progenitor is formed first, allowing the second-born BH progenitor to have high spins as it can get tidally
spun up by its companion \citep{Qin:2018vaa,Chattopadhyay:2019xye,Chattopadhyay:2020lff,Bavera:2020inc}. The probability density functions (PDFs) of the mass and spin parameters for Pop-2 events are plotted in Fig. \ref{appfig:Pop2_mass_spins} in Appendix \ref{appsec:Pop2_params}. For each population, we generate $250,000$ injections per redshift bin for five redshift bins: $z \in [0.02, 0.05]$, $[0.05, 1]$, $[1, 2]$, $[2, 4]$ and $[4, 20]$. The luminosity distance for each injection is obtained by converting the corresponding redshift $z$ using $\Lambda$CDM cosmology, with the values of cosmological parameters corresponding to Planck measurements \citep{Planck:2018vyg} obtained from  \texttt{ASTROPY.PLANCK18} \citep{Astropy:2013muo,Astropy:2018wqo}. $\cos(\iota)$ and $\cos(\delta)$, where $\iota$ and $\delta$ are the inclination angle and the declination respectively, are sampled uniformly between $[-1,1]$.  The right ascension $\alpha$ and the polarization angle $\psi$ are sampled uniformly between $[0, 2\pi]$. $t_c$ and $\phi_c$ are the time and phase of coalescence respectively and are fixed to $0$ for all the injections. All the parameters for both populations have been summarized in Table \ref{tab:pop_par}.

We logarithmically divide the redshift range into $50$ bins and randomly pick injections from each of these bins. The number of injections picked from a redshift bin corresponds to the expected number of NSBH mergers for that redshift range. This number is calculated by using a redshift distribution for the NSBH systems. We follow the analytical redshift distribution given in \citet{Sun:2015bda} that uses the star formation rate (SFR) model described in \citet{Yuksel:2008cu} with a log-normal time delay model proposed in \citet{Wanderman:2014eza} (see Appendix \ref{appsec:det_rate}). The model is calibrated to match the local (i.e., at $z=0$) merger rate density estimated by the LIGO-Virgo collaboration using the detected events. The inferred event-based local merger rate density is in the range $7.8-140$ $\mbox{Gpc}^{-3}$ $\mbox{yr}^{-1}$ \citep{LIGOScientific:2021psn}. We fix the local merger rate density for NSBH systems to $45$ $\mbox{Gpc}^{-3}$ $\mbox{yr}^{-1}$ (which is the median value reported in \citet{LIGOScientific:2021qlt}), resulting in a cosmic NSBH merger rate of $4.0^{+8.5}_{-3.3} \times 10^4$ yr$^{-1}$, where the upper and lower bounds are calculated using the upper and lower bounds on the local merger rate density. Using the redshift distribution and local merger rate density, we construct populations of NSBH events using Pop-1 and Pop-2 parameters corresponding to an observation time of $10$ years. In the following sections, we will present results for both Pop-1 and Pop-2 events and for three local merger rate densities, \textit{low:} $7.8$ $\mbox{Gpc}^{-3}$ $\mbox{yr}^{-1}$, \textit{median:} $45$ $\mbox{Gpc}^{-3}$ $\mbox{yr}^{-1}$ and \textit{high:} $140$ $\mbox{Gpc}^{-3}$ $\mbox{yr}^{-1}$. 
%%%%%%%%%%%%%%%%%%%%%%%%%%%%%%%%%%%%%%%%%%%%%%%%%%

%%%%%%%%%%%%%%%%% SECTION BREAK %%%%%%%%%%%%%%%%%%

\section{Methodology} \label{sec:Method}
In this study, we work with compact binaries where the NS and the BH are assumed to be in a quasi-circular orbit and their spins are assumed to be aligned with the orbital angular momentum of the binary. In this case, the detector response to the GW is given by,
\begin{equation}
\begin{split}
    h^{(A)} (t,\boldsymbol{\mu})&=\,\,F^{(A)}_{+} (\alpha, \delta, \psi, \boldsymbol{\nu})\,h^{(A)}_{+} (t,\mathcal{M}_c,\eta,\chi_1,\chi_2,\iota,D_L) \\
    &+\,\,F^{(A)}_{\times} (\alpha, \delta, \psi, \boldsymbol{\nu})\,h^{(A)}_{\times} (t,\mathcal{M}_c,\eta,\chi_1,\chi_2,\iota,D_L).
\end{split}
\end{equation}
Here, $F^{(A)}_{+}$ and $F^{(A)}_{\times}$ are the antenna pattern functions corresponding to the detector $A$ that depend on the location of the detector, denoted by $\boldsymbol{\nu}$, and the location of the source, expressed using the right ascension $\alpha$, the declination $\delta$ and the polarization angle $\psi$. $h^{(A)}_{+}$ and $h^{(A)}_{\times}$ are the \textit{plus} and \textit{cross} polarizations of the GWs and are a function of variables that describe the properties of the source--- chirp mass $\mathcal{M}_c$ and symmetric mass-ratio $\eta$ are functions of the masses of the two compact objects, $\chi_1$ and $\chi_2$ are the dimensionless spins of the BH and the NS, $\iota$ is the angle between the line of sight of the observer and the total angular momentum of the binary and $D_L$ is the luminosity distance of the binary from the observer. Hence, for a given detector, the detector output is a time-dependent function of $\boldsymbol{\mu} = (\mathcal{M}_c,\eta,\chi_1,\chi_2,\iota,D_L,\alpha, \delta, \psi)$. 

To obtain the errors in measurement of parameter $\boldsymbol{\mu}$, we utilize the Fisher information matrix (FIM) method implemented in \texttt{GWBENCH} \citep{Borhanian:2020ypi}. The method gives the measurement errors in $\boldsymbol{\mu}$ as elements of the covariance matrix $\Sigma$, which is the inverse of the FIM $\Gamma$, defined as
\begin{equation}
    \Sigma_{ij} = \Gamma^{-1}_{ij} = \left(\frac{\partial h}{\partial \theta_{i}},\frac{\partial h}{\partial \theta_{j}}\right)^{-1},
\end{equation}
where $h$ is the GW waveform in the frequency-domain, $\theta_i$ is the $i^{th}$ parameter in $\boldsymbol{\mu}$ and $(\cdot\,,\cdot)$ is the noise-weighted inner product. The square root of the diagonal elements of the covariance matrix, i.e., $\Sigma_{ii}^{1/2}$, gives the uncertainty in the measurement of the $i^{th}$ parameter.
\begin{figure}
\centering
\includegraphics[scale=0.6]{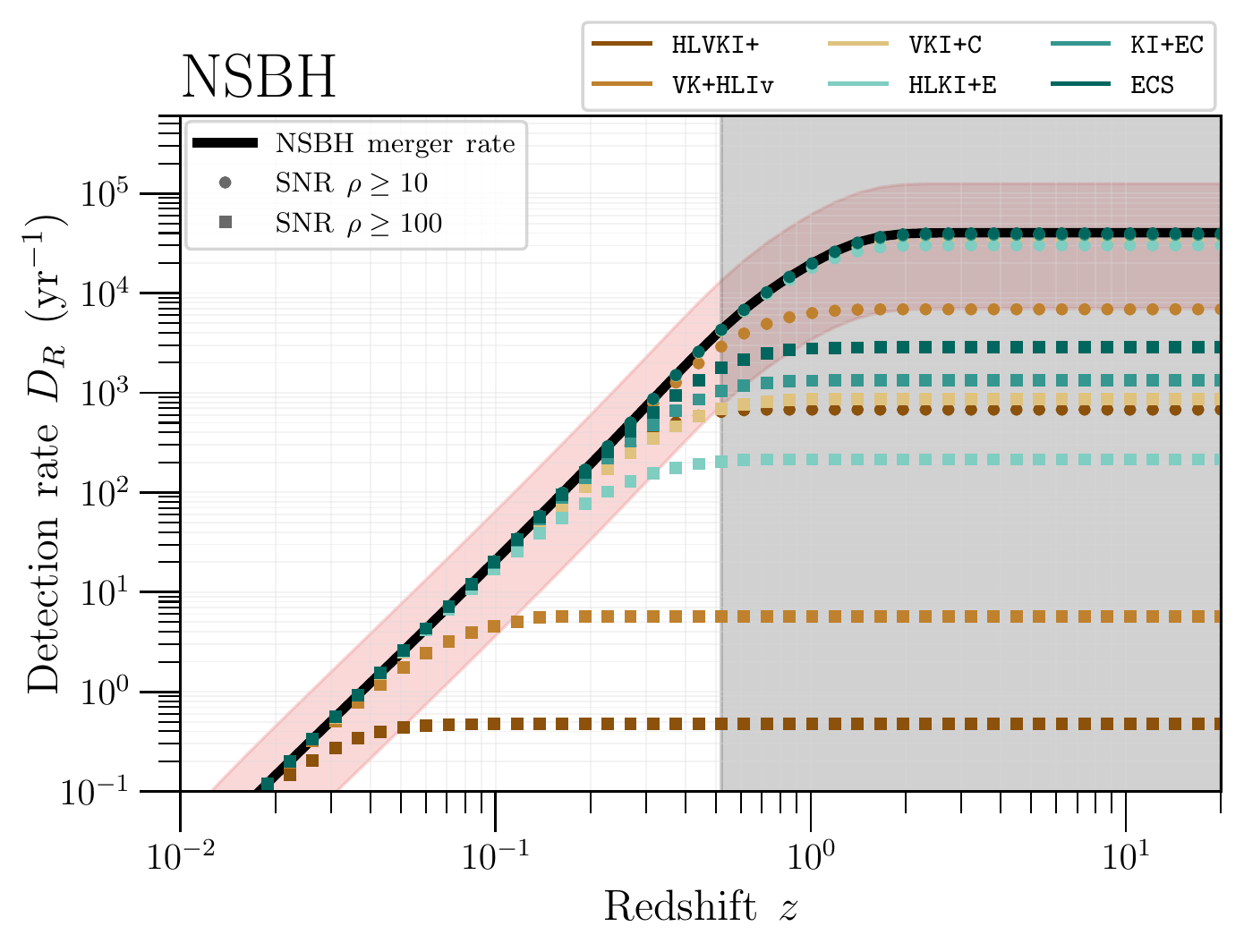}
\caption{\label{fig:eff_rate}The detection rate of NSBH systems as a function of redshift for the six GW detector networks. The black solid line refers to the total NSBH merger rate. The pink-shaded area shows the variation in the total merger rate due to the uncertainty in the value of the local merger rate density. The gray shaded area covers the $z>0.5$ region that is excluded from this analysis.}
\end{figure}

\begin{figure*}
\begin{subfigure}{\linewidth}
  \includegraphics[scale=0.31]{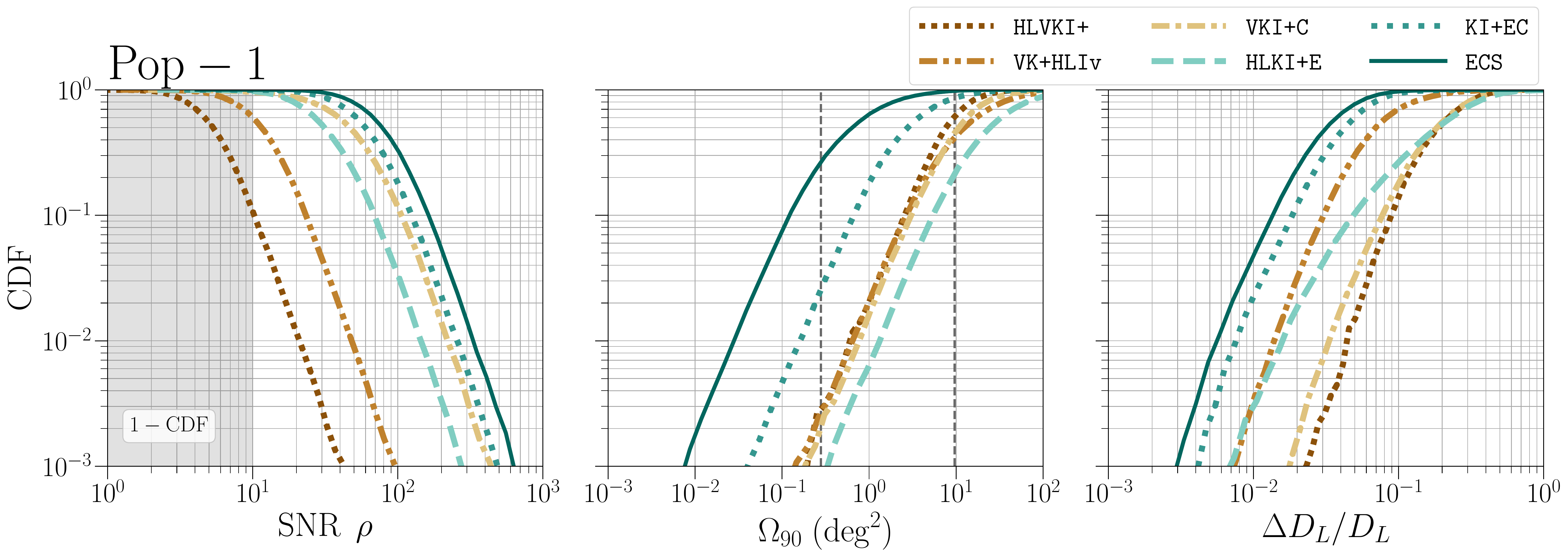}
\end{subfigure}\par\medskip
\begin{subfigure}{\linewidth}
  \includegraphics[scale=0.31]{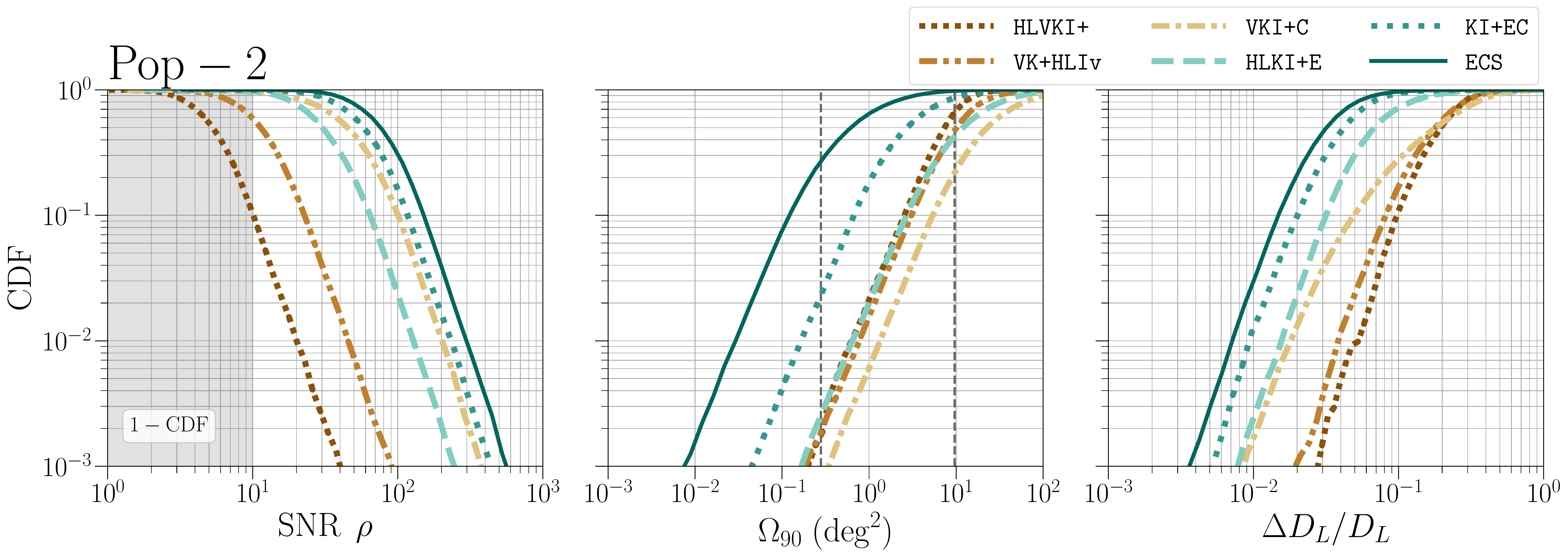}
\end{subfigure}\par\medskip
\caption{\label{fig:mma_cdf} The cumulative density function plots for SNR $\rho$, $90\%$-credible sky area $\Omega_{90}$ and fractional error in luminosity distance $\Delta D_L / D_L$ for the population restricted to $z\leq0.5$. The vertical black dotted lines in the plot for $\Omega_{90}$ correspond to the field of view (FOV) of the Roman Space Telescope (FOV = 0.28 $\mbox{deg}^2$ \citep{Chase:2021ood}) and the Rubin observatory (FOV = 9.6 $\mbox{deg}^2$ \citep{LSST:2008ijt,Chase:2021ood})} 
\end{figure*}

The detection rate is defined as the number of NSBH mergers, up to a given redshift, detected by a GW network per year with an SNR greater than the threshold SNR (see Appendix \ref{appsec:det_rate}). Figure \ref{fig:eff_rate} shows the detection rate of NSBH systems for the six GW detector networks as a function of redshift for threshold SNRs of $10$ and $100$. The pink-shaded area denotes the uncertainty in the local merger rate of NSBH systems. For the golden dark siren study, only systems within $z=0.1$ are considered. For bright sirens, we do not get any KN detections beyond $z\sim0.4$ for the two telescopes we have used \citep{Gupta:2023evt}. Thus, in this work, we will only look at systems located within the redshift of $z = 0.5$. Up to the redshift of $z = 0.5$, our populations contain $\sim 4000$ NSBH systems, all detected by networks with at least one detector as the CE or the ET observatories. The Voyager network, \texttt{VK+HLIv}, is seen to detect $\sim 75\%$ of these events, whereas the \texttt{A+} network, \texttt{HLVKI+}, detects only $15\%$ of all the events. For the detected events, Fig. \ref{fig:mma_cdf} shows the detection and measurement ability of the GW detector networks in the form of cumulative density function (CDF) plots for the SNR $\rho$, the fractional measurement errors in luminosity distance $\Delta D_L/D_L$ and $90\%-$credible sky area $\Omega_{90}$. The corresponding numbers are presented in Table \ref{apptab:mma_sa90_logDL} in Appendix \ref{appsec:logDL_sa90_table}, where we list the number of detections every year that can be localized in the sky to $\Omega_{90} \leq 10$, $1$ and $0.1$ $\mbox{deg}^2$ and the fractional error in luminosity distance is better than $0.1$ and $0.01$. We see that, depending on the network, we can expect to detect $\mathcal{O}(10)$ to $\mathcal{O}(1000)$ events every year where $\Omega_{90} < 1$ $\mbox{deg}^2$, and also when the luminosity distance is constrained to better than $10\%$. Further, $\mathcal{O}(1)$ to $\mathcal{O}(1000)$ events in an observation time of $10$ years can be constrained to better than $0.1$ $\mbox{deg}^2$. The unprecedented measurement ability of next generation ground-based GW detector networks, especially in the measurement of luminosity distances and sky localization, prompt the evaluation of the potential of NSBH systems as golden dark siren candidates.

Following our previous work \citep{Gupta:2023evt}, we focus on kilonovae (KN) as the EM counterpart to GW from NSBH mergers, to assess the case of NSBH as bright sirens for $H_0$ measurement. The KN light curve modeling has been briefly described in Appendix \ref{appsec:kn_numbers} (and in more detail in Section VC of \citet{Gupta:2023evt}). We state the number of KN detections in Table \ref{apptab:r_R_kn_numbers} in Appendix \ref{appsec:kn_numbers}, corresponding to the NSBH GW observations, that can be expected using the Vera C. Rubin Observatory \citep{LSST:2008ijt} and the Nancy Grace Roman Space Telescope \citep{Hounsell:2017ejq}. Depending on the equation of state (EOS) of the NS, we can expect to detect $\mathcal{O}(1)$ to $\mathcal{O}(10)$ KN detections using the $r-$filter in Rubin and $\mathcal{O}(1)$ to $\mathcal{O}(100)$ KN with $R-$filter in Roman, in an observation span of $10$ years. The farthest KN observed with the Rubin ($r-$filter) and the Roman ($R-$filter) were at a distance of $952$ Mpc $(z \sim 0.189)$ and $1.941$ Gpc $(z \sim 0.354)$, and had the peak luminosity (in AB mag) of $24.7$ and $26.2$, respectively. As all of these events are within a redshift of $z\sim 0.4$, we expect the errors in luminosity distances to be in accordance with Fig. \ref{fig:mma_cdf} and Table \ref{apptab:mma_sa90_logDL}, i.e., for most of these events, the luminosity distances will be measured with accuracies better than $10\%$ for the most advanced detector networks. The well-constrained luminosity distance with a detected EM counterpart points to the possibility of using NSBH systems as bright sirens for measuring $H_0$.

To convert the errors in the measurement of luminosity distance into errors in measuring $H_0$, we assume the $\Lambda$CDM cosmology. The relationship between luminosity distance $D_L$ and the redshift $z$ can then be written as
\begin{equation}
\begin{split}
    D_L &= \frac{1+z}{H_0} \int_{1/(1+z)}^{1} \frac{dx}{x^2\sqrt{\Omega_{\Lambda}+\Omega_{m}\,x^{-3}}}\\
    &= \frac{1+z}{H_0} \int_{1/(1+z)}^{1} \frac{dx}{x^2\sqrt{1-\Omega_{m}(1-x^{-3})}} ,
\end{split}
\end{equation}
where $\Omega_{m}$ is the matter density, $\Omega_{\Lambda}$ is the dark energy density, and we have used $\Omega_{\Lambda} = 1\,-\,\Omega_m$.
As luminosity distance is a function of $H_0$ and the matter density, i.e., $D_L = D_L(\boldsymbol{\theta})$ with $\boldsymbol{\theta} = (H_0,\Omega_m)$ the corresponding FIM obtained by combining the estimates from $N$ events can be given as
\begin{equation} \label{eq:FIM_no_prior}
    \Gamma_{ij} = \sum_{k=1}^{N} \frac{1}{\sigma_{D_L}^2} 
    \left. \left(\dfrac{\partial D_{L}}{\partial \theta_i}\right)\left(\dfrac{\partial D_{L}}{\partial \theta_j} \right)\right\rvert_k ,
\end{equation}
i.e., we calculate the FIM for each event and then sum them over for all events. In Eq. (\ref{eq:FIM_no_prior}), $\sigma_{D_L}$ is the absolute error in luminosity distance for a particular GW observation. The expression for luminosity distance is obtained using the \texttt{Planck18} module of \texttt{astropy}. The FIM can then be numerically inverted to obtain the covariance matrix $\Sigma$, whose first and second diagonal elements give the squares of the absolute errors in $H_0$ and $\Omega_m$, respectively.

This method is agnostic of any prior information we might have about $\Omega_{m}$. As we only consider events up to a redshift of $z=0.5$, we do not expect $\Omega_{m}$ to be estimated well, which will also affect the estimates on $H_0$. A possible improvement comes from the realization that, while there is a disagreement between the Planck and the SH0ES measurements of $H_0$, their estimates on $\Omega_m$ are consistent. Planck reports $\Omega_m = 0.315 \pm 0.007$ and SH0ES measurement of $q_0$ is used to give $\Omega_m = 0.327 \pm 0.016$. We can include this information in our analysis by applying a Gaussian prior on $\Omega_m$ with standard deviation given by
\begin{equation}
    \sigma_{\Omega_m} = \sqrt{\sigma_{\mathrm{Planck}}^2 + \sigma_{\mathrm{SH0ES}}^2} = 0.017 .
\end{equation}
A simple way to incorporate this prior in the FIM is to add the FIM term corresponding to $\Omega_m$ with $1/\sigma_{\Omega_m}^2$ \citep{Cutler:1994ys}, i.e.,
\begin{equation}
    \Gamma_{22} = \sum_{k=1}^{N} \frac{1}{\sigma_{D_L}^2} 
    \left. \left(\dfrac{\partial D_{L}}{\partial \Omega_m}\right)^2\,\right\rvert_k + \dfrac{1}{\sigma_{\Omega_m}^2}\,,
\end{equation}
and the rest of the terms in the FIM are calculated the same way as before. The covariance matrix obtained using this FIM is expected to show better constraints for $H_0$ because the application of the prior on $\Omega_m$ restricts the parameter space of $(H_0,\Omega_m)$. In the following sections, we will report the fractional errors in $H_0$ for both cases-- without prior on $\Omega_m$ and with prior on $\Omega_m$, and we will see that the inclusion of prior will significantly improve the measurement of $H_0$. Note that we also obtain bounds on $H_0$ with a broader prior on $\Omega_m$ where the width of the Gaussian prior is taken to be $5$ times the previous value, i.e., $\sigma_{\Omega_m} = 0.085$. However, we do not observe significant changes in the bounds on $H_0$ with this broader prior, and all the general conclusions drawn using the narrow prior still hold. Hence, we only report the results for the narrow prior on $\Omega_m$, i.e., with $\sigma_{\Omega_m} = 0.017$. 

%%%%%%%%%%%%%%%%%%%%%%%%%%%%%%%%%%%%%%%%%%%%%%%%%%

%%%%%%%%%%%%%%%%% SECTION BREAK %%%%%%%%%%%%%%%%%%

\section{NSBH systems as golden dark sirens} \label{sec:dark_sirens}
The detection of GWs from a compact binary merger allows us to constrain the sky-position of the binary using the right ascension and declination measurements \citep[see][Eq. (43)]{Barack:2003fp}. The compact binary can be used as a golden dark siren to measure $H_0$ if the $90\%-$credible sky area is small enough such that, on average, only one galaxy can be present in that patch of the sky. Using Eq. (7) in \citet{Singer:2016eax}, we find that at a redshift of $z=0.1$ ($\sim 476$ Mpc), $\sim25$ suitable galaxies can exist in a sky-area of $1\,\mathrm{deg}^2$. Thus, in a sky-patch of $0.04\,\mathrm{deg}^2$, on average only $\sim 1$ galaxy should be present (we discuss the validity of this assumption in Section \ref{sec:sys}). Consequently, we consider only those NSBH events for the dark siren study for which $\Omega_{90} \leq 0.04\,\mathrm{deg}^2$. 

The combined FIM calculation and inversion to obtain the measurement errors on $H_0$ are done for two cases-- first, without including any prior on $\Omega_m$, and second, with prior on $\Omega_m$, as specified in Section \ref{sec:Method}. We present the fractional errors in $H_0$ using golden dark sirens detected by the six GW detector networks in Fig. \ref{tab:dark_sirens}. The plots are shown for an observation time of $2$ and $5$ years and for three NSBH local merger rate densities. For a given local merger rate density, to pick golden dark sirens for an observation time of $N$ years from our populations which contain events corresponding to an observation time of $10$ years, we randomly pick $N/10$ fraction of golden dark siren events from the population and calculate the fractional error in $H_0$ using the combined FIM for the selected events. This process is repeated $1000$ times, giving us $1000$ instances of fractional errors in $H_0$ corresponding to a particular observation time, local merger rate, and detector network. The markers in Fig. \ref{tab:dark_sirens} denote the median value of $\Delta H_0/H_0$ from these $1000$ iterations and the corresponding error bars show the $68\%-$confidence region. 

The error in $H_0$ decreases with the number of events $N$ as $\sim 1/\sqrt{N}$. Thus, the greater the number of events a detector network detects (with $\Omega_{90} \leq 0.04\,\mathrm{deg}^2$), the better the estimate of $H_0$ will be. Estimating $H_0$ with measurement errors better than $2\%$ is expected to resolve the tension between the Planck and the SH0ES measurements. Among the detected golden dark sirens, there can also be events where the luminosity distance is measured better than the fractional error of $2\%$, which can result in a sub-$2\%$ measurement of $H_0$ just by using a single event. We call these mergers \textit{exceptional events}. The number of golden and exception dark siren events that can be detected by the six GW detector networks in an observation time of $10$ years are mentioned in Table \ref{tab:dark_sirens}. 
\begin{figure*}
\begin{subfigure}{\linewidth}
\centering
  \includegraphics[scale=0.75,trim = 0 14 0 0,clip]{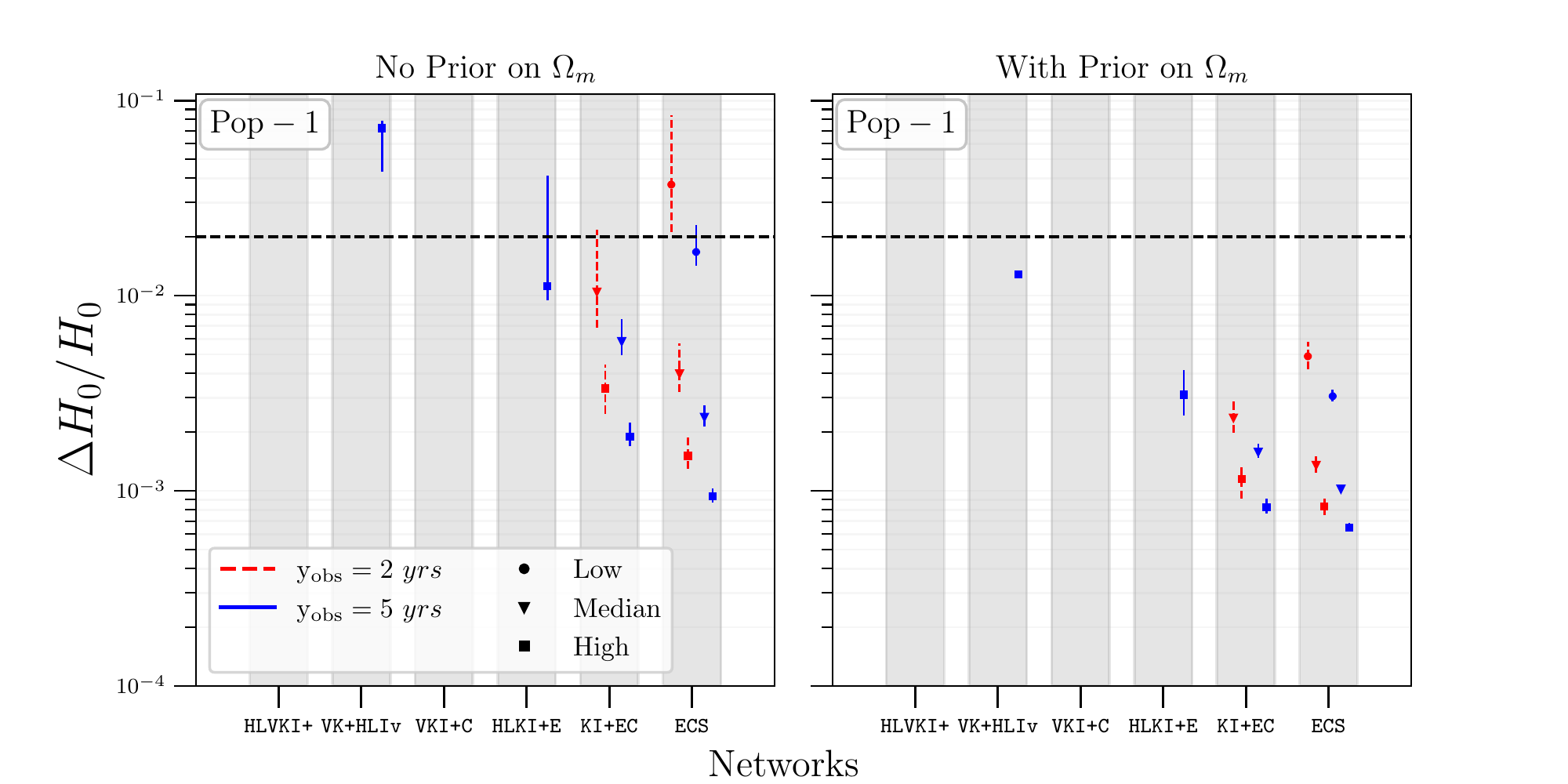}
\end{subfigure}\vspace{-1em}
\begin{subfigure}{\linewidth}
  \centering
  \includegraphics[scale=0.75]{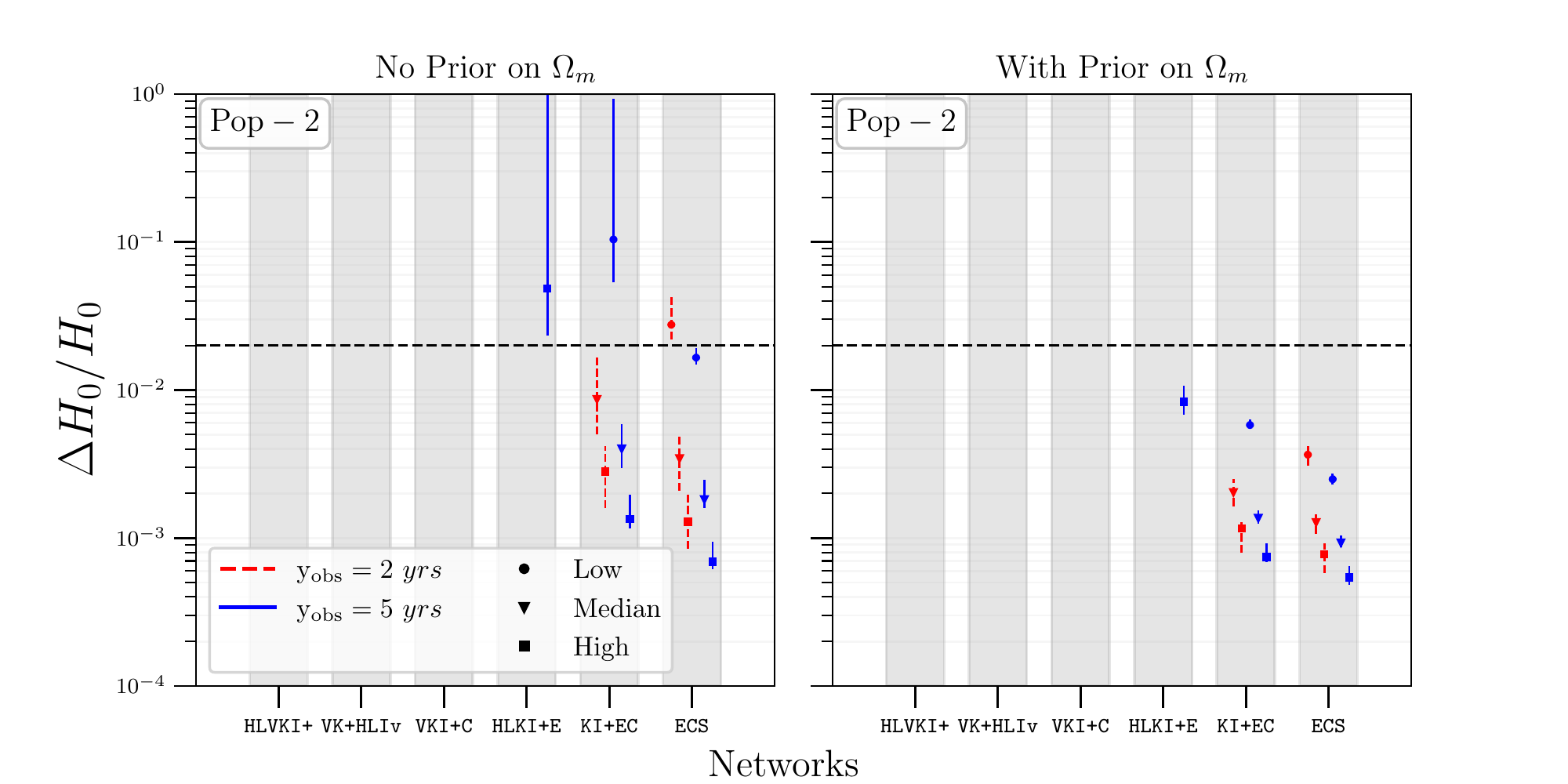}
\end{subfigure}\par
\caption{\label{fig:dark_sirens} The fractional errors in $H_0$ measurement using NSBH systems as golden dark sirens for the observation time of $2$ and $5$ years and for three different values for the local merger rate of NSBH systems, low: $\Dot{n}(0) = 7.8\,\mathrm{Gpc}^{-3}\,\mathrm{yr}^{-1}$, median: $\Dot{n}(0) = 45\,\mathrm{Gpc}^{-3}\,\mathrm{yr}^{-1}$ and high: $\Dot{n}(0) = 140\,\mathrm{Gpc}^{-3}\,\mathrm{yr}^{-1}$. The dotted black line corresponds to the $2\%$ error in $H_0$ measurement which would be enough to resolve the \HL tension.}.
\end{figure*}

From Table \ref{tab:dark_sirens} and Fig. \ref{fig:dark_sirens}, we see that, regardless of the choice of population, local merger rates, or observation times considered in this study, the \texttt{A+} network will not detect any NSBH golden dark siren events. \texttt{VK+HLIv} is expected to detect $0-3$ golden events in $10$ years, depending on the population model and the local merger rate density. While $H_0$ can be constrained using the detections from the Voyager network if the local merger rate is high, we cannot expect to bound $H_0$ better than $2\%$ without including the prior on $\Omega_m$. With the inclusion of the prior for $\Omega_m$, \texttt{VK+HLIv} measures $H_0$ better than $2\%$ for Pop-1 events. Note that when only one golden event is detected in a given observation time, the corresponding FIM is degenerate as it would amount to constraining two quantities, $H_0$ and $\Omega_m$, using only one event. Thus, unless we fix $\Omega_m$, the FIM cannot be inverted to obtain the measurement errors in $H_0$. As we consider $\Omega_m$ to be an unknown, Fig. \ref{fig:dark_sirens} only shows the fractional errors in $H_0$ for those networks, observation times, and local merger rates for which more than one golden event are detected. Thus, no constraints on $H_0$ are presented for the \texttt{VKI+C} network in Fig. \ref{fig:dark_sirens} even though we can expect at most $1$ possible detection in $5$ years. For the $0-6$ dark siren events \texttt{HLKI+E} can detect, the constraints on $H_0$ for high merger rate and $5$ years of observation vary with the choice of population. For Pop-1, even if the $\Omega_m$ prior is not included, $H_0$ can be measured well enough to resolve the tension. For events in Pop-2, we need to include the prior for $\Omega_m$ for measurements better than $2\%$.
\begin{table}
  \centering
  \caption{\label{tab:dark_sirens}The number of golden and exceptional NSBH dark siren detections for both the population models with the six next generation ground-based GW detectors in an observation time of $10$ years.}
  \renewcommand{\arraystretch}{1.4} 
    \begin{tabular}{l|P{1.3cm}P{1.3cm}|P{1.3cm}P{1.3cm}}
    \hhline{-----}
    \multirow{2}{*}{Network} & \multicolumn{2}{c|}{Pop-1} & \multicolumn{2}{c}{Pop-2}\\
    \hhline{~|--|--}
    & \textit{Golden} & \textit{Exceptional} & \textit{Golden} & \textit{Exceptional} \\
    \hhline{-----}
    HLVKI+ & $0$ & $0$ & $0$ & $0$\\
    VK+HLIv & $1^{+2}_{-1}$ & $1^{+2}_{-1}$ & $0^{+1}_{-0}$ & $0^{+1}_{-0}$\\
    VKI+C & $1^{+1}_{-1}$ & $1^{+1}_{-1}$ & $0$ & $0$\\
    HLKI+E & $1^{+5}_{-1}$ & $1^{+5}_{-1}$ & $0^{+3}_{-0}$ & $0^{+2}_{-0}$\\
    KI+EC & $24^{+55}_{-22}$ & $23^{+54}_{-21}$ & $25^{+71}_{-21}$ & $25^{+71}_{-21}$\\
    ECS & $106^{+238}_{-91}$ & $103^{+232}_{-88}$ & $103^{+241}_{-85}$ & $101^{+239}_{-84}$\\
    \hhline{-----}
    \end{tabular}
\end{table}

Two of the most advanced GW detector networks considered in our study, \texttt{KI+EC} and \texttt{ECS}, can expect to detect $\mathcal{O}(10)$ to $\mathcal{O}(100)$ golden events in an observation time of $10$ years. While both the networks can estimate $H_0$ well enough to resolve the tension in an observation span of $2$ years, \texttt{KI+EC} can only do so if the merger rate density is not low. With prior on $\Omega_m$, \texttt{ECS} can resolve the tension in $2$ years even for the case of low merger rate density. With a high merger rate and an observation time of $5$ years, \texttt{ECS} can measure $H_0$ with measurement errors that are as small as $\mathcal{O}(10^{-4})$. 

The number of NSBH events detected as golden and as exceptional events are similar, if not the same, for each detector network and for both populations. This is because we only consider events within the redshift of $z = 0.1$ for the golden dark siren study. The next generation detector networks perform well in terms of measurement ability at such distances, resulting in precise luminosity distance measurements. From Table \ref{tab:dark_sirens}, we note that there is a chance of observing an exceptional event as early as the Voyager era, which can help resolve the \HL tension by itself. Obtaining $\mathcal{O}(1)$ to $\mathcal{O}(10)$ exceptional events \textit{every year} with \texttt{KI+EC} and \texttt{ECS} will also allow one to probe the possible anisotropy in the value of $H_0$ \citep{CalderonBustillo:2020kcg}. 

While we expect to detect $\mathcal{O}(10)$ golden/exceptional events every year with \texttt{ECS}, these constitute only $0.03\%$ of all the NSBH events that ECS is expected to detect in a year. That makes both the occurrence and the observation of such events extremely rare. For comparison with BBH systems, a \texttt{ECS}-like network is expected to detect $\sim 20$ BBH golden dark siren events every year \citep{Borhanian:2020vyr}, which constitutes $0.02\%$ of the BBH events that the network will detect that year \citep{Borhanian:2022czq}. If we only consider events with $z \leq 0.1$, $\sim 90\%$ of the BBH events and $\sim 50\%$ of the NSBH events detected by a \texttt{ECS}-like network are golden dark sirens. Further, \citet{Borhanian:2020vyr} claims that the \texttt{HLVKI+} network can resolve the \HL tension within $2$ years of observation with BBH systems, whereas we see that it is unable to detect any NSBH dark sirens in the span of $5$ years. One must note that in \citet{Borhanian:2020vyr}, the contribution of the uncertainty in $\Omega_{m}$ towards the measurement accuracy of $H_0$ has not been considered. This omission will result in an optimistic forecast of the measurement of $H_0$ using BBH systems.

%%%%%%%%%%%%%%%%%%%%%%%%%%%%%%%%%%%%%%%%%%%%%%%%%%

%%%%%%%%%%%%%%%%% SECTION BREAK %%%%%%%%%%%%%%%%%%

\section{NSBH systems as Bright Sirens} \label{sec:bright_sirens}
The bright siren approach relies on observing an EM counterpart following the GW detection of the merger event. This technique removes the stringent constraint on measuring the sky position using GWs to high accuracy as in the golden dark siren approach, potentially allowing more events to contribute to $H_0$ measurement. However, a limiting factor is the low likelihood of observing an EM counterpart, which in our case is a KN. The generation of a KN is sensitive to the population characteristics, especially the asymmetric mass ratio and the spin of the BH, and the NS EOS. A population with predominantly non-spinning black holes (supported by the current observations \citep{Biscoveanu:2022iue,Zhu:2021jbw}) or one where BHs have retrograde ($\chi_{\mathrm{BH}}<0$) spins, will disfavor KN generation. In contrast, a population that contains BHs with high prograde $(\chi_{\mathrm{BH}}>0)$ spins will encourage it \citep{Kyutoku:2021icp}. Our two population models take both these effects into account-- Pop-1 contains BHs with uniformly sampled spins from $[-0.75,0.75]$, and Pop-2 contains predominantly non-spinning BHs with a small fraction $(\sim 5\%)$ of BHs with high $(\chi_{\mathrm{BH}}>0.75)$ prograde spins.

The EOS of the NS will significantly affect the generation of the KN. In particular, a \textit{softer} EOS like \texttt{APR4} \citep{Akmal:1998cf} will lead to a more compact NS and disfavor tidal disruption. On the other hand, a \textit{stiffer} EOS like \texttt{DD2} \citep{Typel:2009sy} results in a less compact NS, allowing for tidal disruption of the NS before it passes the innermost stable circular orbit, $R_{\mathrm{ISCO}}$, and enabling the generation of the KN. We take this into consideration in our study by reporting the number of KN detections for three EOSs of varying stiffness: \texttt{ALF2} \citep{Alford:2004pf}, \texttt{APR4} and \texttt{DD2}. 

As noted in \citet{Gupta:2023evt}, the $r-$filter in Rubin and the $R-$filter in Roman have similar effective wavelength $(\lambda_{\mathrm{eff}})$ and correspond to the highest numbers of KN detections, so we consider these two filters in the current analysis. The specifications for the two telescopes, including the limiting magnitudes $(m_{\mathrm{lim}})$ are mentioned in Table \ref{tab:telescopes}. Table \ref{apptab:r_R_kn_numbers} contains the number of KN detections for the two telescopes, categorized by the EOS and the $90\%-$credible sky area, for the median local merger rate density. The upper and lower limits of the listed values correspond to the high and low merger rate densities.

The number of bright sirens detected by each EM telescope for the three EOSs is presented in Table \ref{tab:bs_golden}. Among these, we will also have exceptional events-- events for which the measurement error in luminosity distance is $<2\%$ and the merger is followed by a KN detected by the EM telescopes. The number of exceptional events corresponding to each detector network and EOS are also listed in Table \ref{tab:bs_golden}. All the numbers presented are for an observation time of $10$ years and $\Omega_{90} < 10\,\times$ FOV. The number of bright sirens detected in a span of 10 years can range from $\mathcal{O}(1)$ to $\mathcal{O}(100)$. For events in both populations, there is no chance of detecting exceptional bright siren events with \texttt{HLVKI+} and \texttt{VK+HLIv} networks. For \texttt{KI+EC} and \texttt{ECS}, unlike the dark siren case, only $10\%-50\%$ of the bright siren events qualify as exceptional events, with \texttt{ECS} expected to detect $\mathcal{O}(1)$ exceptional events every year (unless \texttt{APR4} is the preferred NS EOS). 

A realistic KN detection scenario would involve a target-of-opportunity (TOO) follow-up strategy to detect KN events based on alerts from GW observations. We use one such TOO strategy in this work, where we consider a combination of $g+i$ filters in the Rubin observatory for the detection of the KN \citep{Andreoni:2021epw, Branchesi:2023mws}. Specifically, the KN is said to be detected if the luminosity is brighter than the limiting magnitude of the $g$ and the $i$ filters on two consecutive nights with $600$ seconds of exposure in each filter. The most optimistic values of limiting magnitudes obtained for an exposure of $600$ s for the $g$ and $i$ filters are $26.62$ and $25.62$, respectively. In addition, we only consider events for which $\Omega_{90} \leq 9.6\,\mbox{deg}^2$, i.e., their localization is smaller than the FOV of the Rubin telescope. One difference, compared to \citet{Branchesi:2023mws} and \citet{Gupta:2023evt}, is that we consider a $100\%$ duty-cycle for Rubin instead of a $50\%$ duty-cycle considered in the other works. This will lead to slightly optimistic estimates. The number of bright siren events and exceptional events observed using this strategy are given in Table \ref{tab:bs_golden_too}. Comparing these values with the number of detections for Rubin in Table \ref{tab:bs_golden}, we see that the TOO strategy leads to $2-3$ times higher number of detections. This is attributed to the longer exposure time (600 s) which leads to the detections of intrinsically dimmer events and bright events that are farther away, compared to the detections made using single 30 s exposures with Rubin. The trend in the detection of exceptional events is similar to that in Table \ref{tab:bs_golden}, with no golden events detected with \texttt{HLVKI+} and \texttt{VK+HLIv}, and $\mathcal{O}(1)$ detections with \texttt{ECS} every year.

\begin{table}
  \centering
  \caption{\label{tab:telescopes}Information about the two EM telescopes considered in this study. An exhaustive collection of such information for other EM telescopes can be found in \citet{Chase:2021ood}.}
  \renewcommand{\arraystretch}{1.2} 
    \begin{tabular}{l P{1cm} P{1cm}}
    \hhline{---}
    Telescope & Rubin & Roman\\
    \hhline{---}
    FOV $(\mathrm{deg^2})$ & 9.6 & 0.28 \\
    Exposure time (s) & $30$ s  & $67$ s \\
    Filter & $r$ & $R$  \\
    $\lambda_{\mathrm{eff}}$ (\r{A}) & 6156  & 6160 \\
    $m_{\mathrm{lim}}$ (AB mag) & 24.7 & 26.2 \\
    \hhline{---}
    \end{tabular}
\end{table}
\begin{table*}
  \centering
  \caption{\label{tab:bs_golden}Number of bright sirens and exceptional events detected by the two telescopes, using the $r-$filter in Rubin and the $R-$filter in Roman, in an observation span of $10$ years. The KN considered correspond to events that can be localized to a sky patch smaller than $10\times$FOV using GW observations, where FOV is the field of view of the particular telescope. }
  \renewcommand{\arraystretch}{1.6} 
    \begin{tabular}{l |P{1cm} P{0.6cm} P{1cm}|P{1cm} P{0.6cm} P{1cm}|P{1cm} P{0.6cm} P{1cm}|P{1cm} P{0.6cm} P{1cm}}
    \hhline{-------------}
    Type & \multicolumn{6}{c|}{Bright Sirens} & \multicolumn{6}{c}{Exceptional Events} \\
    \hhline{-|-|-|-|-|-|-|-|-|-|-|-|-}
    Filter & \multicolumn{3}{c|}{Rubin $r-$filter} & \multicolumn{3}{c|}{Roman $R-$filter} & \multicolumn{3}{c|}{Rubin $r-$filter} & \multicolumn{3}{c}{Roman $R-$filter} \\
    \hhline{-|-|-|-|-|-|-|-|-|-|-|-|-}
    EOS & ALF2 & APR4 & DD2 & ALF2 & APR4 & DD2 & ALF2 & APR4 & DD2 & ALF2 & APR4 & DD2 \\
    \hhline{-------------}
    \multicolumn{13}{c}{\textit{Pop-1}}\\
    \hhline{-------------}
    HLVKI\texttt{+} & $9^{+16}_{-8}$ & $4^{+5}_{-3}$ & $16^{+28}_{-15}$ & $3^{+4}_{-3}$ & $1^{+0}_{-1}$ & $6^{+7}_{-6}$ & $0$ & $0$ & $0$ & $0$ & $0$ & $0$\\
    VK\texttt{+}HLIv & $14^{+25}_{-13}$ & $4^{+9}_{-3}$ & $23^{+42}_{-22}$ & $8^{+16}_{-8}$ & $3^{+5}_{-3}$ & $19^{+42}_{-19}$ & $0$ & $0$ & $0$ & $0$ & $0$ & $0$\\
    VKI\texttt{+}C & $14^{+25}_{-13}$ & $4^{+9}_{-3}$ & $23^{+42}_{-22}$ & $3^{+5}_{-3}$ & $1^{+1}_{-1}$ & $9^{+18}_{-9}$ & $1^{+1}_{-1}$ & $0$ & $1^{+1}_{-1}$ & $1^{+1}_{-1}$ & $0$ & $1^{+1}_{-1}$\\
    HLKI\texttt{+}E & $14^{+25}_{-13}$ & $4^{+9}_{-3}$ & $23^{+42}_{-22}$ & $15^{+28}_{-14}$ & $5^{+7}_{-4}$ & $26^{+61}_{-25}$ & $1^{+2}_{-1}$ & $0$ & $2^{+3}_{-2}$ & $1^{+2}_{-1}$ & $0$ & $2^{+4}_{-2}$\\
    KI\texttt{+}EC & $14^{+25}_{-13}$ & $4^{+9}_{-3}$ & $23^{+42}_{-22}$ & $56^{+100}_{-45}$ & $21^{+36}_{-15}$ & $100^{+193}_{-84}$ & $2^{+6}_{-2}$ & $0^{+1}_{-0}$ & $5^{+13}_{-5}$ & $3^{+8}_{-2}$ & $0^{+3}_{-0}$ & $6^{+17}_{-5}$\\
    ECS & $14^{+25}_{-13}$ & $4^{+9}_{-3}$ & $23^{+42}_{-22}$ & $88^{+157}_{-68}$ & $33^{+60}_{-25}$ & $156^{+293}_{-128}$ & $3^{+13}_{-3}$ & $0^{+5}_{-0}$ & $8^{+24}_{-8}$ & $7^{+24}_{-6}$ & $1^{+7}_{-1}$ & $16^{+44}_{-15}$\\
    \hhline{-------------}
    \multicolumn{13}{c}{\textit{Pop-2}}\\
    \hhline{-------------}
    HLVKI\texttt{+} & $16^{+18}_{-14}$ & $0$ & $31^{+41}_{-26}$ & $3^{+5}_{-2}$ & $0$ & $6^{+12}_{-4}$ & $0$ & $0$ & $0$ & $0$ & $0$ & $0$\\
    VK\texttt{+}HLIv & $19^{+30}_{-16}$ & $0$ & $37^{+82}_{-31}$ & $16^{+21}_{-12}$ & $0$ & $29^{+43}_{-24}$ & $0$ & $0$ & $0$ & $0$ & $0$ & $0$\\
    VKI\texttt{+}C & $20^{+30}_{-17}$ & $0$ & $38^{+83}_{-32}$ & $7^{+10}_{-5}$ & $0$ & $14^{+21}_{-11}$ & $0$ & $0$ & $0^{+2}_{-0}$ & $0$ & $0$ & $0^{+1}_{-0}$\\
    HLKI\texttt{+}E & $20^{+30}_{-17}$ & $0$ & $38^{+84}_{-32}$ & $22^{+42}_{-17}$ & $0$ & $44^{+84}_{-36}$ & $1^{+6}_{-1}$ & $0$ & $4^{+11}_{-4}$ & $1^{+3}_{-1}$ & $0$ & $2^{+4}_{-2}$\\
    KI\texttt{+}EC & $20^{+30}_{-17}$ & $0$ & $38^{+84}_{-32}$ & $79^{+173}_{-64}$ & $0$ & $160^{+330}_{-132}$ & $5^{+15}_{-3}$ & $0$ & $13^{+31}_{-10}$ & $6^{+19}_{-4}$ & $0$ & $19^{+39}_{-15}$\\
    ECS & $20^{+30}_{-17}$ & $0$ & $38^{+84}_{-32}$ & $115^{+262}_{-92}$ & $0$ & $247^{+512}_{-202}$ & $7^{+17}_{-5}$ & $0$ & $18^{+42}_{-15}$ & $16^{+40}_{-13}$ & $0$ & $38^{+81}_{-31}$\\
    \hhline{-------------}
    \end{tabular}
\end{table*}
\begin{figure*}
\begin{subfigure}{\linewidth}
\centering
  \includegraphics[scale=0.6,trim = 0 15 0 0,clip]{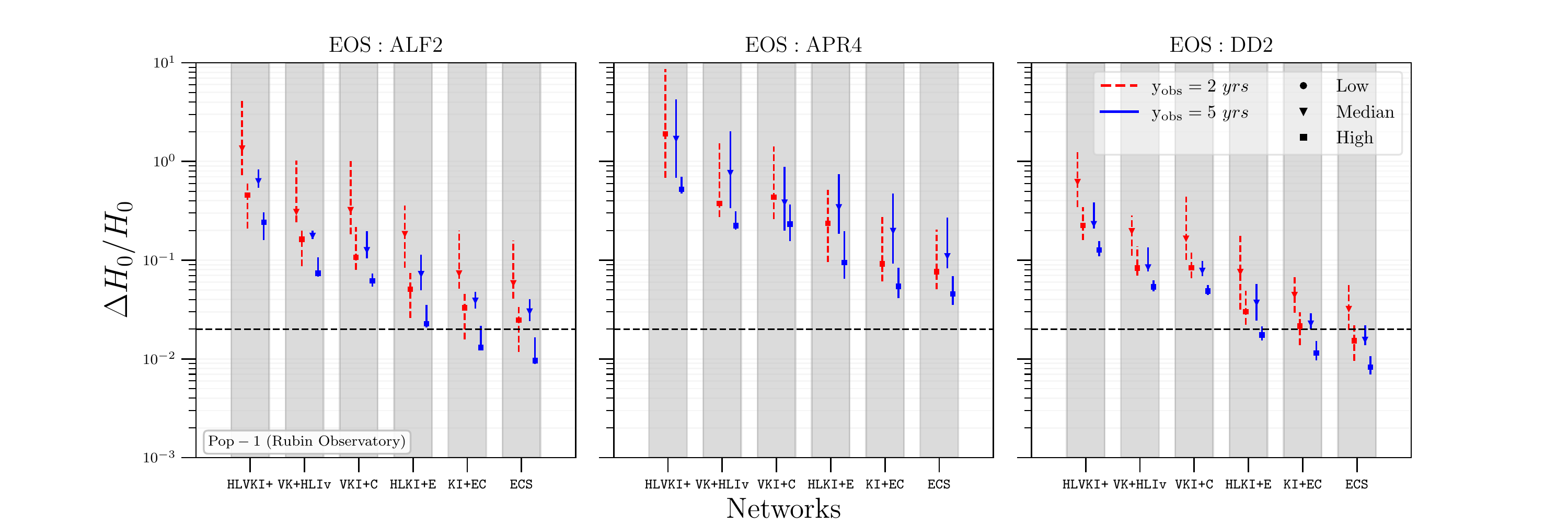}
\end{subfigure}\vspace{-1em}
\begin{subfigure}{\linewidth}
  \centering
  \includegraphics[scale=0.6,trim = 0 15 0 0,clip]{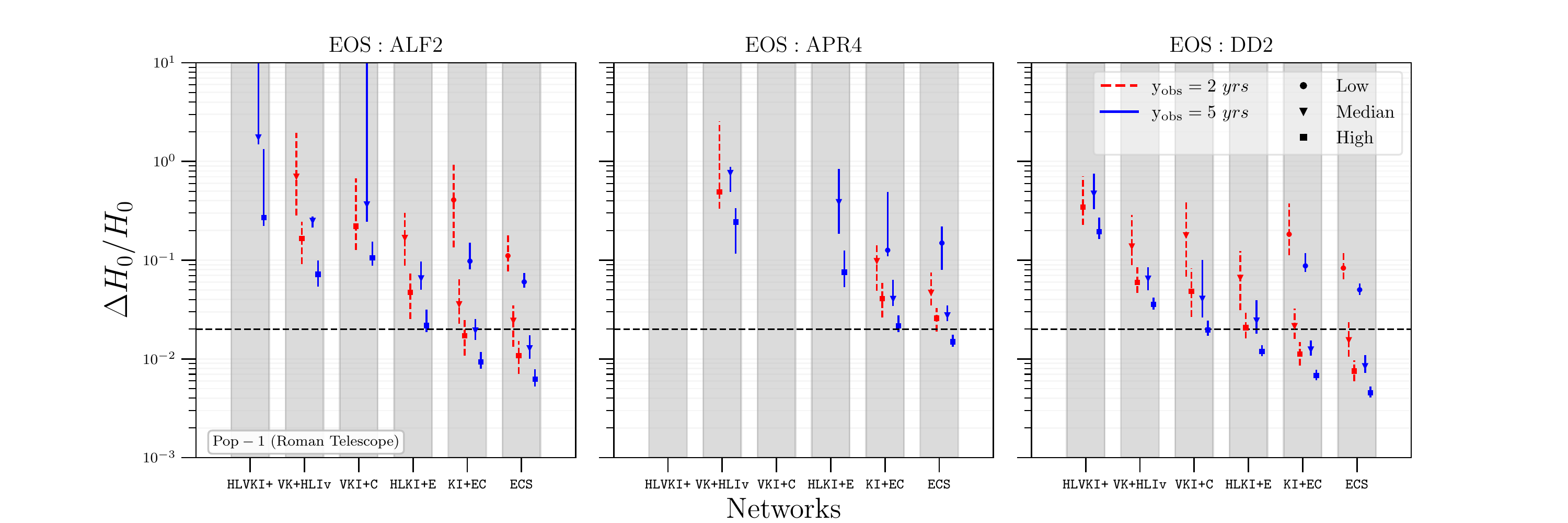}
\end{subfigure}\vspace{-1em}
\begin{subfigure}{\linewidth}
\centering
  \includegraphics[scale=0.6,trim = 0 15 0 0,clip]{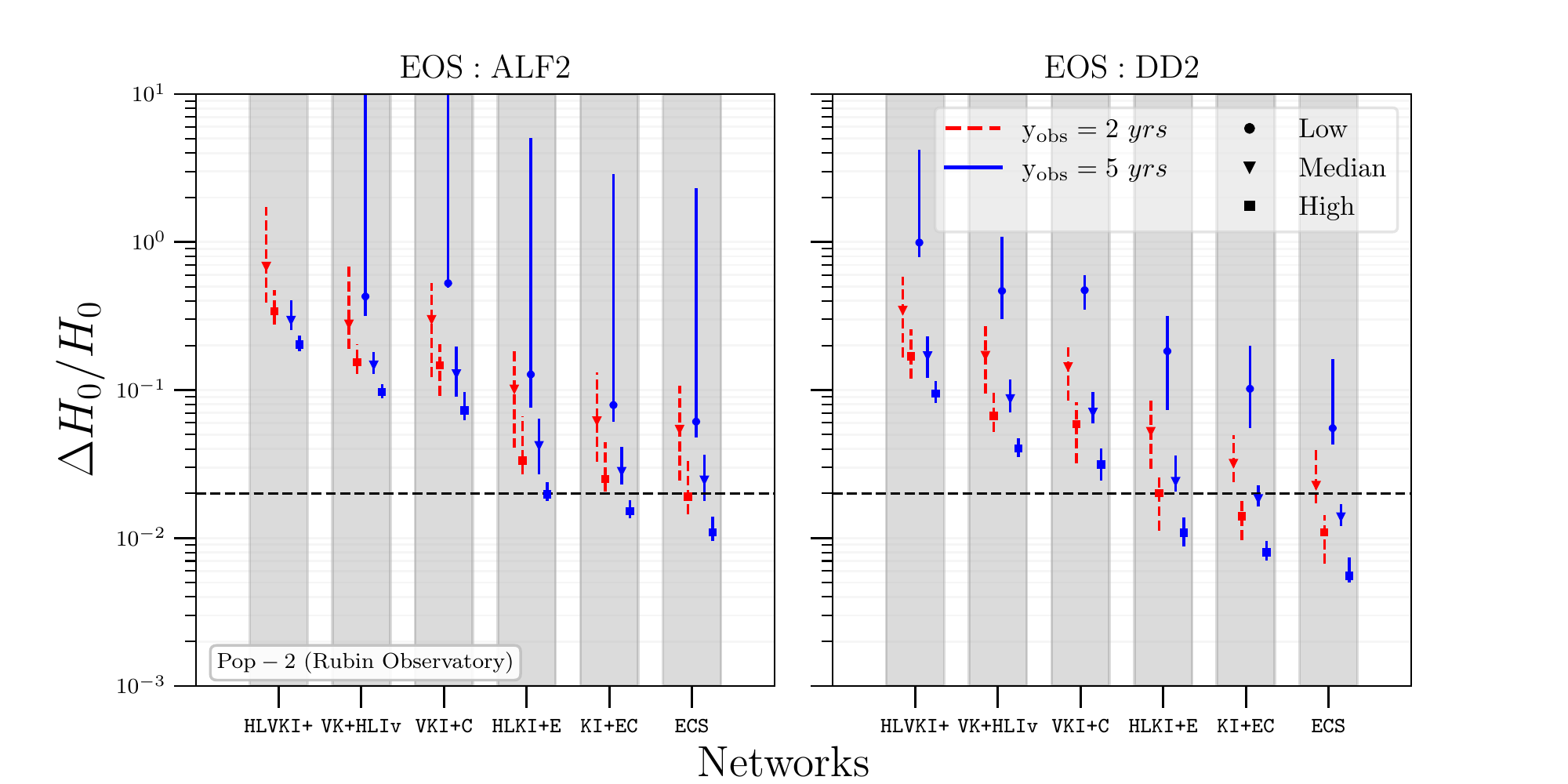}
\end{subfigure}\vspace{-1em}
\begin{subfigure}{\linewidth}
  \centering
  \includegraphics[scale=0.6]{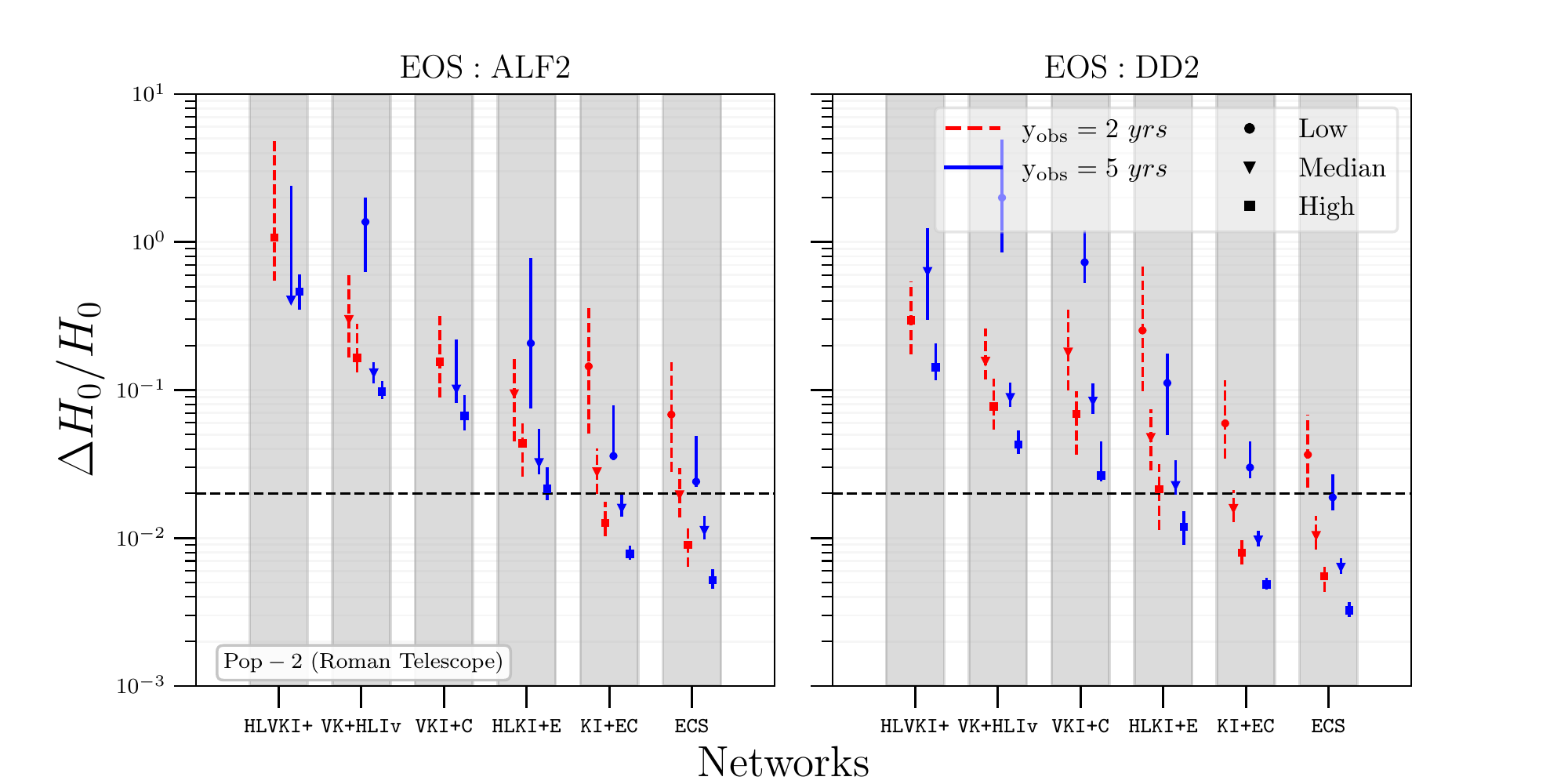}
\end{subfigure}\vspace{0em}
\caption{\label{fig:bs_Om0} The fractional errors in $H_0$ measurement using NSBH systems as bright sirens for the case when prior on $\Omega_m$ is \textit{not} included. The top two panels show the errors for events in Pop-1 and the bottom two panels show the errors for Pop-2.}
\end{figure*}
\begin{figure*}
\begin{subfigure}{\linewidth}
\centering
  \includegraphics[scale=0.6,trim = 0 15 0 0,clip]{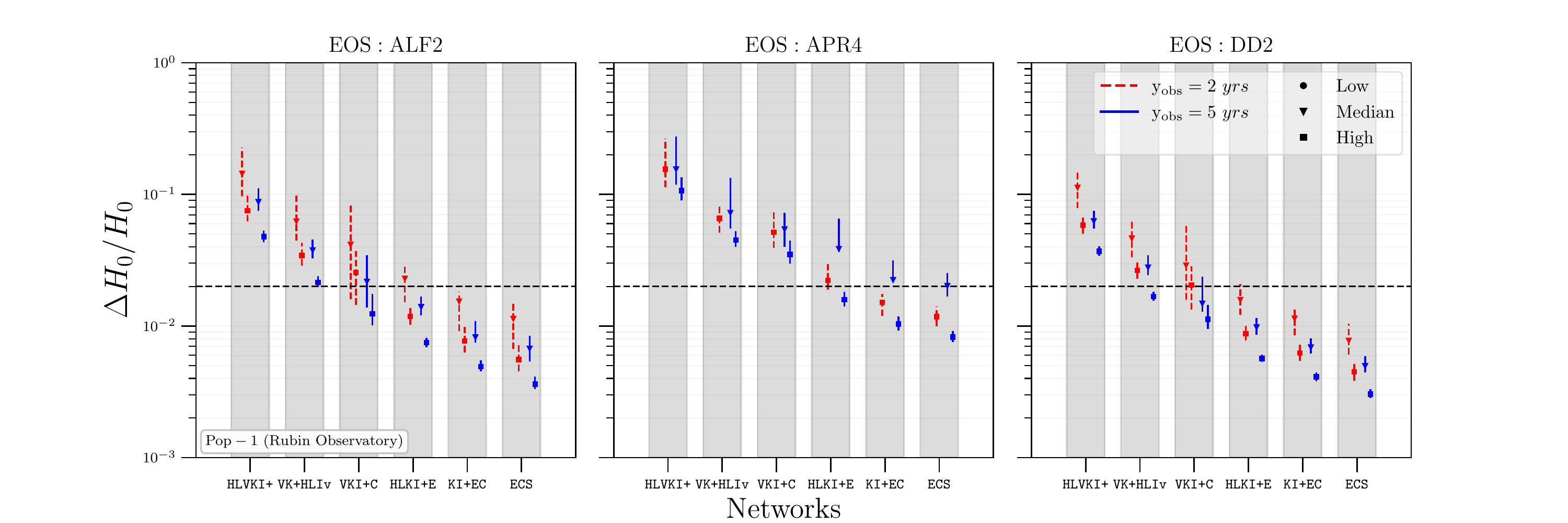}
\end{subfigure}\vspace{-1em}
\begin{subfigure}{\linewidth}
  \centering
  \includegraphics[scale=0.6,trim = 0 15 0 0,clip]{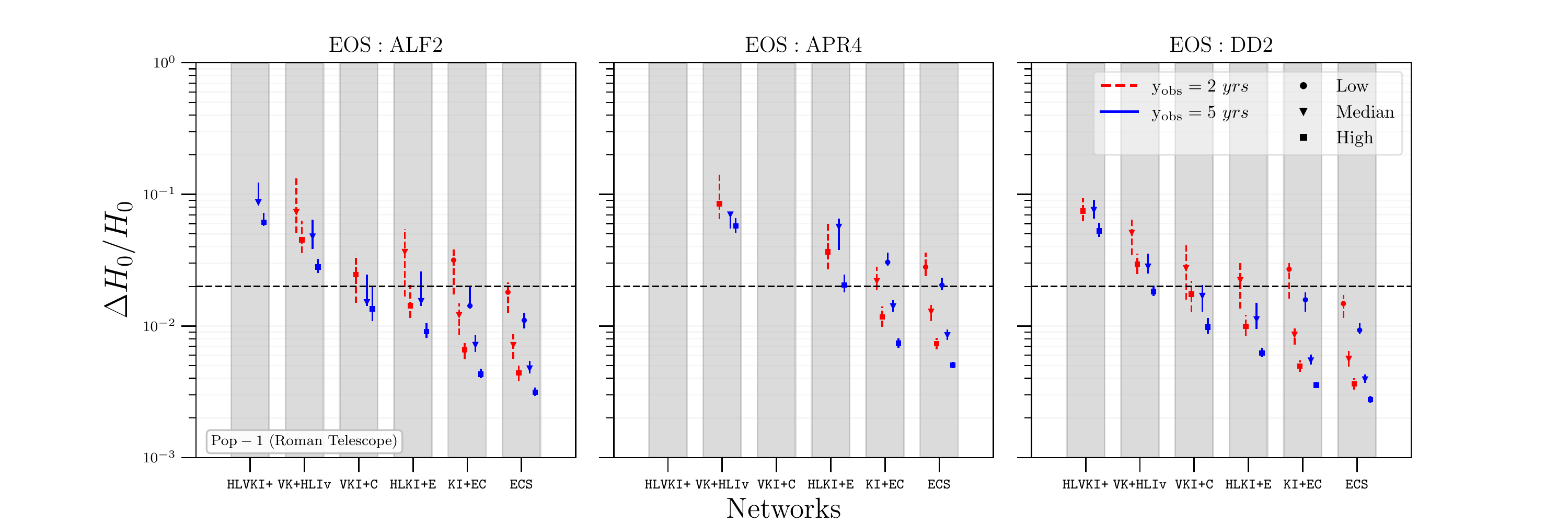}
\end{subfigure}\vspace{-1em}
\begin{subfigure}{\linewidth}
\centering
  \includegraphics[scale=0.6,trim = 0 15 0 0,clip]{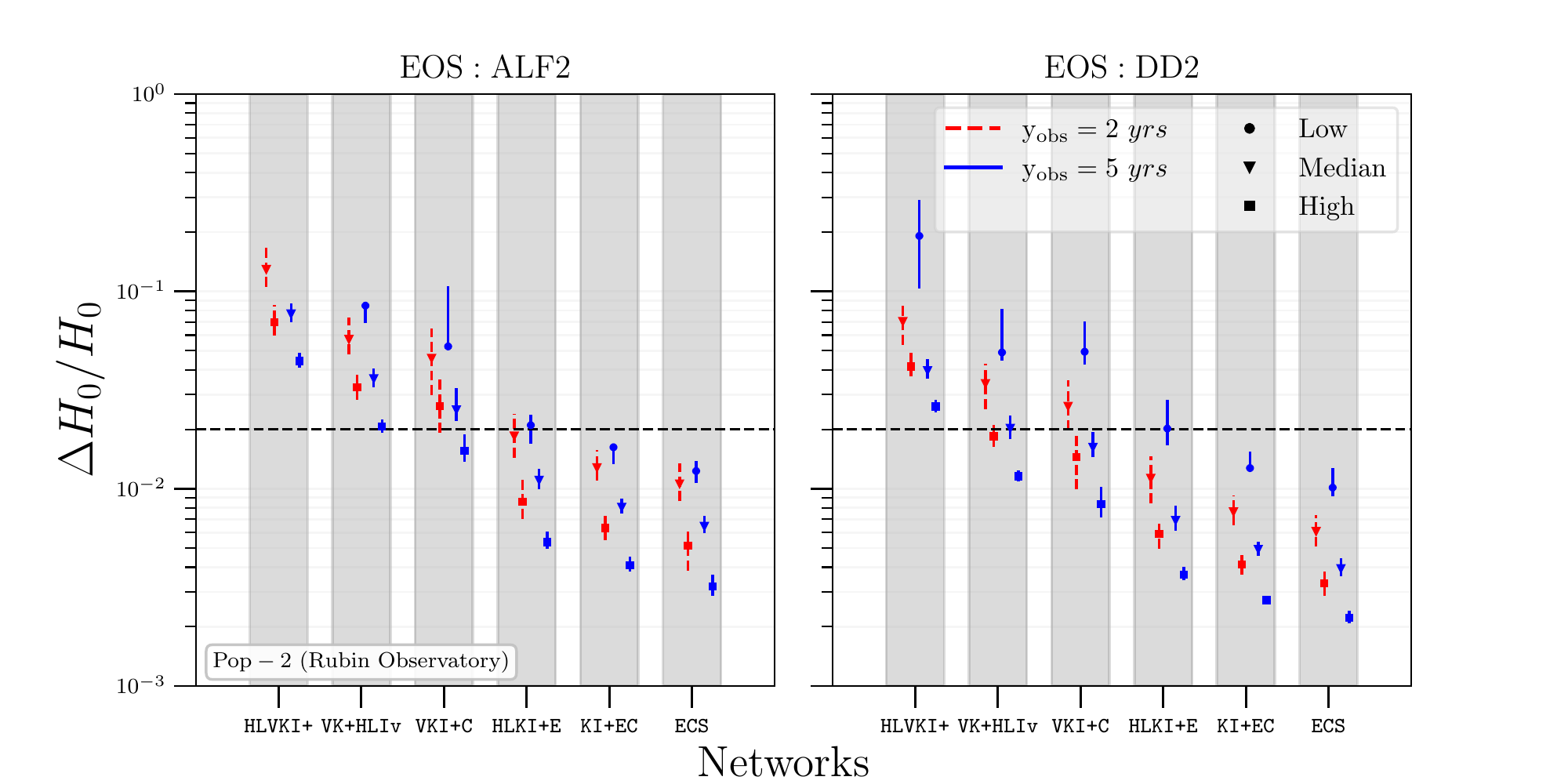}
\end{subfigure}\vspace{-1em}
\begin{subfigure}{\linewidth}
  \centering
  \includegraphics[scale=0.6]{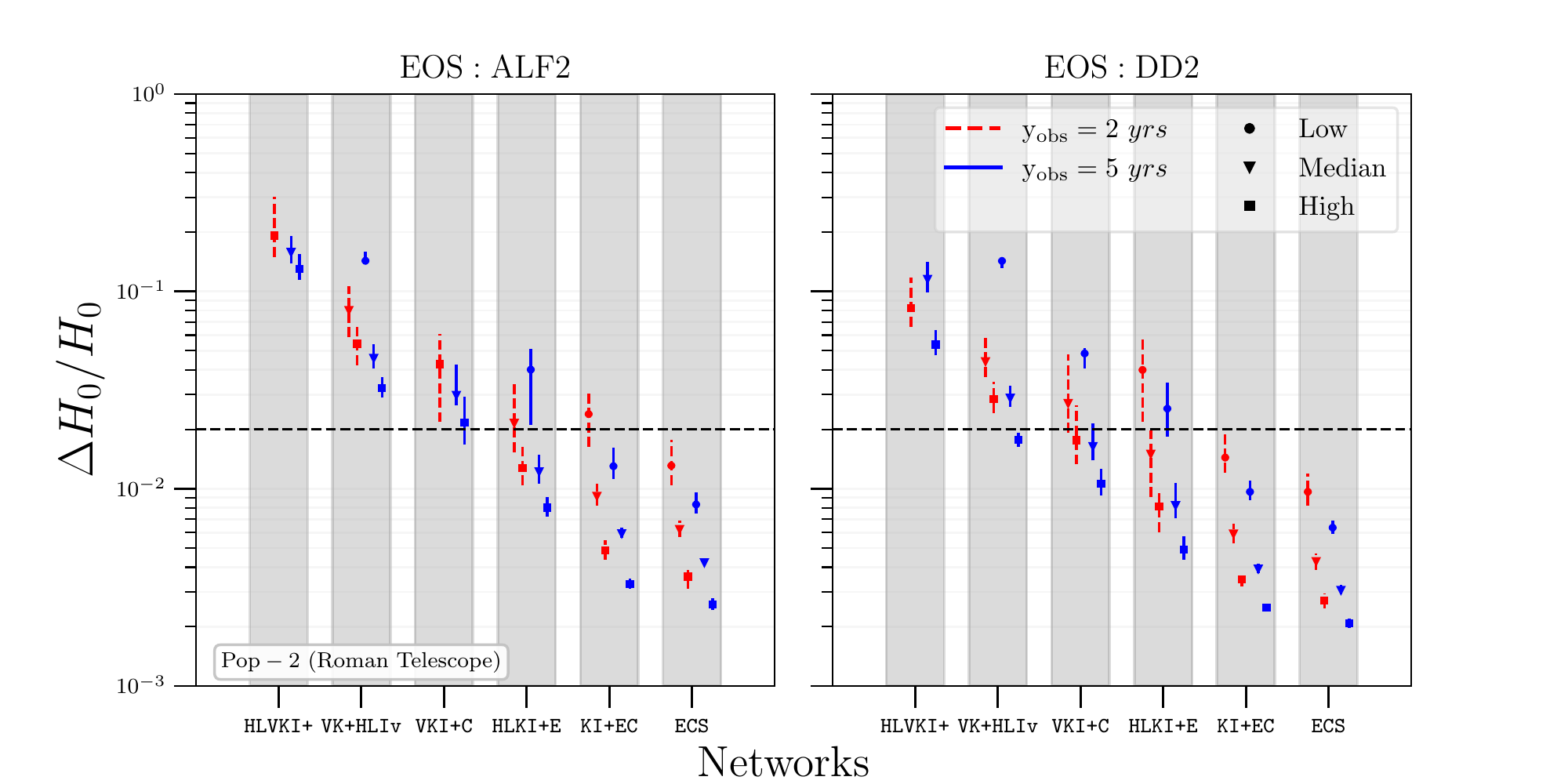}
\end{subfigure}\vspace{0em}
\caption{\label{fig:bs_Om0_rest} The fractional errors in $H_0$ measurement using NSBH systems as bright sirens for the case when prior on $\Omega_m$ is included. The top two panels show the errors for events in Pop-1 and the bottom two panels show the errors for Pop-2.}
\end{figure*}
\begin{figure*}
\begin{subfigure}{\linewidth}
\centering
  \includegraphics[scale=0.6,trim = 0 15 0 0,clip]{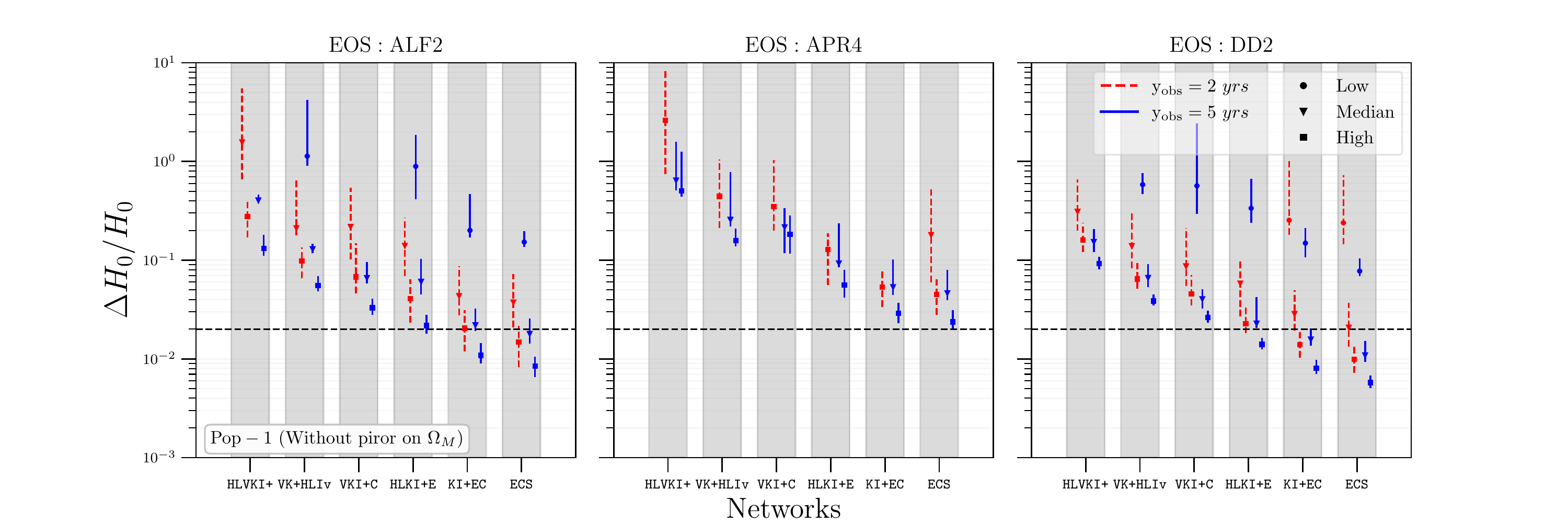}
\end{subfigure}\vspace{-1em}
\begin{subfigure}{\linewidth}
  \centering
  \includegraphics[scale=0.6,trim = 0 15 0 0,clip]{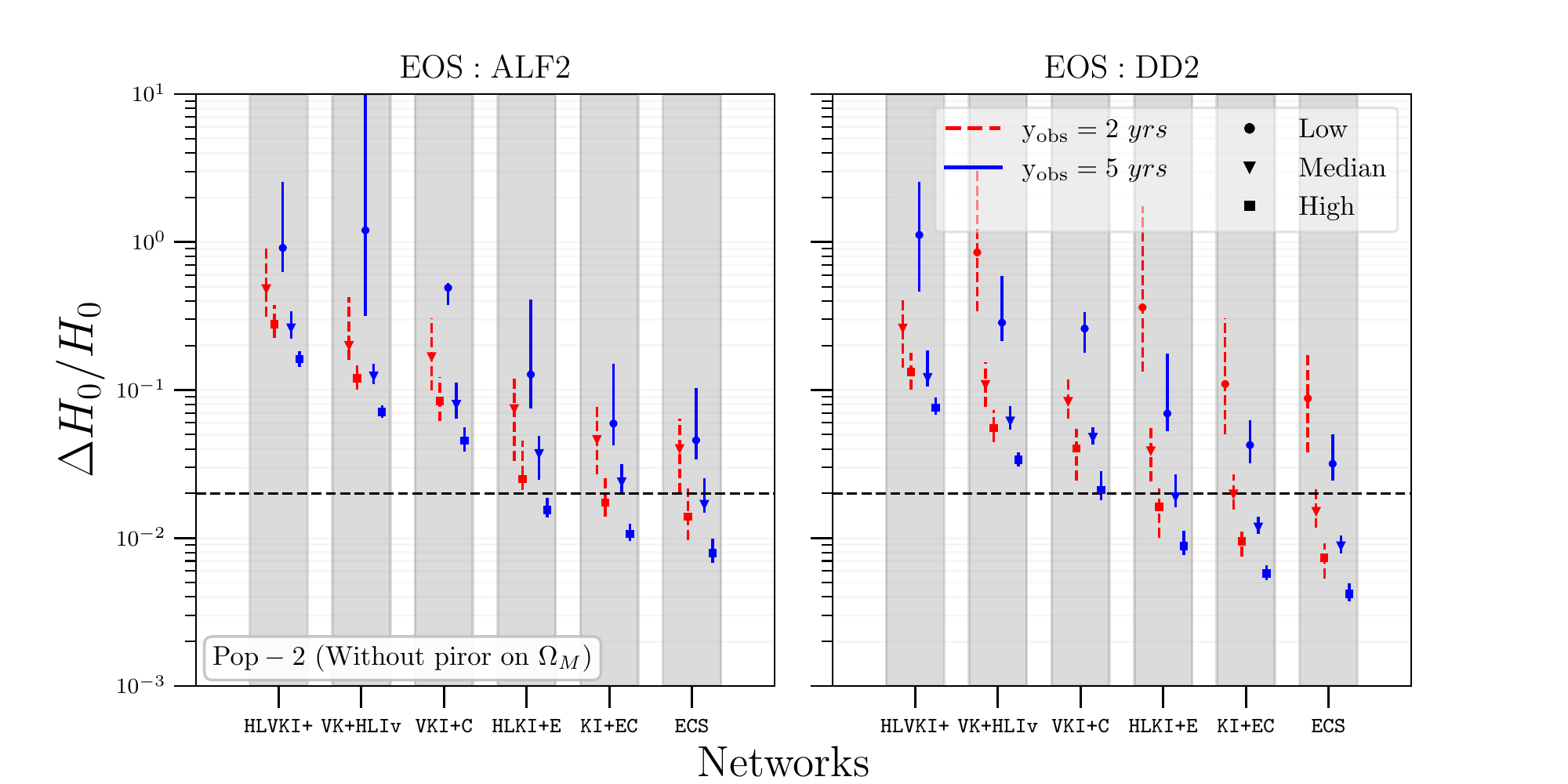}
\end{subfigure}\vspace{-1em}
\begin{subfigure}{\linewidth}
\centering
  \includegraphics[scale=0.6,trim = 0 15 0 0,clip]{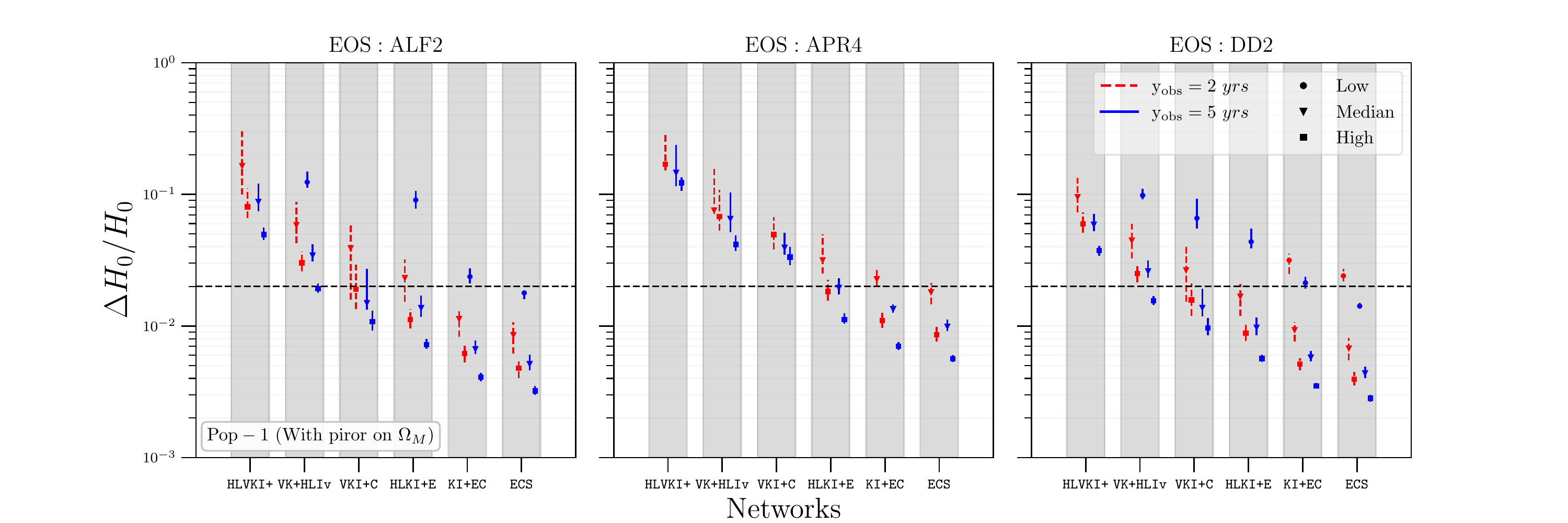}
\end{subfigure}\vspace{-1em}
\begin{subfigure}{\linewidth}
  \centering
  \includegraphics[scale=0.6]{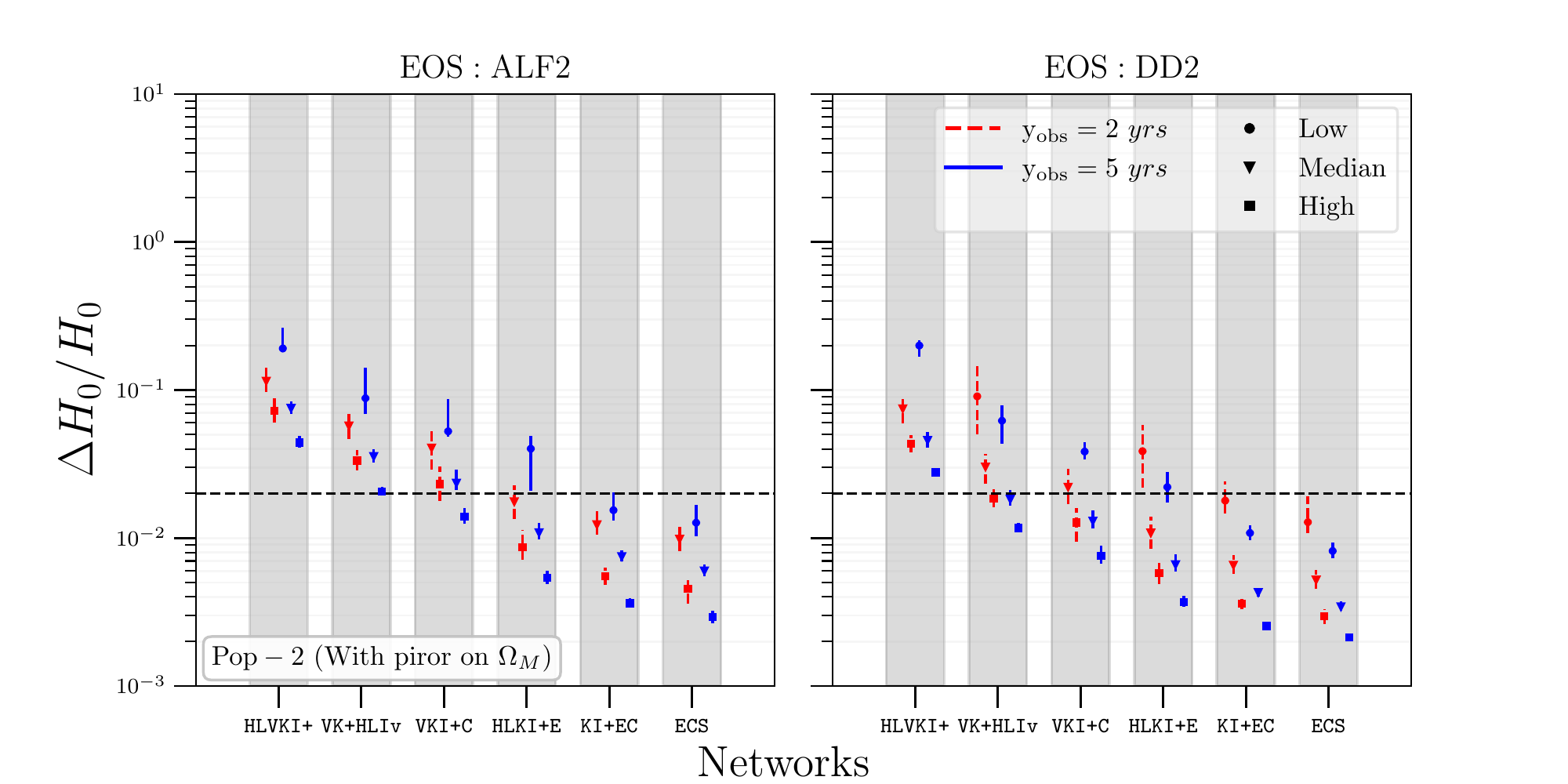}
\end{subfigure}\vspace{0em}
\caption{\label{fig:bs_too} The fractional errors in $H_0$ measurement using NSBH systems as bright sirens when the $g+i$ TOO strategy is followed with the Rubin observatory. The top two panels show the errors for events in Pop-1 and Pop-2 when prior on $\Omega_m$ is not included, and the bottom two panels show the errors for the two populations when the prior on $\Omega_m$ is included.}
\end{figure*}

\begin{table}
  \centering
  \caption{\label{tab:bs_golden_too}Number of bright sirens and exceptional events detected using the $g+i$ TOO strategy with Rubin in an observation span of $10$ years. The KN considered correspond to events that can be localized better than the FOV of Rubin using GW observations.}
  \renewcommand{\arraystretch}{1.7} 
    \begin{tabular}{l |P{0.7cm} P{0.7cm} P{0.7cm}|P{0.7cm} P{0.7cm} P{0.7cm}}
    \hhline{-------}
    Type & \multicolumn{3}{c|}{Bright Sirens} & \multicolumn{3}{c}{Exceptional Events} \\
    \hhline{-|-|-|-|-|-|-}
    EOS & ALF2 & APR4 & DD2 & ALF2 & APR4 & DD2\\
    \hhline{-------}
    \multicolumn{7}{c}{\textit{Pop-1}}\\
    \hhline{-------}
    HLVKI\texttt{+} & $12^{+24}_{-11}$ & $6^{+4}_{-5}$ & $23^{+38}_{-21}$ & $0$ & $0$ & $0$\\
    VK\texttt{+}HLIv & $23^{+45}_{-19}$ & $8^{+12}_{-7}$ & $40^{+72}_{-33}$ & $0$ & $0$ & $0$\\
    VKI\texttt{+}C & $15^{+32}_{-13}$ & $6^{+7}_{-5}$ & $31^{+58}_{-26}$ & $1^{+1}_{-1}$ & $0$ & $1^{+1}_{-1}$\\
    HLKI\texttt{+}E & $25^{+58}_{-22}$ & $10^{+22}_{-9}$ & $43^{+91}_{-37}$ & $1^{+2}_{-1}$ & $0$ & $2^{+3}_{-2}$\\
    KI\texttt{+}EC & $39^{+81}_{-33}$ & $12^{+29}_{-10}$ & $60^{+126}_{-51}$ & $4^{+12}_{-4}$ & $0^{+3}_{-0}$ & $6^{+19}_{-6}$\\
    ECS & $40^{+82}_{-33}$ & $12^{+29}_{-10}$ & $65^{+131}_{-54}$ & $8^{+23}_{-8}$ & $2^{+9}_{-2}$ & $13^{+35}_{-13}$\\
    \hhline{-------}
    \multicolumn{7}{c}{\textit{Pop-2}}\\
    \hhline{-------}
    HLVKI\texttt{+} & $18^{+23}_{-15}$ & $0$ & $34^{+51}_{-28}$ & $0$ & $0$ & $0$\\
    VK\texttt{+}HLIv & $25^{+47}_{-21}$ & $0$ & $58^{+111}_{-49}$ & $0$ & $0$ & $0$\\
    VKI\texttt{+}C & $24^{+35}_{-21}$ & $0$ & $46^{+82}_{-39}$ & $0$ & $0$ & $0^{+2}_{-0}$\\
    HLKI\texttt{+}E & $28^{+53}_{-24}$ & $0$ & $68^{+134}_{-58}$ & $1^{+6}_{-1}$ & $0$ & $4^{+10}_{-4}$\\
    KI\texttt{+}EC & $32^{+78}_{-26}$ & $0$ & $89^{+201}_{-72}$ & $5^{+18}_{-3}$ & $0$ & $14^{+38}_{-11}$\\
    ECS & $32^{+80}_{-26}$ & $0$ & $90^{+212}_{-73}$ & $7^{+24}_{-5}$ & $0$ & $23^{+59}_{-19}$\\
    \hhline{-------}
    \end{tabular}
\end{table}

To obtain the combined estimates on the measurement errors in $H_0$ corresponding to an observation span and merger rate density, we follow the same process used for dark sirens in Section \ref{sec:dark_sirens} for each EOS and the two telescopes. However, instead of picking events where $\Omega_{90} \leq 0.04\,\mathrm{deg}^2$, we choose events for which the corresponding KN are detected by a given EM telescope. Figure \ref{fig:bs_Om0} shows the estimates on fractional errors in the measurement of $H_0$ using bright sirens without including a prior on $\Omega_m$, whereas Fig. \ref{fig:bs_Om0_rest} shows the errors when prior for $\Omega_m$ is included. We also calculate the bounds that can be put on $H_0$ using the $g+i$ TOO strategy, shown in Fig. \ref{fig:bs_too}.

Firstly, we do not get any bright sirens for \texttt{APR4}, which is the softest of the three EOSs, for events in Pop-2. Here, we note that the numerical fits used to generate the KN light curves are only valid for $\chi_{\mathrm{BH}} < 0.75$. Thus, even though Pop-2 has NSBH events with high BH spins favorable for KN generation, they have not been considered for the bright siren study. Hence, the numbers reported in this work might underestimate the number of detections. 

Even for Pop-1, \texttt{APR4} gives the least KN detections and the worst bounds on $H_0$. For this EOS, with no prior for $\Omega_m$, only \texttt{KI+EC} and \texttt{ECS} seem to have a chance at resolving the \HL tension with Pop-1 events and the Roman telescope. Even for the two remaining EOSs, \texttt{HLVKI+} cannot measure $H_0$ better than an error of $2\%$, even in an observation span of $5$ years. For \texttt{VK+HLIv} and \texttt{VKI+C}, the $2\%$ mark lies outside the $68\%$ confidence interval of the fractional error in $H_0$, making it improbable that these networks will resolve the \HL tension in $5$ years. \texttt{HLKI+E} will be capable of resolving the tension in $5$ years of observation span, with \texttt{KI+EC} and \texttt{ECS} capable of doing the same in $2$ years if the local NSBH merger rate is high. Moreover, for Pop-2 events, the \HL tension can be resolved with the \texttt{ECS} network and the Roman telescope in $5$ years of observation even if the local merger rate density of NSBH systems is on the lower end. Inclusion of the prior for $\Omega_m$ improves the errors on $H_0$ drastically, allowing the Voyager network to resolve the tension in $5$ years. With Roman, both \texttt{KI+EC} and \texttt{ECS} can constrain $H_0$ to better than $2\%$ in only two years, even if the merger rate density of NSBH systems is low. When the $g+i$ TOO strategy is considered, we see that the $H_0$ measurement accuracy improves slightly compared to the case of Rubin, which is due to the greater number of KN detections.

If using BNS systems as bright sirens, \citet{Chen:2020zoq} finds that in $5$ years of observation, the \texttt{HLV+} network (with $50\%$ duty-cycle) with the Rubin observatory will constrain $H_0$ to $\sim 2\%$ with $12$ bright siren observations every year. In comparison, for the best-case scenario, NSBH systems will result in a $\sim 3\%$ bound on $H_0$ with $\sim 3$ bright siren detections every year using \texttt{HLVKI+} ($100\%$ duty-cycle). With a median local merger rate $\sim 8$ times that of NSBH systems, BNS systems are expected to outperform NSBHs in bright siren measurements of $H_0$, unless the NSBH population contains highly spinning or precessing objects \citep{Vitale:2018wlg}. 

Compared to the dark siren case, we get $\sim 2$ times as many bright siren events detected with the Roman telescope. While these numbers are encouraging, it is crucial to note that the bright siren method relies heavily on the EOS used to describe the NS and the characteristics of the NSBH population. A soft EOS like \texttt{APR4} paired with Pop-2 results in zero KN detections, ruling out the bright siren approach for the resolution of $H_0$ using NSBH mergers. On the other hand, using a stiff EOS like \texttt{DD2} with either of the two populations and a high local merger rate can resolve the \HL tension with \texttt{VK+HLIv} only in $2$ years. Thus, the bright siren approach in resolving the \HL tension is contingent on the NS EOS, the KN generation mechanism, the NSBH population characteristics, and the local merger rate density. Our treatment tries to account for some of these uncertainties by using multiple population models, EOSs, and local merger rate densities and estimating the range of measurement errors on $H_0$ using NSBH systems as bright sirens. 
%%%%%%%%%%%%%%%%%%%%%%%%%%%%%%%%%%%%%%%%%%%%%%%%%%

%%%%%%%%%%%%%%%%% SECTION BREAK %%%%%%%%%%%%%%%%%%

\section{Gray Sirens} \label{sec:gray_sirens}
In Sections \ref{sec:dark_sirens} and \ref{sec:bright_sirens}, we evaluated the potential of NSBH systems as dark sirens and bright sirens, respectively. While networks like \texttt{HLVKI+} and \texttt{VKI+C} could not constrain $H_0$ in an observation span of $5$ years using the dark siren approach, the bright siren method enabled the measurement of $H_0$ with these detectors to better than $10\%$ in the same observation span. Events detected by \texttt{KI+EC} and \texttt{ECS} performed well both as dark and as bright sirens. We also saw that the bright siren approach depends significantly on the EOS, with the worst bounds on $H_0$ obtained for \texttt{APR4}. The dark siren and the bright siren approaches have the same end product-- measurement of $H_0$. Thus, the bounds on $H_0$ from both approaches can be combined to obtain a more precise measurement of $H_0$. This allows us to assess the utility of NSBH systems as \textit{gray sirens}. 

\begin{figure*}
\begin{subfigure}{\linewidth}
\centering
  \includegraphics[scale=0.6,trim = 0 15 0 0,clip]{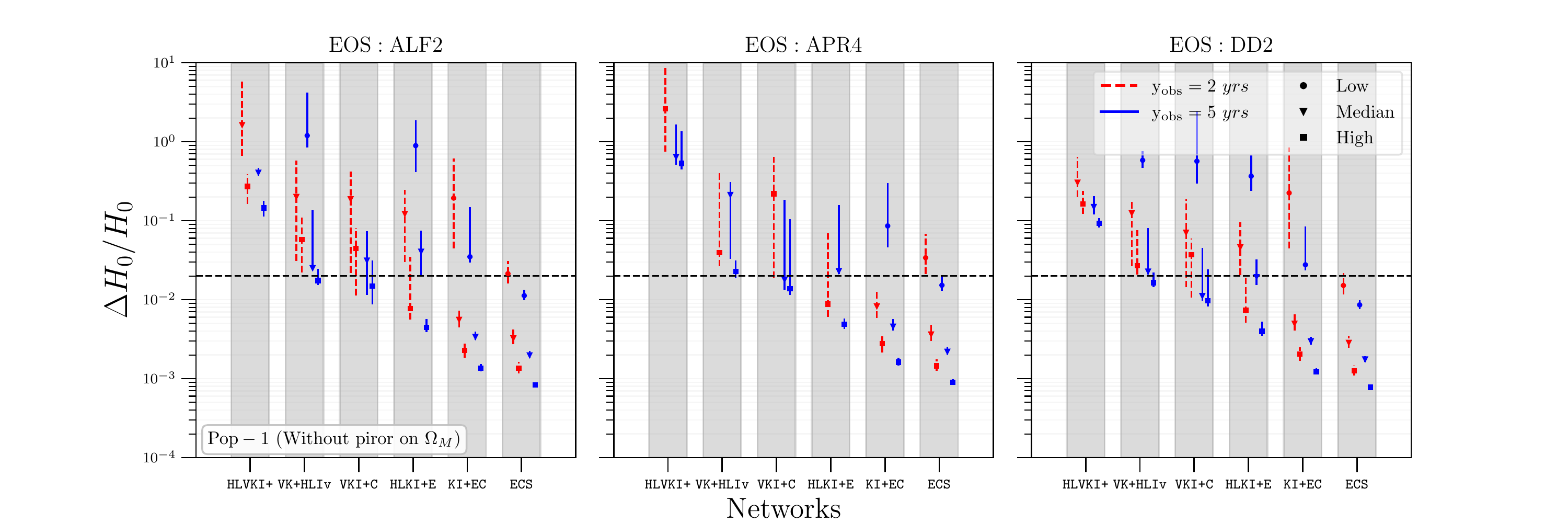}
\end{subfigure}\vspace{-1em}

\begin{subfigure}{\linewidth}
\centering
  \includegraphics[scale=0.6]{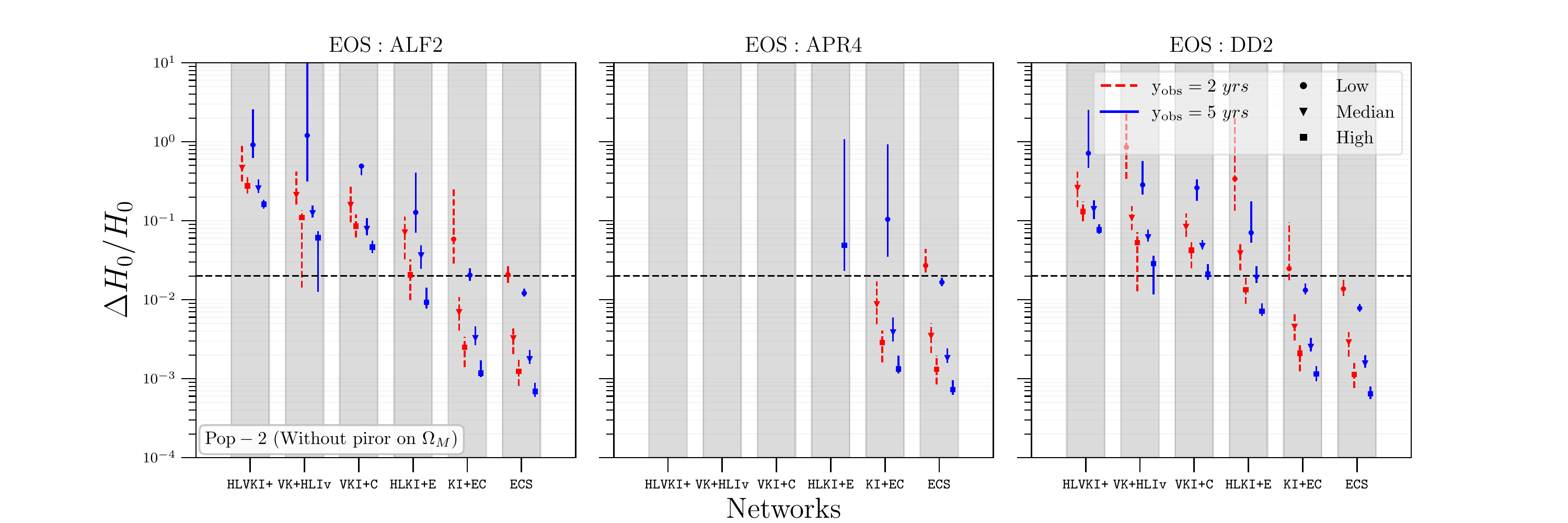}
\end{subfigure}\vspace{-1em}
\caption{\label{fig:gs_Om0} The fractional errors in $H_0$ measurement using NSBH systems as golden gray sirens for the case when prior on $\Omega_m$ is \textit{not} included. Here, we have only considered the contribution of those bright siren events that were detected using the $g+i$ TOO strategy with Rubin.}
\end{figure*}

\begin{figure*}
\begin{subfigure}{\linewidth}
\centering
  \includegraphics[scale=0.6,trim = 0 15 0 0,clip]{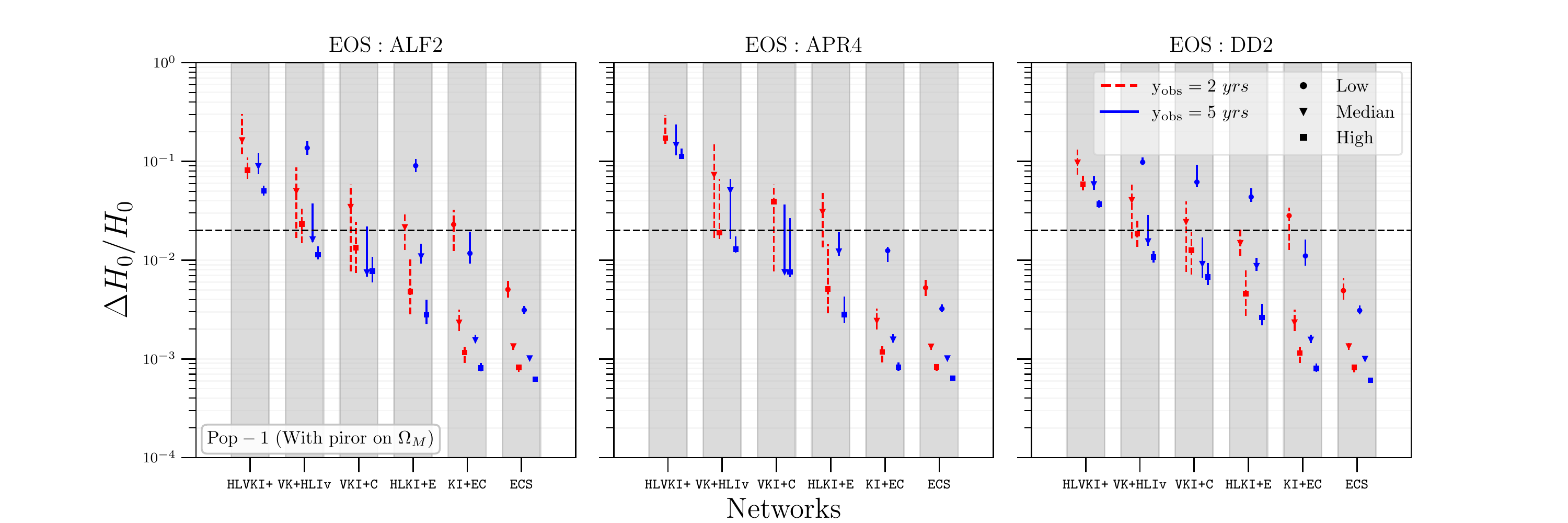}
\end{subfigure}\vspace{-1em}
\begin{subfigure}{\linewidth}
\centering
  \includegraphics[scale=0.6]{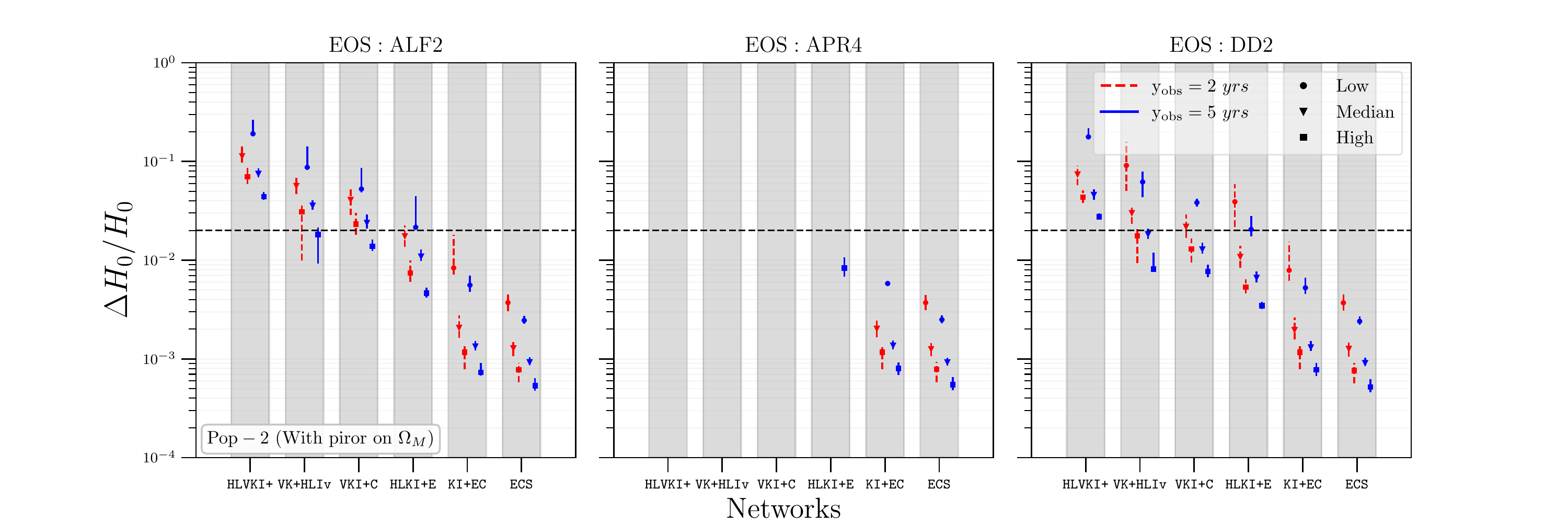}
\end{subfigure}\vspace{-1em}
\caption{\label{fig:gs_Om0_rest} The fractional errors in $H_0$ measurement using NSBH systems as gray sirens for the case when prior on $\Omega_m$ is included. The top two panels show the errors for events in Pop-1 and the bottom two panels show the errors for Pop-2. Same as Fig. \ref{fig:gs_Om0}, we have only considered the contribution of those bright siren events that were detected using the $g+i$ TOO strategy with Rubin.}
\end{figure*}

 This is accomplished by picking events from our populations that either qualify as golden dark sirens or as bright sirens. Using the selected events, the combined FIM is calculated and used to get measurement errors on $H_0$ in the same way as was done in Sections \ref{sec:dark_sirens} and \ref{sec:bright_sirens}. Figures \ref{fig:gs_Om0} and \ref{fig:gs_Om0_rest} present the bounds on $H_0$ measurement using the golden gray siren approach with the $g+i$ TOO strategy, without and with prior on $\Omega_m$, respectively. The results corresponding to the $r-$filter of Rubin and the $R-$filter of Roman are given in Figs. \ref{appfig:gs_Rubin} and \ref{appfig:gs_Roman} in Appendix \ref{appsec:gray_sirens}.


In Section \ref{sec:bright_sirens}, we noted that for Pop-2 events and \texttt{APR4} as the EOS we did not expect any KN events, rendering the bright siren approach moot. However, the NSBH dark siren events can still be used to constrain $H_0$. This is evident in Figs. \ref{fig:gs_Om0} and \ref{fig:gs_Om0_rest}, where $H_0$ is constrained better than $2\%$ with \texttt{KI+EC} and \texttt{ECS} when the prior for $\Omega_m$ is not included, and also by \texttt{HLKI+E} when the prior is included. For the stiffer EOS, we see that all detector networks except \texttt{HLVKI+} have a chance to resolve the \HL tension. The Voyager network might be able to resolve it in $5$ years even for median local merger rate. \texttt{KI+EC} and \texttt{ECS} will be able to resolve it in $2-5$ years regardless of the merger rate.

\begin{table*}
  \centering
  \caption{\label{tab:gs_numbers}The number of detections using the golden gray siren approach, i.e., combining the golden dark sirens and the bright sirens in an observation span of 10 years. For the bright sirens detected with the Rubin $r-$filter and Roman $R-$filter, we only consider events for which $\Omega_{90}\leq10\times$FOV of the respective telescope. For bright sirens detected with Rubin telescope following the $g+i$ TOO strategy, we consider events for which $\Omega_{90}\leq\,$FOV of Rubin.}
  \renewcommand{\arraystretch}{1.6} 
    \begin{tabular}{l |P{1.1cm} P{1.1cm} P{1.1cm}|P{1.1cm} P{1.1cm} P{1.1cm}|P{1.1cm} P{1.1cm} P{1.1cm}}
    \hhline{----------}
    Filter & \multicolumn{3}{c|}{Rubin $r-$filter} & \multicolumn{3}{c|}{Roman $R-$filter} & \multicolumn{3}{c}{Rubin $g+i$ TOO}\\
    \hhline{----------}
    EOS & ALF2 & APR4 & DD2 & ALF2 & APR4 & DD2 & ALF2 & APR4 & DD2\\
    \hhline{----------}
    \multicolumn{10}{c}{\textit{Pop-1}}\\
    \hhline{----------}
    HLVKI\texttt{+} & $9^{+16}_{-8}$ & $4^{+5}_{-3}$ & $16^{+28}_{-15}$ & $3^{+4}_{-3}$ & $1^{+0}_{-1}$ & $6^{+7}_{-6}$ &  $12^{+24}_{-11}$ & $6^{+4}_{-5}$ & $23^{+38}_{-21}$\\
    VK\texttt{+}HLIv & $15^{+27}_{-14}$ & $5^{+11}_{-4}$ & $24^{+44}_{-23}$ & $9^{+18}_{-9}$ & $4^{+7}_{-4}$ & $20^{+44}_{-20}$ & $24^{+47}_{-20}$ & $9^{+14}_{-8}$ & $41^{+74}_{-34}$\\
    VKI\texttt{+}C & $15^{+26}_{-14}$ & $5^{+10}_{-4}$ & $24^{+43}_{-23}$ & $4^{+6}_{-4}$ & $2^{+2}_{-2}$ & $10^{+19}_{-10}$ & $16^{+33}_{-14}$ & $7^{+8}_{-6}$ & $32^{+59}_{-27}$\\
    HLKI\texttt{+}E & $15^{+30}_{-14}$ & $5^{+14}_{-4}$ & $24^{+47}_{-23}$ & $16^{+33}_{-15}$ & $6^{+12}_{-5}$ & $27^{+66}_{-26}$ & $26^{+63}_{-23}$ & $11^{+27}_{-10}$ & $44^{+96}_{-38}$\\
    KI\texttt{+}EC & $38^{+80}_{-35}$ & $28^{+64}_{-25}$ & $47^{+97}_{-44}$ & $80^{+155}_{-67}$ & $45^{+91}_{-37}$ & $124^{+248}_{-106}$ & $63^{+136}_{-55}$ & $36^{+84}_{-32}$ & $84^{+181}_{-73}$\\
    ECS & $119^{+260}_{-103}$ & $110^{+247}_{-94}$ & $126^{+276}_{-110}$ & $193^{+392}_{-158}$ & $139^{+298}_{-116}$ & $259^{+527}_{-216}$ & $145^{+317}_{-123}$ & $118^{+267}_{-101}$ & $168^{+365}_{-142}$\\
    \hhline{----------}
    \multicolumn{10}{c}{\textit{Pop-2}}\\
    \hhline{----------}
    HLVKI\texttt{+} & $16^{+18}_{-14}$ & $0$ & $31^{+41}_{-26}$ & $3^{+5}_{-2}$ & $0$ & $6^{+12}_{-4}$ & $18^{+23}_{-15}$ & $0$ & $34^{+51}_{-28}$\\
    VK\texttt{+}HLIv & $19^{+31}_{-16}$ & $0^{+1}_{-0}$ & $37^{+83}_{-31}$ & $16^{+22}_{-12}$ & $0^{+1}_{-0}$ & $29^{+44}_{-24}$ & $25^{+48}_{-21}$ & $0^{+1}_{-0}$ & $58^{+112}_{-49}$\\
    VKI\texttt{+}C & $20^{+30}_{-17}$ & $0$ & $38^{+83}_{-32}$ & $7^{+10}_{-5}$ & $0$ & $14^{+21}_{-11}$ & $24^{+35}_{-21}$ & $0$ & $46^{+82}_{-39}$\\
    HLKI\texttt{+}E & $20^{+33}_{-17}$ & $0^{+3}_{-0}$ & $38^{+87}_{-32}$ & $22^{+45}_{-17}$ & $0^{+3}_{-0}$ & $44^{+87}_{-36}$ & $28^{+56}_{-24}$ & $0^{+3}_{-0}$ & $68^{+137}_{-58}$\\
    KI\texttt{+}EC & $45^{+101}_{-38}$ & $25^{+71}_{-21}$ & $63^{+155}_{-53}$ & $104^{+244}_{-85}$ & $25^{+71}_{-21}$ & $185^{+401}_{-153}$ & $57^{+149}_{-47}$ & $25^{+71}_{-21}$ & $114^{+272}_{-93}$\\
    ECS & $123^{+270}_{-102}$ & $103^{+241}_{-85}$ & $140^{+319}_{-116}$ & $218^{+502}_{-177}$ & $103^{+241}_{-85}$ & $349^{+747}_{-286}$ & $135^{+320}_{-111}$ & $103^{+241}_{-85}$ & $192^{+447}_{-157}$\\
    \hhline{----------}
    \end{tabular}
\end{table*}

Table \ref{tab:gs_numbers} gives the number of NSBH detections that have been considered for the golden gray siren method, i.e., each event is either a golden dark siren, a bright siren, or both. As is evident when comparing the number of golden dark sirens in Table \ref{tab:dark_sirens} and bright sirens in Tables \ref{tab:bs_golden} and \ref{tab:bs_golden_too} to the golden gray siren events in Table \ref{tab:gs_numbers}, there are very few $(\sim \mathcal{O}(1))$ events that qualify as both golden dark sirens and bright sirens. So, the square root of the sum of squares of the errors obtained from the golden dark and the bright siren approach is a good approximation of the expected measurement errors using the golden gray siren method.

Another important thing to note is the dependence of $H_0$ accuracy on the EOS using the gray siren approach. From Figs. \ref{fig:gs_Om0} and \ref{fig:gs_Om0_rest}, we see that the effect of EOS is apparent for \texttt{HLVKI+}, \texttt{VK+HLIv}, \texttt{VKI+C} and \texttt{HLKI+E}, where $H_0$ is estimated best for \texttt{DD2} and worst for \texttt{APR4}. This is due to the low number of golden dark siren measurements for these networks. For the two most advanced networks, \texttt{KI+EC} and \texttt{ECS}, the precision for $H_0$ measurement does not significantly change with EOS. This is because the $H_0$ estimates for these networks are dominated by the golden dark siren measurements, due to the exceptionally well-measured distance estimates. 

An important use of the numbers reported in Table \ref{tab:gs_numbers} and the Figs. \ref{fig:gs_Om0} and \ref{fig:gs_Om0_rest} is to account for the change in the measurement errors of $H_0$ with the inclusion of factors that we have ignored. This is done by using the fact that the bounds on $H_0$ change approximately as $1/\sqrt{N}$, where $N$ is the number of detections. For e.g., with the assumed $100\%$ duty-cycle, the \texttt{A+} network with \texttt{DD2} as the EOS and Pop-2 events has a measurement error of $2.7\%$ associated with the high local merger rate density, in an observation span of $5$ years (see the plot in the second row and third column of Fig. \ref{fig:gs_Om0_rest}). Now, if we assume a $50\%$ duty-cycle, the measurement error increases to $2.7\%\times \sqrt{2} = 3.8\%$. As can be verified from the same plot, this is close to the error corresponding to a high merger rate and an observation span of $2$ years with $100\%$ duty-cycle. Following these steps, one can calculate the expected constraints on $H_0$ using the golden gray siren approach with NSBH systems for different merger rates, observation times, duty cycles, the efficiency of the two telescopes in the detection of the EM counterpart, change in the exposure times for the telescopes, NS EOS, etc.

%%%%%%%%%%%%%%%%%%%%%%%%%%%%%%%%%%%%%%%%%%%%%%%%%%

%%%%%%%%%%%%%%%%% SECTION BREAK %%%%%%%%%%%%%%%%%%
\section{Systematic uncertainties} \label{sec:sys}
Several systematic uncertainties are associated with the measurement of $H_0$ using GWs. Some of these include uncertainty due to instrumental calibration errors \citep{Sun:2020wke,Huang:2022rdg} and peculiar velocity fields \citep{Chen:2017rfc,Mukherjee:2019qmm,Nicolaou:2019cip,Howlett:2019mdh}. The calibration errors can lead to a systematic error and uncertainty of $< 7\%$ in the amplitude of the GW \citep{Sun:2020wke} and are expected to reduce in future networks. The uncertainty in redshift measurement due to peculiar velocity associated with the host galaxy is dominant only for nearby galaxies. For instance, for GW170817 which was $z \sim 0.01$ away the uncertainty in peculiar velocity was $150$ km/s, leading to a $\sim 5\%$ uncertainty in redshift measurement \citep{LIGOScientific:2017adf}. Most of the events considered in this study lie farther away compared to GW170817. For e.g., for dark sirens detected by \texttt{ECS}, $\sim 90\%$ of the events lie beyond $z=0.05$, i.e., $5$ times as far away as GW170817. The nearest event considered in the study is twice as far as GW170817. Thus, we do not expect the effect of peculiar velocity uncertainties to considerably change the overall expectation regarding the resolution of $H_0$ with a particular GW network.

The golden dark siren approach assumes that within a redshift of $0.1$, on average, only one galaxy will lie in the sky area of $0.04$ $\mathrm{deg}^2$. This assumption relies on the galaxy catalog used. It is possible that even within the sky area of $0.04$ $\mathrm{deg}^2$, there exist multiple faint galaxies (that are not part of the catalog) that might host the golden dark siren. While the statistical dark siren approach also relies on galaxy catalogs, unique identification of the host galaxy is not necessary. As long as galaxies belonging to the same galaxy cluster as the host have been considered, the approach will ascertain the correct value of $H_0$. One way to mitigate the effects of catalog incompleteness in the golden dark siren approach is to perform a targeted follow-up of the ascertained sky area with a telescope. In principle, a long enough observation will allow the detection of any faint galaxies that lie in the sky patch. Then, if only one galaxy is observed, as was the initial assumption, the host will be identified uniquely. We are currently working towards a more detailed analysis of the systematic biases associated with the golden dark siren approach and the possibility of uniquely identifying the host.

A source of systematic uncertainty in the bright siren approach is due to the viewing angle dependence of the detection of EM counterparts \citep{Chen:2020dyt}. In particular, the models used to generate luminosity curves for KN for events in this study assume isotropic emission. Viewing angle dependence can change the peak luminosity associated with a KN, which will affect the number of detections that have been reported in Tables \ref{tab:bs_golden}, \ref{tab:bs_golden_too}, \ref{tab:gs_numbers} and \ref{apptab:r_R_kn_numbers}. While the possible aspherical nature of KN emission is not accounted for by the KN model that has been used, the effect on the $H_0$ constraints can be calculated, if the model describing the dependence of viewing angle on KN luminosity is known, with the $1/\sqrt{N}$ dependence of $H_0$ constraints and the numbers reported in this work.

%%%%%%%%%%%%%%%%%%%%%%%%%%%%%%%%%%%%%%%%%%%%%%%%%%

%%%%%%%%%%%%%%%%% SECTION BREAK %%%%%%%%%%%%%%%%%%

\section{Conclusions} \label{sec:concl}
The tension between the early-Universe and late-Universe measurements of $H_0$ has been an active area of research in cosmology and GW physics. The measurement of luminosity distance from GW observation along with the value of the corresponding redshift from host-galaxy identification can provide an independent estimate of $H_0$ and help resolve the tension in $H_0$. The host can be identified by using an EM counterpart (if there is one), which is called the bright siren method, or without the EM counterpart, by using events that are very well-localized in the sky, which is referred to as the golden dark siren method. In this work, we assessed the potential of NSBH systems in resolving the $H_0$ tension.

With the improved estimation of binary parameters facilitated by the activation of higher order modes due to the presence of unequal masses, NSBH systems can be used as golden dark sirens to resolve the \HL tension. While there is no hope of resolving the \HL tension with golden dark sirens with the \texttt{HLVKI+} network, the Voyager network (\texttt{VK+HLIv}) can measure $H_0$ to less than $2\%$ measurement uncertainty in an observation span of $5$ years. More advanced networks like \texttt{ECS}, which contains a triangular Einstein Telescope and two Cosmic Explorers, can measure $H_0$ to a precision better than $1\%$ with only $2$ years of observation (see Fig. \ref{fig:dark_sirens}). Moreover, the \texttt{ECS} network will observe $\mathcal{O}(10)$ exceptional dark siren events, i.e., events that can individually measure $H_0$ with measurement uncertainty less than $2\%$, every year.

NSBH mergers can also be accompanied by EM counterparts, like short gamma-ray bursts and kilonovae (KN), rendering them as possible bright siren candidates for $H_0$ measurement. In Table \ref{apptab:r_R_kn_numbers} in Appendix \ref{appsec:kn_numbers}, we reported the number of expected KN detections using two telescopes: the Vera C. Rubin Observatory and the Nancy Grace Roman Space Telescope, for three NS EOS of varying stiffness: \texttt{ALF2}, \texttt{APR4} and \texttt{DD2}, in an observation span of 10 years. We found that the constraints on $H_0$ using the bright siren approach depend on population characteristics and the NS EOS. While $H_0$ could not be measured with any of the networks for Pop-2 events with the \texttt{APR4} EOS, networks with the Einstein Telescope or the Cosmic Explorer detectors could resolve the tension with Pop-1 events.  For the other two EOS, it seemed unlikely that \texttt{HLVKI+} will be able to resolve the tension in $5$ years, with the best measurement of $H_0$ having a fractional error of $2.6\%$. With Voyager sensitivity, \texttt{VK+HLIv} can resolve the tension in 5 years and \texttt{ECS} will be able to measure $H_0$ to better than $1\%$ precision only in $2$ years. We also expect \texttt{ECS} to detect $\mathcal{O}(1)$ exceptional bright siren events every year. These are events that can individually measure $H_0$ to better than $2\%$ precision. 

In both dark and bright siren approaches, we expect to detect $\mathcal{O}(10)$ events every year with \texttt{ECS} that contribute towards the measurement of $H_0$. Due to the effectiveness of NSBH both as dark and as bright sirens, we refer to them as gray sirens. In Section \ref{sec:gray_sirens}, we combined the golden dark siren and bright siren events to calculate constraints on $H_0$. While it is still unlikely that \texttt{HLVKI+} will be able to resolve the \HL tension in $5$ years, \texttt{VK+HLIv} is expected to resolve it in the same duration. Even for \texttt{APR4}, networks with the Einstein Telescope or the Cosmic Explorer detectors will be able to resolve the $H_0$ tension. With \texttt{ECS}, it would be possible to detect 10--30 NSBH golden gray (golden dark, bright, or both) siren events every year. We also noted that $H_0$ estimates corresponding to the next generation networks that contain both the Einstein Telescope and (one or two) Cosmic Explorer detectors are not contingent on the EOS of the NS as they will be dominated by the measurement of $H_0$ from golden dark siren measurements.

Thus, NSBH systems can be important sources for independent calculation of $H_0$ and can resolve the \HL tension as early as the Voyager era ($\sim$mid 2030s). While our analysis relies on the FIM, it is the first comprehensive study that presents the expectation regarding the measurement of $H_0$ with NSBH systems and next generation detectors using both dark and bright siren approaches. In the current work, we account for several uncertainties associated with NSBH systems by including two different population models, three local merger rates, two observation spans, and six ground-based GW detectors. We also give the associated number of events used and bounds on $H_0$ for all possible cases to allow the reader to extrapolate approximate bounds on $H_0$ for a general scenario. For the bright and gray siren methods, we generated KN light curves for three EOS using numerical relativity fits to portray a realistic picture of the scope of these methods. A possible direction for future work can be to use multiple KN models that include the viewing angle dependence to gauge its effect on the number of KN detections. Another important extension would be to repeat this analysis within a Bayesian framework. Historically, using Bayesian inference for a large number of events has been tedious, leading to the inclination toward FIM analysis for large-scale studies. With the introduction of new techniques for parameter estimation \citep{Zackay:2018qdy,Leslie:2021ssu,Roulet:2022kot}, the goal of implementing a fully-Bayesian study might be within reach.   

\section*{Acknowledgements}

We would like to thank Bangalore Sathyaprakash, Arnab Dhani, Rahul Kashyap and Hsin-Yu Chen for useful discussions and comments on the work. We also wish to thank Rahul Kashyap for providing the scripts that were used to generate the kilonova light curves, Debatri Chattopadhyay for helping with the construction of Pop-2, Marica Branchesi for bringing the $g+i$ TOO strategy to our attention, and V. Ashley Villar and Nandini Hazra for helping with the computation of the limiting magnitudes for the Rubin filters. This research was supported by NSF grant numbers PHY-2012083, AST-2006384 and PHY-2207638.

%%%%%%%%%%%%%%%%%%%%%%%%%%%%%%%%%%%%%%%%%%%%%%%%%%
\section*{Data Availability}

The data used in this study has been extracted from our data used in the previous work \citep{Gupta:2023evt}. It will be made available upon reasonable request.



\bibliographystyle{mnras}
\bibliography{bibliography.bib} % if your bibtex file is called example.bib

%\begin{thebibliography}{99}
%\end{thebibliography}

%%%%%%%%%%%%%%%%%%%%%%%%%%%%%%%%%%%%%%%%%%%%%%%%%%

%%%%%%%%%%%%%%%%% APPENDICES %%%%%%%%%%%%%%%%%%%%%

\appendix

\section{Masses and spins for systems in Pop-2}
\label{appsec:Pop2_params}
We consider two populations of NSBH systems in this study. Pop-1 considers broad distributions over BH and NS masses and BH spins. The BH masses are informed by the \texttt{POWER+PEAK} distribution used in \citet{LIGOScientific:2021psn}. The NS mass is picked from a uniform distribution between $[1,\,2.9]\msun$ so as to not restrict ourselves to the galactic population of NS and allow for a broad distribution. The spin of the BH and the NS are picked from uniform distributions between $[-0.75,0.75]$ and $[-0.05,0.05]$, respectively.

\begin{figure*}
\centering
\includegraphics[scale=0.85]{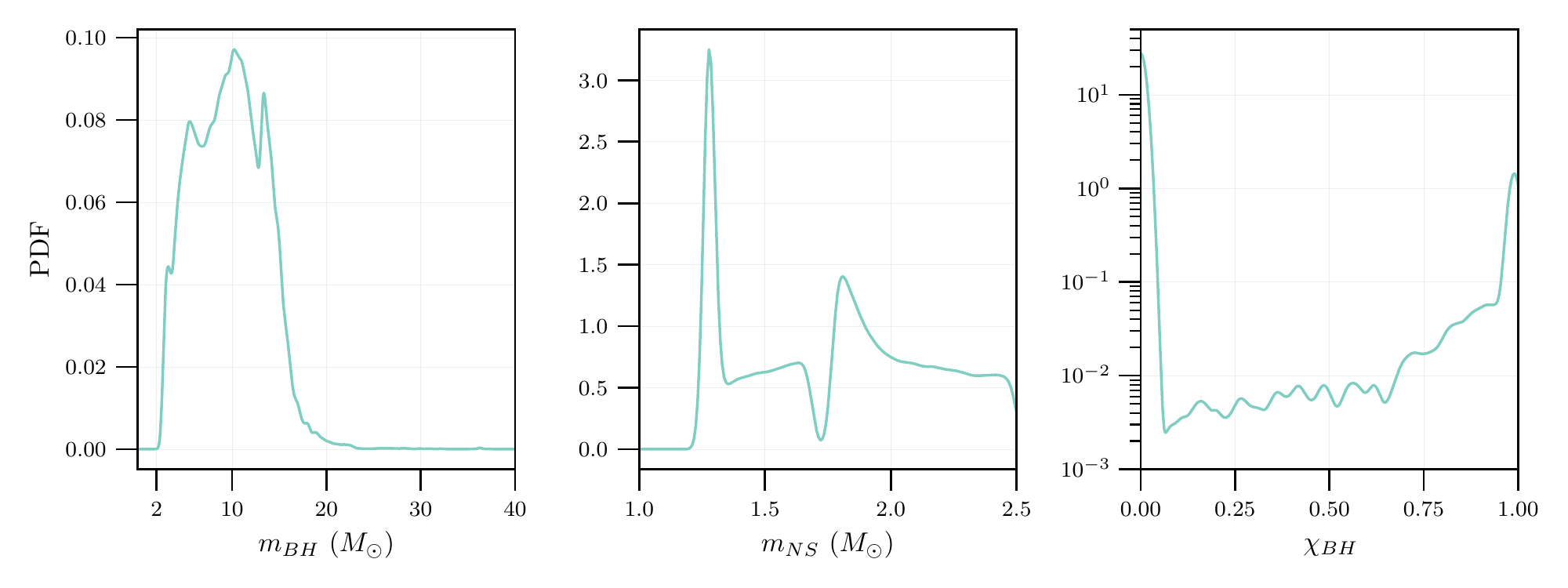}
\caption{\label{appfig:Pop2_mass_spins}The probability density function (PDF) plots for the masses of the BH and the NS and spins of BH in Pop-2.}
\end{figure*}

Due to the absence of a large catalog of NSBH events, we use a minimum set of assumptions to fix the characteristics of Pop-1. For Pop-1, as the masses and spins of the BH and the NS are sampled independently, they ignore
any correlations that may exist between their properties
due to physical processes involved in the binary formation
and evolution. To account for this, Pop-2 uses the mass profiles for the NS and the BH in NSBH systems from a population synthesis model that conforms to the isolated binary formation channel for the formation of NSBH systems. Specifically, we use the fiducial model from \citet{Broekgaarden:2021iew}. The NSs are assumed to be non-spinning. The BH spins are calculated from the fits described in Eq. (2) and (3) of \citet{Chattopadhyay:2022cnp}. The resulting mass profiles for the BH and the NS and the spins of the BH are shown in Fig. \ref{appfig:Pop2_mass_spins}.

The fits for the BH spin only apply to systems where the NS progenitor is formed first, allowing the second-born BH progenitor to have high spins as it can get tidally spun-up by its companion. For systems where the NS progenitor is formed second, the BH spin is taken to be 0. In Pop-2, all BHs have spins greater than or equal to $0$ and $\sim 5\%$ systems have $\chi_{BH} > 0.75$. However, as the numerical scripts used to generate KN light curves are only valid for BH spins less than $0.75$, the remaining $5\%$ of the events in Pop-2 are ignored for the bright siren and gray siren study.

\section{Redshift distribution and detection rate} \label{appsec:det_rate}
The merger rate density $\Dot{n}(z)$ is proportional to convolution of the star formation rate $\psi(z)$ with the probability density of the time delay distribution (the time between the formation of progenitor stars and the eventual merger of compact binaries) $P(t_d)$,
\begin{equation}
    \Dot{n}(z) \propto \int_{t_{d,min}}^{t_{d,max}} \psi (z_f(z,t_d))\,P(t_d)\,\,dt_d,
\end{equation}
where $t_{d,min}$ and $t_{d,max}$ are the minimum and maximum delay times. We use the star formation rate model proposed in \citet{Yuksel:2008cu} and $P(t_d)$ to be the log-normal time delay distribution introduced in \citet{Wanderman:2014eza}, which gives the redshift distribution as \citep{Zhu:2020ffa,Sun:2015bda},
\begin{multline}
    \Dot{n}(z) \propto \Bigg[\,(1+z)^{4.131\eta}\,+\, \left(\frac{1+z}{22.37}\right)^{-0.5789\eta} +\,\left(\frac{1+z} {2.978}\right)^{-4.735\eta} \\ +\,\left(\frac{1+z}{2.749}\right)^{-10.77\eta} +\,\left(\frac{1+z}{2.867}\right)^{-17.51\eta}\,+\,\left(\frac{1+z}{3.04}\right)^{-\frac{0.08148+z^{0.574}}{0.08682}\eta} \Bigg]^{1/\eta},
\end{multline}
where $\eta = -5.51$. $\Dot{n}(z)$ is normalized such that $\Dot{n}(0)$ is equal to the local merger rate density.

The detection rate is calculated using the efficiency of a network $\epsilon(z,\rho_{*})$, which is defined as the fraction of events at that redshift that are detected by the network with an SNR greater than the threshold SNR $\rho_{*}$. Using the network efficiency and the merger rate density, the detection rate $D_R$ in the detector frame is given by
\begin{equation}
D_R (z) = \int_{0}^{z} \epsilon (z',\rho_{*}) \frac{\Dot{n} (z')}{(1+z')} \frac{d V}{d z'}\,dz',
\end{equation}
where $dV/dz'$ is the comoving volume element and the $(1+z')$ term in the denominator converts the detection rate from the source frame to the detector frame.

%%%%%%%%%%%%%%%%%%%%%%%%%%%%%%%%%%%%%%%%%%%%%%%%%%%%%%%%%%%%%%%%%%%%%%%%%%%%%%%%%%%%

\section{Measurement errors in luminosity distance and sky-position} \label{appsec:logDL_sa90_table}
\begin{table*}
  \centering
  \caption{\label{apptab:mma_sa90_logDL}For the events for which $z\leq0.5$, the table lists the number of detections per year for the six detector networks with $90\%$-credible sky area $\Omega_{90} < 10$, $1$, $0.1$ and $0.01$ $\mbox{deg}^2$ and fractional error in luminosity distance $\Delta D_L / D_L < 0.1$ and $0.01$.}
  \renewcommand{\arraystretch}{1.5} 
    \begin{tabular}{l |P{2.1cm} P{2.1cm} P{2.2cm} P{2cm}|P{2.1cm} P{2.2cm}}
    \hhline{-|----|--}
    Metric & \multicolumn{4}{c|}{$\Omega_{90}\mbox{ (deg)}^2$} & \multicolumn{2}{c}{$\Delta D_L / D_L$} \\
    \hhline{-|----|--}
    Quality & $\leq 10$ & $\leq 1$ & $\leq 0.1$ & $\leq 0.01$ & $\leq 0.1$ & $\leq 0.01$ \\
    \hhline{-------}
    \multicolumn{7}{c}{\textit{Pop-1}}\\
    \hhline{-------}
    HLVKI\texttt{+} & $2.9^{+6.1}_{-2.4} \times 10^{2}$ & $8.7^{+19.5}_{-7.3}$ & $2.0^{+2.0}_{-2.0} \times 10^{-1}$ & $0$ & $6.0^{+12.3}_{-4.7} \times 10$ & $0.0^{+0.1}_{-0.0}$ \\
    VK\texttt{+}HLIv & $1.0^{+2.2}_{-0.9} \times 10^{3}$ & $3.1^{+7.1}_{-2.6} \times 10$ & $6.0^{+13.0}_{-6.0} \times 10^{-1}$ & $0$ & $3.9^{+8.2}_{-3.2} \times 10^{2}$ & $3.0^{+9.0}_{-3.0} \times 10^{-1}$ \\
    VKI\texttt{+}C & $6.8^{+14.6}_{-5.6} \times 10^{2}$ & $1.9^{+4.1}_{-1.6} \times 10$ & $2.0^{+6.0}_{-2.0} \times 10^{-1}$ & $0$ & $8.7^{+18.4}_{-7.2} \times 10^{2}$ & $9.4^{+19.6}_{-7.4}$ \\
    HLKI\texttt{+}E & $1.4^{+3.0}_{-1.2} \times 10^{3}$ & $6.0^{+13.4}_{-4.9} \times 10$ & $1.9^{+2.3}_{-1.9}$ & $0$ & $2.5^{+5.3}_{-2.0} \times 10^{3}$ & $9.9^{+22.1}_{-8.4}$ \\
    KI\texttt{+}EC & $3.2^{+6.7}_{-2.6} \times 10^{3}$ & $5.6^{+12.1}_{-4.7} \times 10^{2}$ & $1.4^{+3.1}_{-1.2} \times 10$ & $2.0^{+4.0}_{-2.0} \times 10^{-1}$ & $3.5^{+7.4}_{-2.9} \times 10^{3}$ & $7.0^{+13.7}_{-5.4} \times 10$ \\
    ECS & $3.7^{+7.8}_{-3.1} \times 10^{3}$ & $2.2^{+4.8}_{-1.9} \times 10^{3}$ & $2.3^{+5.0}_{-1.9} \times 10^{2}$ & $5.1^{+9.8}_{-4.2}$ & $3.6^{+7.7}_{-3.0} \times 10^{3}$ & $1.4^{+2.9}_{-1.1} \times 10^{2}$ \\
    \hhline{-------}
    \multicolumn{7}{c}{\textit{Pop-2}}\\
    \hhline{-------}
    HLVKI\texttt{+} & $3.0^{+6.4}_{-2.5} \times 10^{2}$ & $9.1^{+20.3}_{-7.7}$ & $0.0^{+0.3}_{-0.0}$ & $0$ & $4.4^{+8.3}_{-3.7} \times 10$ & $0.0^{+0.1}_{-0.0}$ \\
    VK\texttt{+}HLIv & $1.1^{+2.3}_{-0.9} \times 10^{3}$ & $3.2^{+6.8}_{-2.6} \times 10$ & $4.0^{+15.0}_{-4.0} \times 10^{-1}$ & $0$ & $3.5^{+7.4}_{-2.9} \times 10^{2}$ & $0.0^{+0.3}_{-0.0}$ \\
    VKI\texttt{+}C & $7.1^{+15.1}_{-5.9} \times 10^{2}$ & $1.8^{+3.8}_{-1.5} \times 10$ & $3.0^{+5.0}_{-3.0} \times 10^{-1}$ & $0$ & $9.2^{+19.4}_{-7.6} \times 10^{2}$ & $4.8^{+8.3}_{-4.0}$ \\
    HLKI\texttt{+}E & $1.5^{+3.1}_{-1.2} \times 10^{3}$ & $5.9^{+12.2}_{-4.8} \times 10$ & $9.0^{+35.0}_{-8.0} \times 10^{-1}$ & $0$ & $2.5^{+5.3}_{-2.1} \times 10^{3}$ & $7.1^{+16.3}_{-6.2}$ \\
    KI\texttt{+}EC & $3.2^{+6.8}_{-2.6} \times 10^{3}$ & $5.7^{+12.0}_{-4.7} \times 10^{2}$ & $1.2^{+3.1}_{-1.0} \times 10$ & $0.0^{+0.3}_{-0.0}$ & $3.5^{+7.4}_{-2.9} \times 10^{3}$ & $3.7^{+7.4}_{-3.1} \times 10$ \\
    ECS & $3.7^{+7.9}_{-3.1} \times 10^{3}$ & $2.3^{+4.9}_{-1.9} \times 10^{3}$ & $2.3^{+4.8}_{-1.9} \times 10^{2}$ & $4.5^{+11.7}_{-3.8}$ & $3.6^{+7.7}_{-3.0} \times 10^{3}$ & $8.6^{+18.1}_{-7.2} \times 10$ \\
    \hhline{-------}
    \end{tabular}
\end{table*}
For the NSBH events from the two populations that lie within the redshift of $0.5$, Table \ref{apptab:mma_sa90_logDL} provides the number of detections every year for which the fractional error in luminosity distance is less than $10\%$ and $1\%$, and the sky-position is determined better than $10$, $1$ and $0.1$ $\mbox{deg}^2$. The numbers reported correspond to the median NSBH local merger rate density of $45$ $\mbox{Gpc}^{-3}$ $\mbox{yr}^{-1}$. The lower and upper bounds correspond to the lower and upper bound in the event-based local merger rate density for NSBH systems \citep{LIGOScientific:2021psn}.

%%%%%%%%%%%%%%%%%%%%%%%%%%%%%%%%%%%%%%%%%%%%%%%%%%%%%%%%%%%%%%%%%%%%%%%%%%%%%%

\section{Kilonova modelling and number of detections} \label{appsec:kn_numbers}
\begin{table*}
  \centering
  \caption{\label{apptab:r_R_kn_numbers}The number of KN detections with the $r-$filter of Rubin Telescope and $R-$filter of the Roman Observatory for both Pop-1 and Pop-2 for an observation time of 10 years. The events are categorized based on if they can be localized in the sky, using GW observations, better than the FOV of the EM telescope, 10 times the FOV of the EM telescope, or $100\,\,\mbox{deg}^2$. They have been further divided into 3 columns based on the EOS that was used to generate the KN light curves.}
  \renewcommand{\arraystretch}{1.5} 
    \begin{tabular}{l |P{1.3cm} P{1.3cm} P{1.3cm}|P{1.3cm} P{1.3cm} P{1.3cm}|P{1.3cm} P{1.3cm} P{1.3cm}}
    \hhline{----------}
    \multicolumn{10}{c}{Pop-1} \\
    \hhline{----------}
    \multicolumn{10}{c}{Rubin $r-$filter} \\
    \hhline{----------}
    \multirow{2}{*}{Network} & \multicolumn{3}{c|}{$\Omega_{90}<$ FOV} & \multicolumn{3}{c|}{$\Omega_{90}<$ $10\,\times\,$FOV} & 
    \multicolumn{3}{c}{$\Omega_{90}<$ $100 \,\mbox{deg}^2$}\\
    \hhline{~|---|---|---}
    & ALF2 & APR4 & DD2 & ALF2 & APR4 & DD2 & ALF2 & APR4 & DD2 \\
    \hhline{-|-|-|-|-|-|-|-|-|-}
    HLVKI\texttt{+} & $9^{+13}_{-7}$ & $4^{+4}_{-3}$ & $16^{+23}_{-13}$ & $9^{+16}_{-8}$ & $4^{+5}_{-3}$ & $16^{+28}_{-15}$ & $9^{+16}_{-8}$ & $4^{+5}_{-3}$ & $16^{+28}_{-15}$ \\
    VK\texttt{+}HLIv & $14^{+21}_{-12}$ & $4^{+7}_{-3}$ & $23^{+33}_{-19}$ & $14^{+25}_{-13}$ & $4^{+9}_{-3}$ & $23^{+42}_{-22}$ & $14^{+25}_{-13}$ & $4^{+9}_{-3}$ & $23^{+42}_{-22}$ \\
    HLKI\texttt{+}E & $14^{+21}_{-11}$ & $4^{+7}_{-3}$ & $23^{+33}_{-18}$ & $14^{+25}_{-13}$ & $4^{+9}_{-3}$ & $23^{+42}_{-22}$ & $14^{+25}_{-13}$ & $4^{+9}_{-3}$ & $23^{+42}_{-22}$ \\
    VKI\texttt{+}C & $14^{+17}_{-7}$ & $4^{+5}_{-3}$ & $23^{+29}_{-14}$ & $14^{+25}_{-13}$ & $4^{+9}_{-3}$ & $23^{+42}_{-22}$ & $14^{+25}_{-13}$ & $4^{+9}_{-3}$ & $23^{+42}_{-22}$ \\
    KI\texttt{+}EC & $14^{+24}_{-13}$ & $4^{+9}_{-3}$ & $23^{+40}_{-22}$ & $14^{+25}_{-13}$ & $4^{+9}_{-3}$ & $23^{+42}_{-22}$ & $14^{+25}_{-13}$ & $4^{+9}_{-3}$ & $23^{+42}_{-22}$ \\
    ECS & $14^{+25}_{-13}$ & $4^{+9}_{-3}$ & $23^{+42}_{-22}$ & $14^{+25}_{-13}$ & $4^{+9}_{-3}$ & $23^{+42}_{-22}$ & $14^{+25}_{-13}$ & $4^{+9}_{-3}$ & $23^{+42}_{-22}$ \\
    \hhline{----------}
    \multicolumn{10}{c}{Roman $R-$filter} \\
    \hhline{----------}
    \multirow{2}{*}{Network} & \multicolumn{3}{c|}{$\Omega_{90}<$ FOV} & \multicolumn{3}{c|}{$\Omega_{90}<$ $10\,\times\,$FOV} & 
    \multicolumn{3}{c}{$\Omega_{90}<$ $100 \,\mbox{deg}^2$}\\
    \hhline{~|---|---|---}
    & ALF2 & APR4 & DD2 & ALF2 & APR4 & DD2 & ALF2 & APR4 & DD2 \\
    \hhline{-|-|-|-|-|-|-|-|-|-}
    HLVKI\texttt{+} & $0$ & $0$ & $0$ & $3^{+4}_{-3}$ & $1^{+0}_{-1}$ & $6^{+7}_{-6}$ & $16^{+30}_{-15}$ & $7^{+6}_{-6}$ & $29^{+52}_{-27}$ \\
    VK\texttt{+}HLIv & $0$ & $0$ & $0$ & $8^{+16}_{-8}$ & $3^{+5}_{-3}$ & $19^{+42}_{-19}$ & $66^{+131}_{-51}$ & $23^{+50}_{-17}$ & $118^{+242}_{-97}$ \\
    HLKI\texttt{+}E & $0^{+1}_{-0}$ & $0$ & $0^{+1}_{-0}$ & $15^{+28}_{-14}$ & $5^{+7}_{-4}$ & $26^{+61}_{-25}$ & $96^{+168}_{-74}$ & $39^{+68}_{-30}$ & $169^{+313}_{-138}$ \\
    VKI\texttt{+}C & $0$ & $0$ & $0$ & $3^{+5}_{-3}$ & $1^{+1}_{-1}$ & $9^{+18}_{-9}$ & $84^{+151}_{-64}$ & $30^{+56}_{-22}$ & $155^{+293}_{-126}$ \\
    KI\texttt{+}EC & $3^{+5}_{-3}$ & $1^{+1}_{-1}$ & $6^{+8}_{-6}$ & $56^{+100}_{-45}$ & $21^{+36}_{-15}$ & $100^{+193}_{-84}$ & $97^{+170}_{-75}$ & $39^{+68}_{-30}$ & $171^{+318}_{-140}$ \\
    ECS & $26^{+52}_{-20}$ & $12^{+22}_{-9}$ & $50^{+109}_{-41}$ & $88^{+157}_{-68}$ & $33^{+60}_{-25}$ & $156^{+293}_{-128}$ & $97^{+170}_{-75}$ & $39^{+68}_{-30}$ & $171^{+318}_{-140}$ \\
    \hhline{----------}
    \multicolumn{10}{c}{Pop-2} \\
    \hhline{----------}
    \multicolumn{10}{c}{Rubin $r-$filter} \\
    \hhline{----------}
    \multirow{2}{*}{Network} & \multicolumn{3}{c|}{$\Omega_{90}<$ FOV} & \multicolumn{3}{c|}{$\Omega_{90}<$ $10\,\times\,$FOV} & 
    \multicolumn{3}{c}{$\Omega_{90}<$ $100 \,\mbox{deg}^2$}\\
    \hhline{~|---|---|---}
    & ALF2 & APR4 & DD2 & ALF2 & APR4 & DD2 & ALF2 & APR4 & DD2 \\
    \hhline{-|-|-|-|-|-|-|-|-|-}
    HLVKI\texttt{+} & $16^{+16}_{-11}$ & $0$ & $31^{+33}_{-19}$ & $16^{+18}_{-14}$ & $0$ & $31^{+41}_{-26}$ & $16^{+18}_{-14}$ & $0$ & $31^{+41}_{-26}$ \\
    VK\texttt{+}HLIv & $19^{+22}_{-14}$ & $0$ & $37^{+58}_{-28}$ & $19^{+30}_{-16}$ & $0$ & $37^{+82}_{-31}$ & $19^{+30}_{-16}$ & $0$ & $37^{+82}_{-31}$ \\
    HLKI\texttt{+}E & $20^{+24}_{-15}$ & $0$ & $38^{+61}_{-30}$ & $20^{+30}_{-17}$ & $0$ & $38^{+84}_{-32}$ & $20^{+30}_{-17}$ & $0$ & $38^{+84}_{-32}$ \\
    VKI\texttt{+}C & $20^{+20}_{-15}$ & $0$ & $38^{+47}_{-23}$ & $20^{+30}_{-17}$ & $0$ & $38^{+83}_{-32}$ & $20^{+30}_{-17}$ & $0$ & $38^{+83}_{-32}$ \\
    KI\texttt{+}EC & $20^{+30}_{-17}$ & $0$ & $38^{+82}_{-32}$ & $20^{+30}_{-17}$ & $0$ & $38^{+84}_{-32}$ & $20^{+30}_{-17}$ & $0$ & $38^{+84}_{-32}$ \\
    ECS & $20^{+30}_{-17}$ & $0$ & $38^{+84}_{-32}$ & $20^{+30}_{-17}$ & $0$ & $38^{+84}_{-32}$ & $20^{+30}_{-17}$ & $0$ & $38^{+84}_{-32}$ \\
    \hhline{----------}
    \multicolumn{10}{c}{Roman $R-$filter} \\
    \hhline{----------}
    \multirow{2}{*}{Network} & \multicolumn{3}{c|}{$\Omega_{90}<$ FOV} & \multicolumn{3}{c|}{$\Omega_{90}<$ $10\,\times\,$FOV} & 
    \multicolumn{3}{c}{$\Omega_{90}<$ $100 \,\mbox{deg}^2$}\\
    \hhline{~|---|---|---}
    & ALF2 & APR4 & DD2 & ALF2 & APR4 & DD2 & ALF2 & APR4 & DD2 \\
    \hhline{-|-|-|-|-|-|-|-|-|-}
    HLVKI\texttt{+} & $0$ & $0$ & $0$ & $3^{+5}_{-2}$ & $0$ & $6^{+12}_{-4}$ & $32^{+42}_{-27}$ & $0$ & $59^{+92}_{-51}$ \\
    VK\texttt{+}HLIv & $0$ & $0$ & $0$ & $16^{+21}_{-12}$ & $0$ & $29^{+43}_{-24}$ & $93^{+222}_{-76}$ & $0$ & $196^{+435}_{-163}$ \\
    HLKI\texttt{+}E & $0$ & $0$ & $0^{+2}_{-0}$ & $22^{+42}_{-17}$ & $0$ & $44^{+84}_{-36}$ & $122^{+275}_{-99}$ & $0$ & $260^{+542}_{-213}$ \\
    VKI\texttt{+}C & $0$ & $0$ & $0$ & $7^{+10}_{-5}$ & $0$ & $14^{+21}_{-11}$ & $115^{+262}_{-92}$ & $0$ & $243^{+511}_{-197}$ \\
    KI\texttt{+}EC & $7^{+9}_{-5}$ & $0$ & $12^{+18}_{-9}$ & $79^{+173}_{-64}$ & $0$ & $160^{+330}_{-132}$ & $124^{+281}_{-100}$ & $0$ & $267^{+555}_{-218}$ \\
    ECS & $50^{+94}_{-41}$ & $0$ & $100^{+182}_{-83}$ & $115^{+262}_{-92}$ & $0$ & $247^{+512}_{-202}$ & $124^{+281}_{-100}$ & $0$ & $267^{+555}_{-218}$ \\
    \hhline{----------}
    \end{tabular}
\end{table*}

For the bright siren study, we generate KN light curves for all NSBH events situated at $z \leq 0.5$. These KN light curves were generated using numerical recipes from \citet{Kruger:2020gig} and \citet{Raaijmakers:2021slr}. We select the events from the population with $q\leq4$ and $0\leq\chi_{BH}\leq0.75$, that are detected by each of the six networks with SNR $ \geq 10$. The restrictions on the masses and the spins of the parameters are due to the range of validity of the fits used. The remnant mass, i.e., the estimated baryon mass outside the BH approximately $10$ ms after the merger, can be calculated using the formula in \citet{Foucart:2018rjc}. This gives the \textit{normalized} remnant mass which, when multiplied by the baryonic mass of the initial NS, gives the remnant mass outside the BH. The mass of the dynamical ejecta is obtained using the fits from \citet{Kruger:2020gig}. The mass of the disk surrounding the BH is calculated by subtracting the mass of the dynamical ejecta from the remnant mass. From the obtained disk mass, a fraction can become gravitationally unbound, referred to as disk wind, and is computed using \citep{Raaijmakers:2021slr}
\begin{equation}
    \xi = \frac{M_{ej}}{M_{disk}} = \xi_1 + \frac{\xi_2-\xi_1}{1+e^{1.5(q-3)}},
\end{equation}
where $\xi_1 \in (0.04,0.32)$ and $\xi_2 \in (0.04,0.14)$. We set $\xi_1$ and $\xi_2$ to the average values of the upper and lower bounds used in \citet{Raaijmakers:2021slr}, i.e., $\xi_1 = 0.18$ and $\xi_2 = 0.29$, respectively. The velocity of the dynamical ejecta is approximated using \citet{Raaijmakers:2021slr} and the velocity of the disk wind is set to $0.1c$ \citep{De:2020jdt,Siegel:2017nub}. The opacities for the dynamical ejecta lie in the range $(1$--$10)$ cm$^2$ g$^{-1}$ due to the Lanthanide rich r-process nucleosynthesis. The disk, on the other hand, becomes relatively optically thin after interactions with neutrinos, with opacity in the range $(0.1$--$1)$ cm$^2$ g$^{-1}$. For this work, we fix dynamical and disk matter opacity to $8$ and $0.5$, respectively. The luminosity curves for both the dynamical ejecta and the unbound disk mass are determined by integrating the heating function (which accounts for the heating due to $\beta-$decay) implemented by using the numerical fit from \citet{Korobkin:2012uy}. The total bolometric luminosity curve for the system is obtained by adding the dynamical ejecta and the unbound disk mass at each time. To obtain the bandwise luminosity curves from bolometric ones, we use Planck's law to calculate the spectral flux density $f_{\nu}$, which is converted to AB magnitude (AB mag) using
\begin{equation}
    m_{\mathrm{AB}} = -2.5\,\mbox{log}_{10}\,f_{\nu}\,-\,48.6.
\end{equation}
The BNS counterpart of this KN model has been used and compared with other models in \citet{Villar:2017wcc, Kashyap:2019ypm,Wu:2021ibi}. The model used in this study uses the same heating function and light curve model as these works. The only difference comes in the amount of ejected mass and velocities of the ejecta, which is different in the NSBH case compared to the BNS case, and has been chosen specifically for NSBH mergers from numerical fits as discussed above.

Once the KN light curves are evaluated, we claim the detection of a KN if the peak luminosity of the light curve is brighter than the limiting magnitude of a particular filter for a given telescope. 
Using this criterion, we mention the number of KN detections associated with Pop-1 and Pop-2 GW events in Table \ref{apptab:r_R_kn_numbers}. We only list the number of detections for the $r-$filter of the Rubin observatory and the $R-$filter of the Roman telescope as we obtain the most number of detections for these two filters (see Tables XIV and XV in \citet{Gupta:2023evt} for the number of KN detections for each filter for the two populations).

The numbers of detections in Table \ref{apptab:r_R_kn_numbers} are categorized with respect to the $90\%-$credible sky area of the associated GW event. This is because the GW networks supply the telescopes with the sky position of the event to allow the telescopes to slew in position and detect the EM counterpart. At a time, an EM telescope can only observe the sky-area that is equivalent to its FOV. Many EM telescopes can also slew to cover multiple patches in the sky, observing an area much larger than their FOV. Consequently, in Table \ref{apptab:r_R_kn_numbers}, we categorize the number of KN detections according to if $\Omega_{90} < $ FOV and $\Omega_{90}<10\,\times\,$FOV for a given telescope. As Rubin has a large FOV, it will not need to slew to observe most of the KN it detects. From a pragmatic viewpoint, this increases the chances that Rubin will be able to detect these KN. On the other hand, the FOV of Roman is $\sim 30$ times smaller than that of Rubin. Consequently, Roman will need to cover $\mathcal{O}(10)$ sky patches to observe the KN. Targeted follow-ups to GW observations with both telescopes are required to fully realize the multimessenger prospects of NSBH mergers. Another categorization in Table \ref{apptab:r_R_kn_numbers} is with respect to the EOS. We have considered three NS EOS for this study--- \texttt{ALF2}, \texttt{APR4} and \texttt{DD2}. Among the three, \texttt{APR4} is the softest EOS and leads to more compact NS, whereas \texttt{DD2} is the stiffest EOS and leads to less compact NS. Consequently, we see that most KN detections are recorded corresponding to the \texttt{DD2} EOS, whereas the least number of detections are obtained with \texttt{APR4}.  

From Table \ref{apptab:r_R_kn_numbers}, we see that the number of KN observations following NSBH GW detections can range between $0-\mathcal{O}(10)$ every year. Comparing this to $\sim 40,000$ estimated NSBH mergers in the Universe (see Fig. \ref{fig:eff_rate}), we only expect to detect KN for less than $0.1\%$ of the NSBH mergers. Even within the redshift of $z\leq0.5$, with the $\sim 4,000$ NSBH mergers that are expected to take place, less than $1\%$ of the KN are expected to be detected. These estimates are consistent with other works in literature that have looked at the probability of the generation of KN as a result of NSBH mergers \citep{Zhu:2020ffa,Fragione:2021cvv,Biscoveanu:2022iue}.

%%%%%%%%%%%%%%%%%%%%%%%%%%%%%%%%%%%%%%%%%%%%%%%%%%

\section{Measurement errors for $H_0$ using gray sirens with Rubin and Roman} \label{appsec:gray_sirens}
\begin{figure*}
\begin{subfigure}{\linewidth}
\centering
  \includegraphics[scale=0.6,trim = 0 15 0 0,clip]{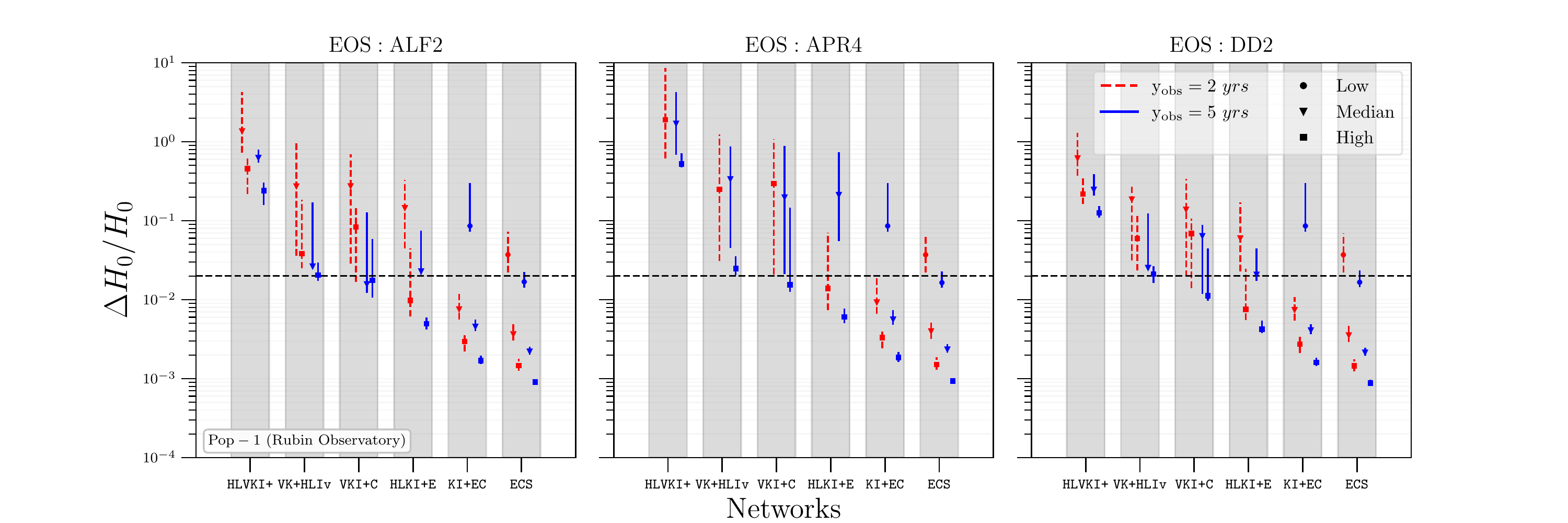}
\end{subfigure}\vspace{-1em}
\begin{subfigure}{\linewidth}
  \centering
  \includegraphics[scale=0.6,trim = 0 15 0 0,clip]{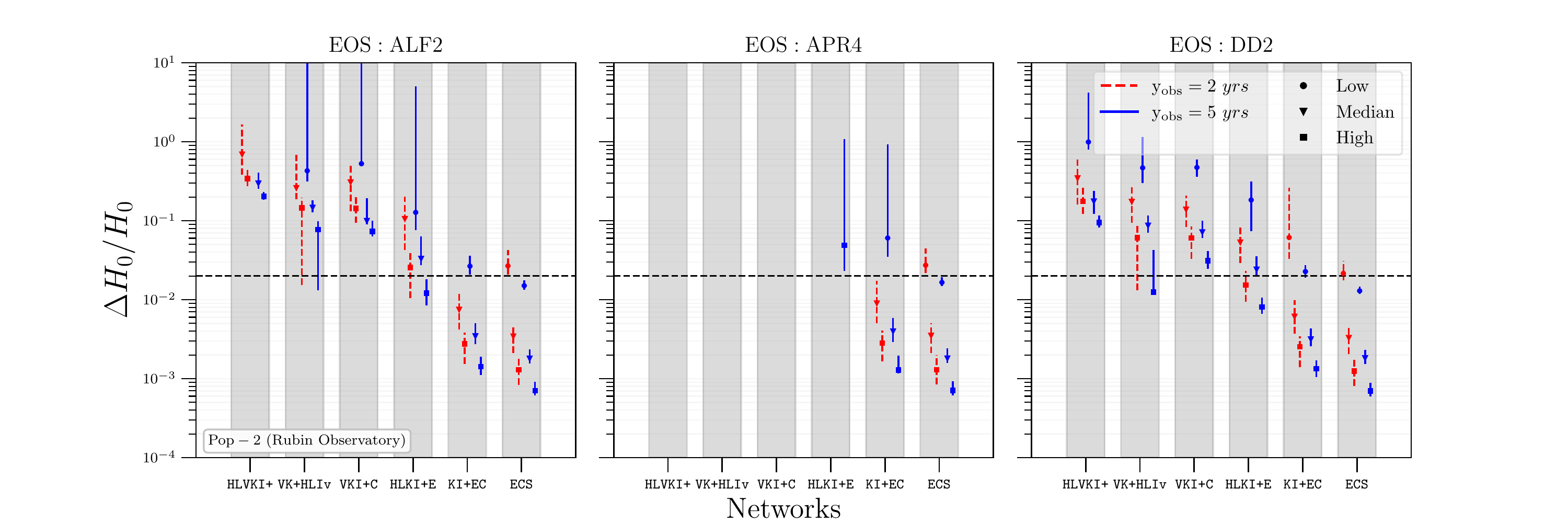}
\end{subfigure}\vspace{-1em}
\begin{subfigure}{\linewidth}
\centering
  \includegraphics[scale=0.6,trim = 0 15 0 0,clip]{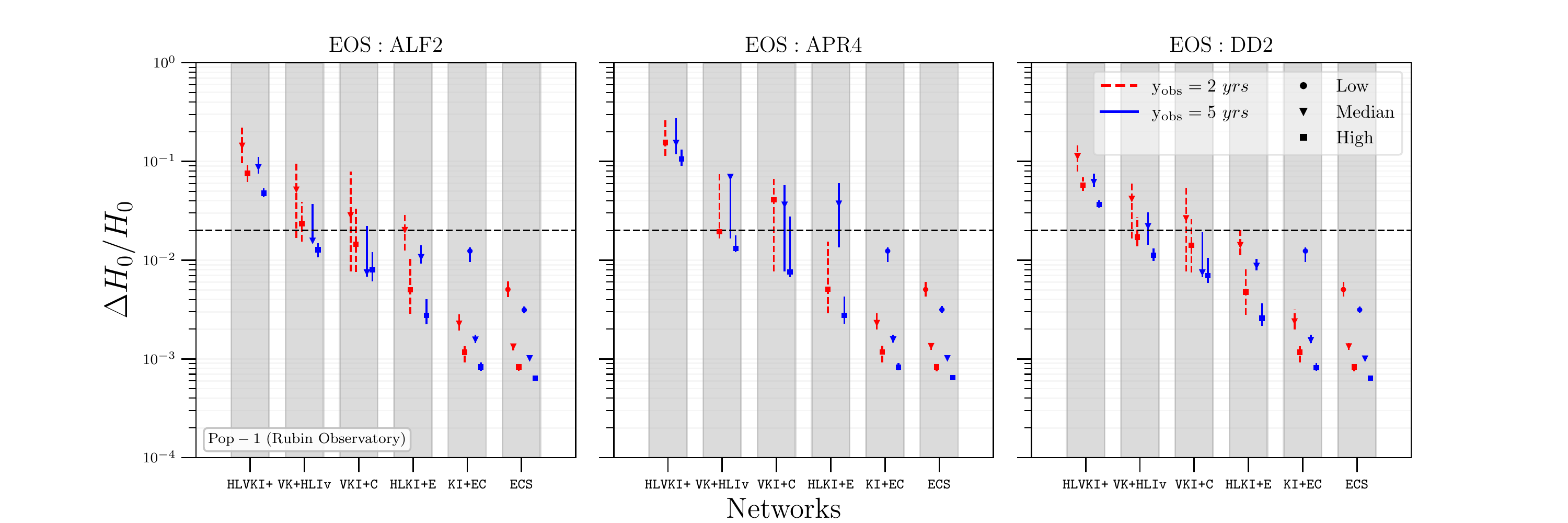}
\end{subfigure}\vspace{-1em}
\begin{subfigure}{\linewidth}
  \centering
  \includegraphics[scale=0.6]{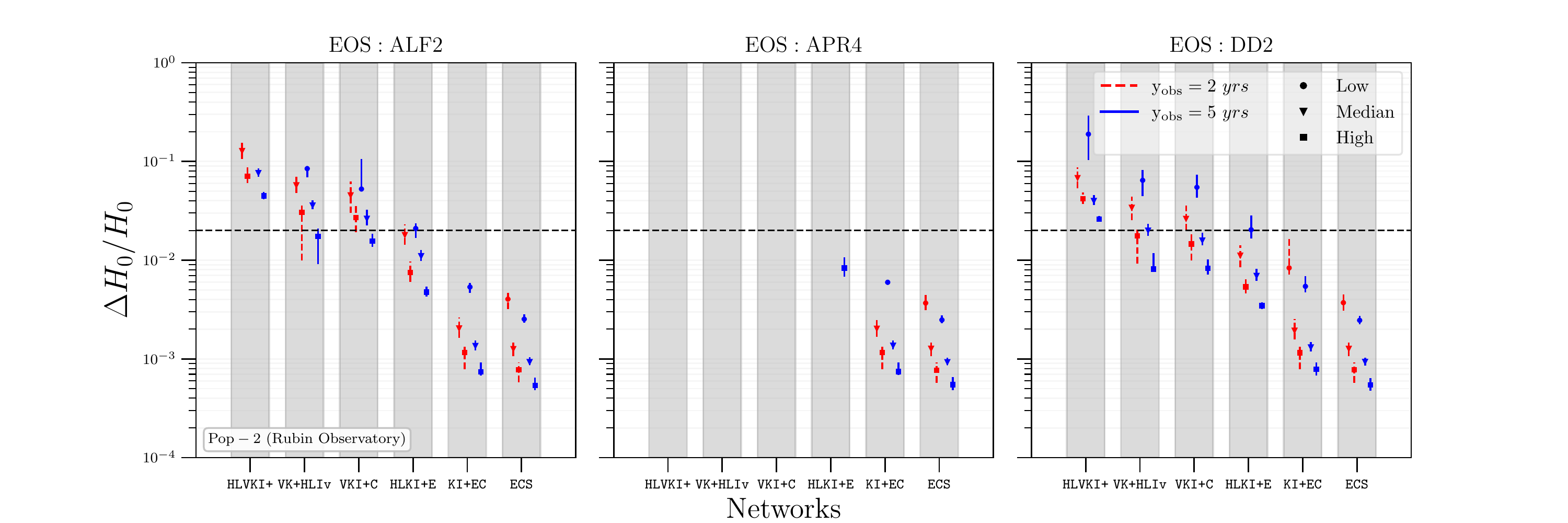}
\end{subfigure}\vspace{0em}
\caption{\label{appfig:gs_Rubin} The measurement accuracy of $H_0$ using NSBH systems as golden gray sirens where the KN are detected using $r-$filter of the Rubin telescope. The top two panels show the errors for events in Pop-1 and Pop-2 when prior on $\Omega_m$ is not included, and the bottom two panels show the errors when prior on $\Omega_m$ is included.}
\end{figure*}

\begin{figure*}
\begin{subfigure}{\linewidth}
  \centering
  \includegraphics[scale=0.6,trim = 0 15 0 0,clip]{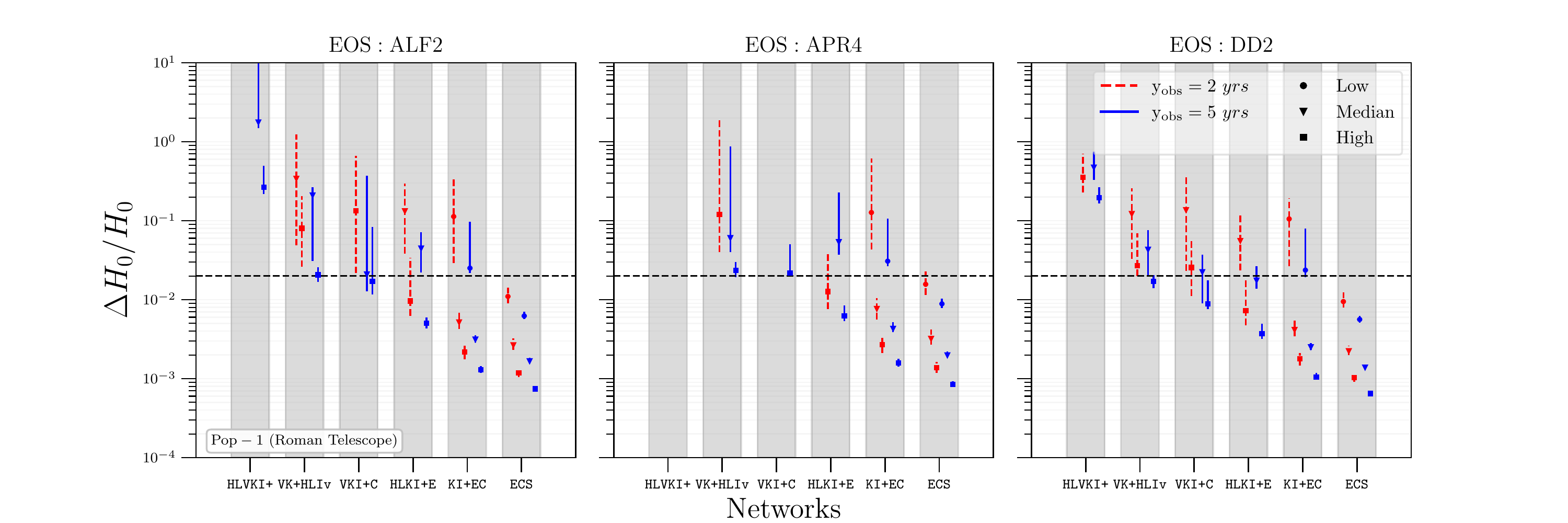}
\end{subfigure}\vspace{-1em}
\begin{subfigure}{\linewidth}
  \centering
  \includegraphics[scale=0.6,trim = 0 15 0 0,clip]{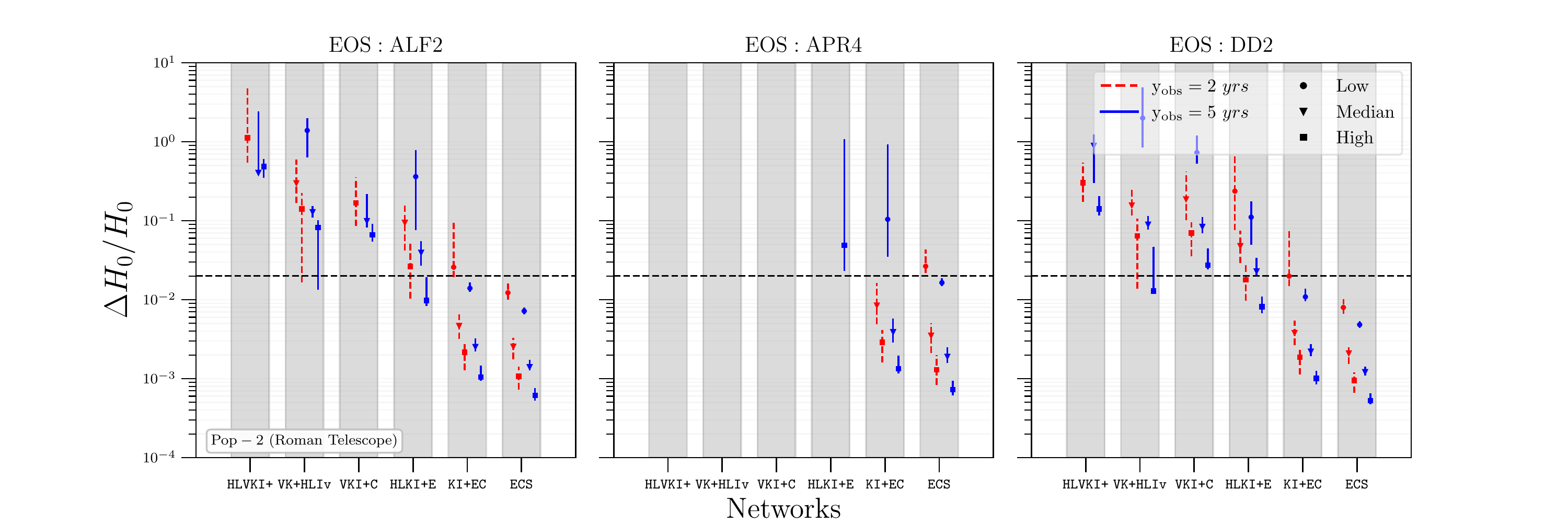}
\end{subfigure}\vspace{-0.5em}
\begin{subfigure}{\linewidth}
  \centering
  \includegraphics[scale=0.6,trim = 0 15 0 0,clip]{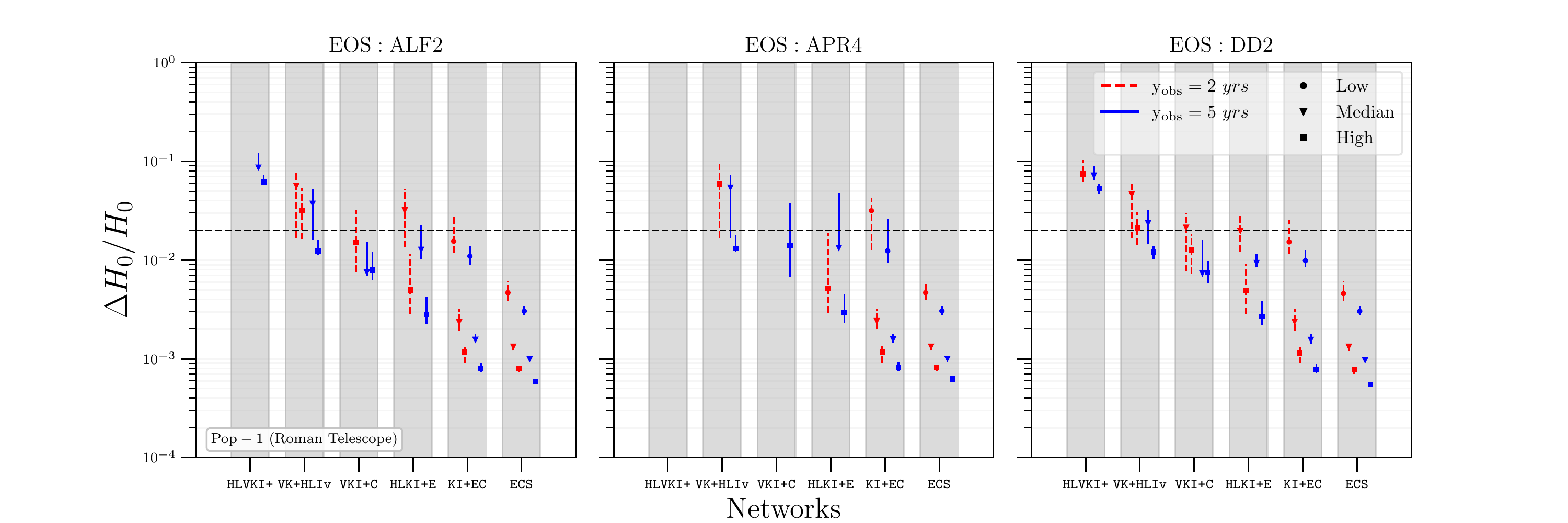}
\end{subfigure}\vspace{-1em}
\begin{subfigure}{\linewidth}
  \centering
  \includegraphics[scale=0.6]{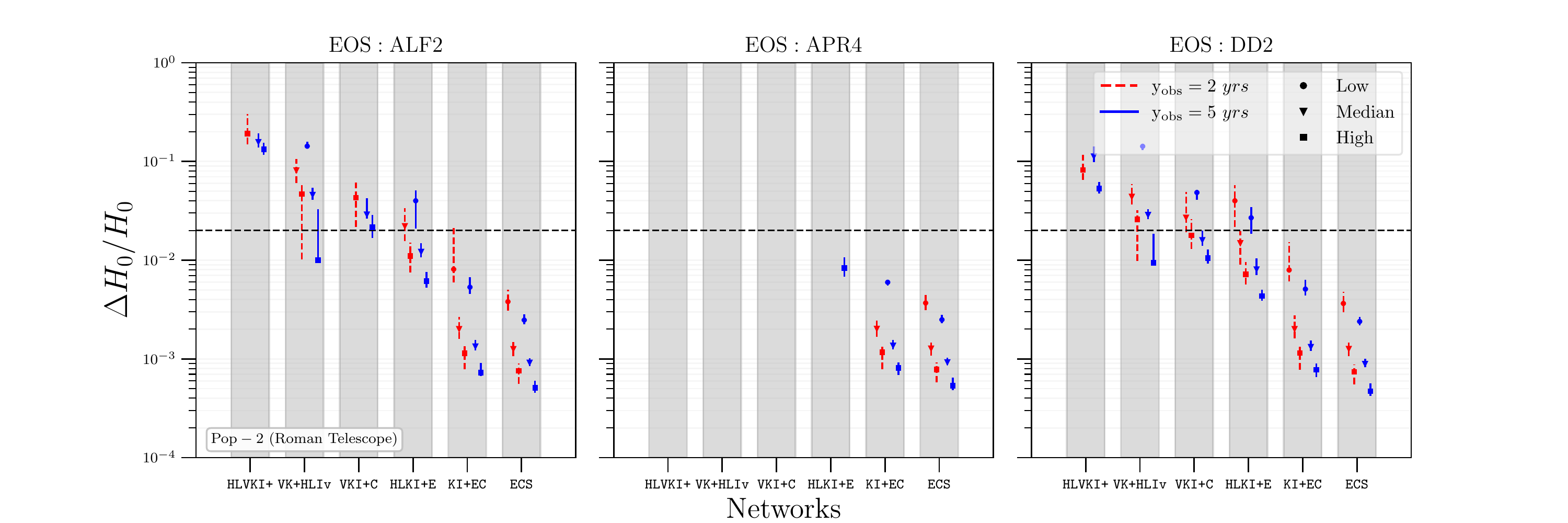}
\end{subfigure}\vspace{0em}
\caption{\label{appfig:gs_Roman} The fractional errors in $H_0$ measurement using NSBH systems as gray sirens where only those NSBH bright sirens were included that were observed using the $R-$filter of the Roman telescope. The top two panels show the measurement accuracy when prior on $\Omega_m$ is not included, and the bottom two panels show the measurement accuracy when the prior on $\Omega_m$ is included.}
\end{figure*}

In this section, we report the precision in $H_0$ measurements using the gray siren approach when the $r-$filter of Rubin and the $R-$filter of Roman are used. The estimates for three different NS EOS for Rubin and Roman are given in Figs. \ref{appfig:gs_Rubin} and \ref{appfig:gs_Roman}, respectively. The corresponding number of NSBH events detected in an observation span of 10 years are mentioned in Table \ref{tab:gs_numbers}.

The general conclusions regarding the capabilities of different detector networks toward the resolution of \HL tension are similar to the ones discussed in Section \ref{sec:gray_sirens} for the $g+i$ TOO strategy. In particular, the inclusion of the prior on $\Omega_m$ has a considerable effect on the measurement of $H_0$, especially on the less sensitive networks. For e.g., Voyager is unable to measure $H_0$ to a precision that is better than $2\%$ with Pop-2 events without using a prior on $\Omega_m$, but it is able to resolve the tension when the prior is included. For the less sensitive networks, the estimates on $H_0$ precision using the gray siren approach rely heavily on the bright siren measurements (due to a low number of golden dark siren measurements, c.f. Table \ref{tab:dark_sirens}). As the number of bright siren detections with $g+i$ is greater than those from $r-$filter of Rubin and $R-$filter of Roman (c.f. Tables \ref{tab:bs_golden} and \ref{fig:bs_too}), the estimates in Figs. \ref{appfig:gs_Rubin} and \ref{appfig:gs_Roman} are slightly worse than those for the $g+i$ TOO strategy, presented in Figs. \ref{fig:gs_Om0} and \ref{fig:gs_Om0_rest}. However, as mentioned in Section \ref{sec:gray_sirens}, for \texttt{KI+EC} and \texttt{ECS}, as the golden dark siren measurements dominate the $H_0$ precision, the estimates for all three bright siren approaches are similar. As an additional consequence,  here too we see that the effect of the NS EOS on $H_0$ measurement is nullified for \texttt{KI+EC} and \texttt{ECS}, and is only evident in the other, less sensitive, detector networks.

% Don't change these lines
\bsp	% typesetting comment
\label{lastpage}
\end{document}